\input harvmac
\input amssym
\let\includefigures=\iftrue
%
%
%
\newfam\black
\input rotate
\input epsf
\noblackbox
\includefigures
\message{If you do not have epsf.tex (to include figures),} \message{change the
option at the top of the tex file.}
\def\figin{\epsfcheck\figin}\def\figins{\epsfcheck\figins}
\def\epsfcheck{\ifx\epsfbox\UnDeFiNeD
\message{(NO epsf.tex, FIGURES WILL BE IGNORED)}
\gdef\figin##1{\vskip2in}\gdef\figins##1{\hskip.5in}
\else\message{(FIGURES WILL BE INCLUDED)}%
\gdef\figin##1{##1}\gdef\figins##1{##1}\fi}
\def\DefWarn#1{}

\def\figinsert{\goodbreak\midinsert}
\def\ifig#1#2#3{\DefWarn#1\xdef#1{fig.~\the\figno}
\writedef{#1\leftbracket fig.\noexpand~\the\figno}%
\figinsert\figin{\centerline{#3}}\medskip\centerline{\vbox{\baselineskip12pt
\advance\hsize by -1truein\noindent\footnotefont{\bf Fig.~\the\figno:} #2}}
\bigskip\endinsert\global\advance\figno by1}
\else
\def\ifig#1#2#3{\xdef#1{fig.~\the\figno}
\writedef{#1\leftbracket fig.\noexpand~\the\figno}%
\global\advance\figno by1} \fi


\lref\LianGK{ B.~H.~Lian and G.~J.~Zuckerman, ``New Selection Rules And
Physical States In 2-D Gravity: Conformal Gauge,'' Phys.\ Lett.\ B {\bf 254},
417 (1991).
}
\lref\DouglasUP{ M.~R.~Douglas, I.~R.~Klebanov, D.~Kutasov, J.~Maldacena,
E.~Martinec and N.~Seiberg, ``A new hat for the $c = 1$ matrix model,''
arXiv:hep-th/0307195.
}
\lref\FujitsuAE{ A.~Fujitsu, ``Ope.Math: Operator Product Expansions In Free
Field Realizations Of Conformal Field Theory,'' Comput.\ Phys.\ Commun.\  {\bf
79}, 78 (1994).
}
\lref\BouwknegtVA{ P.~Bouwknegt, J.~G.~McCarthy and K.~Pilch, `BRST analysis of
physical states for 2-D (super)gravity coupled to (super)conformal matter,''
arXiv:hep-th/9110031.
}
\lref\WittenZD{ E.~Witten, ``Ground ring of two-dimensional string theory,''
Nucl.\ Phys.\ B {\bf 373}, 187 (1992) [arXiv:hep-th/9108004].
}
\lref\DornSV{ H.~Dorn and H.~J.~Otto, ``Some conclusions for noncritical string
theory drawn from two and three point functions in the Liouville sector,''
arXiv:hep-th/9501019.
}
\lref\ZamolodchikovAA{ A.~B.~Zamolodchikov and A.~B.~Zamolodchikov, ``Structure
constants and conformal bootstrap in Liouville field theory,'' Nucl.\ Phys.\ B
{\bf 477}, 577 (1996) [arXiv:hep-th/9506136].
}
\lref\KutasovQX{ D.~Kutasov, E.~J.~Martinec and N.~Seiberg, ``Ground rings and
their modules in 2-D gravity with $c \le 1$ matter,'' Phys.\ Lett.\ B {\bf
276}, 437 (1992) [arXiv:hep-th/9111048].
}

\lref\PeriwalGF{ V.~Periwal and D.~Shevitz, ``Unitary Matrix
Models As Exactly Solvable String Theories,'' Phys.\ Rev.\ Lett.\
{\bf 64}, 1326 (1990).
}

\lref\GrossHE{ D.~J.~Gross and E.~Witten, ``Possible Third Order
Phase Transition In The Large N Lattice Gauge Theory,'' Phys.\
Rev.\ D {\bf 21}, 446 (1980).
}

\lref\CrnkovicWD{ C.~Crnkovic, M.~R.~Douglas and G.~W.~Moore,
``Loop equations and the topological phase of multi-cut matrix
models,'' Int.\ J.\ Mod.\ Phys.\ A {\bf 7}, 7693 (1992)
[arXiv:hep-th/9108014].
}

\lref\JohnsonHY{ C.~V.~Johnson, ``Non-perturbative string
equations for type 0A,'' arXiv:hep-th/0311129.
}

\lref\SeibergEB{ N.~Seiberg, ``Notes On Quantum Liouville Theory And Quantum
Gravity,'' Prog.\ Theor.\ Phys.\ Suppl.\  {\bf 102}, 319 (1990).
}
\lref\DiFrancescoNW{ P.~Di Francesco, P.~Ginsparg and J.~Zinn-Justin, ``2-D
Gravity and random matrices,'' Phys.\ Rept.\  {\bf 254}, 1 (1995)
[arXiv:hep-th/9306153].
}
\lref\KlebanovWG{ I.~R.~Klebanov, J.~Maldacena and N.~Seiberg, ``Unitary and
complex matrix models as 1-d type 0 strings,'' arXiv:hep-th/0309168.
}
\lref\ZamolodchikovAH{ A.~B.~Zamolodchikov and A.~B.~Zamolodchikov, ``Liouville
field theory on a pseudosphere,'' arXiv:hep-th/0101152.
}
\lref\TeschnerRV{ J.~Teschner, ``Liouville theory revisited,'' Class.\ Quant.\
Grav.\ {\bf 18}, R153 (2001) [arXiv:hep-th/0104158].
}
\lref\MooreIR{ G.~W.~Moore, N.~Seiberg and M.~Staudacher, ``From loops to
states in 2-D quantum gravity,'' Nucl.\ Phys.\ B {\bf 362}, 665 (1991).
}
\lref\FateevIK{ V.~Fateev, A.~B.~Zamolodchikov and A.~B.~Zamolodchikov,
``Boundary Liouville field theory. I: Boundary state and boundary  two-point
function,'' arXiv:hep-th/0001012.
}
\lref\TeschnerMD{ J.~Teschner, ``Remarks on Liouville theory with boundary,''
arXiv:hep-th/0009138.
}
\lref\FriedanGE{ D.~Friedan, E.~J.~Martinec and S.~H.~Shenker, ``Conformal
Invariance, Supersymmetry And String Theory,'' Nucl.\ Phys.\ B {\bf 271}, 93
(1986).
}
\lref\FriedanRX{ D.~Friedan, ``Notes On String Theory And Two-Dimensional
Conformal Field Theory,'' EFI-85-99-CHICAGO
{\it To appear in Proc. of Workshop on Unified String Theories, Santa Barbara,
CA, Jul 29 - Aug 16, 1985}
}
\lref\LustTJ{ D.~Lust and S.~Theisen, ``Lectures On String Theory,'' Lect.\
Notes Phys.\ {\bf 346}, 1 (1989).
}
\lref\BouwknegtAM{ P.~Bouwknegt, J.~G.~McCarthy and K.~Pilch, ``Ground ring for
the 2-D NSR string,'' Nucl.\ Phys.\ B {\bf 377}, 541 (1992)
[arXiv:hep-th/9112036].
}
\lref\ItohIY{ K.~Itoh and N.~Ohta, ``BRST cohomology and physical states in 2-D
supergravity coupled to $c \le 1$ matter,'' Nucl.\ Phys.\ B {\bf 377}, 113
(1992) [arXiv:hep-th/9110013].
}
\lref\RashkovJX{ R.~C.~Rashkov and M.~Stanishkov, ``Three-point correlation
functions in N=1 Super Lioville Theory,'' Phys.\ Lett.\ B {\bf 380}, 49 (1996)
[arXiv:hep-th/9602148].
}
\lref\PoghosianDW{ R.~H.~Poghosian, ``Structure constants in the N = 1
super-Liouville field theory,'' Nucl.\ Phys.\ B {\bf 496}, 451 (1997)
[arXiv:hep-th/9607120].
}
\lref\FukudaBV{ T.~Fukuda and K.~Hosomichi, ``Super Liouville theory with
boundary,'' Nucl.\ Phys.\ B {\bf 635}, 215 (2002) [arXiv:hep-th/0202032].
}
\lref\CrnkovicMR{ C.~Crnkovic and G.~W.~Moore, ``Multicritical Multicut Matrix
Models,'' Phys.\ Lett.\ B {\bf 257}, 322 (1991).
}
\lref\CrnkovicMS{ C.~Crnkovic, M.~R.~Douglas and G.~W.~Moore, ``Physical
Solutions For Unitary Matrix Models,'' Nucl.\ Phys.\ B {\bf 360}, 507 (1991).
}
\lref\CrnkovicWD{ C.~Crnkovic, M.~R.~Douglas and G.~W.~Moore, ``Loop equations
and the topological phase of multi-cut matrix models,'' Int.\ J.\ Mod.\ Phys.\
A {\bf 7}, 7693 (1992) [arXiv:hep-th/9108014].
}
\lref\PandaGE{ S.~Panda and S.~Roy, ``BRST cohomology ring in $c(M) < 1$ NSR
string theory,'' Phys.\ Lett.\ B {\bf 358}, 229 (1995) [arXiv:hep-th/9507054].
}
\lref\MartinecKA{ E.~J.~Martinec, ``The annular report on non-critical string
theory,'' arXiv:hep-th/0305148.
}
\lref\KostovCG{ I.~K.~Kostov, ``Strings with discrete target space,'' Nucl.\
Phys.\ B {\bf 376}, 539 (1992) [arXiv:hep-th/9112059].
}
\lref\KostovHN{ I.~K.~Kostov, ``Multiloop correlators for closed strings with
discrete target space,'' Phys.\ Lett.\ B {\bf 266}, 42 (1991).
}
\lref\KostovXT{ I.~K.~Kostov, ``Loop Amplitudes For Nonrational String
Theories,'' Phys.\ Lett.\ B {\bf 266}, 317 (1991).
}

\lref\DijkgraafDH{ R.~Dijkgraaf and C.~Vafa, ``A perturbative
window into non-perturbative physics,'' arXiv:hep-th/0208048.
}

\lref\CachazoRY{ F.~Cachazo, M.~R.~Douglas, N.~Seiberg and
E.~Witten, ``Chiral rings and anomalies in supersymmetric gauge
theory,'' JHEP {\bf 0212}, 071 (2002) [arXiv:hep-th/0211170].
}

\lref\DiFrancescoXZ{ P.~Di Francesco, H.~Saleur and J.~B.~Zuber, ``Generalized
Coulomb Gas Formalism For Two-Dimensional Critical Models Based On SU(2) Coset
Construction,'' Nucl.\ Phys.\ B {\bf 300}, 393 (1988).
}
\lref\AlexandrovUN{ S.~Alexandrov, ``$(m,n)$ ZZ branes and the $c = 1$ matrix
model,'' arXiv:hep-th/0310135.
}
\lref\KazakovYH{ V.~A.~Kazakov and A.~Marshakov, ``Complex curve of the two
matrix model and its tau-function,'' J.\ Phys.\ A {\bf 36}, 3107 (2003)
[arXiv:hep-th/0211236].
}
\lref\ChaichianYZ{ M.~Chaichian and P.~P.~Kulish, ``On The Method Of Inverse
Scattering Problem And Backlund Transformations For Supersymmetric Equations,''
Phys.\ Lett.\ B {\bf 78}, 413 (1978).
}
\lref\DHokerZY{ E.~D'Hoker, ``Classical And Quantal Supersymmetric Liouville
Theory,'' Phys.\ Rev.\ D {\bf 28}, 1346 (1983).
}
\lref\AhnEV{ C.~Ahn, C.~Rim and M.~Stanishkov, ``Exact one-point function of
N=1 super-Liouville theory with boundary,'' Nucl.\ Phys.\ B {\bf 636}, 497
(2002) [arXiv:hep-th/0202043].
}
\lref\DiFrancescoNK{ P.~Di Francesco, P.~Mathieu and D.~Senechal, ``Conformal
Field Theory,''
}
\lref\DiFrancescoJD{ P.~Di Francesco and D.~Kutasov, ``Unitary Minimal Models
Coupled To 2-D Quantum Gravity,'' Nucl.\ Phys.\ B {\bf 342}, 589 (1990).
}
\lref\GoulianQR{ M.~Goulian and M.~Li, ``Correlation Functions In Liouville
Theory,'' Phys.\ Rev.\ Lett.\  {\bf 66}, 2051 (1991).
}
\lref\DiFrancescoUD{ P.~Di Francesco and D.~Kutasov, ``World sheet and
space-time physics in two-dimensional (Super)string theory,'' Nucl.\ Phys.\ B
{\bf 375}, 119 (1992) [arXiv:hep-th/9109005].
}
\lref\DiFrancescoZK{ P.~Di Francesco and J.~B.~Zuber, ``Fusion potentials. 1,''
J.\ Phys.\ A {\bf 26}, 1441 (1993) [arXiv:hep-th/9211138].
}
\lref\ImbimboXT{ C.~Imbimbo, ``New modular representations and
fusion algebras from quantized SL(2,R)
arXiv:hep-th/9301031.
}
\lref\ImbimboFW{ C.~Imbimbo, ``Polynomial Fusion Rings From
Sl(2,R) Chern-Simons Theory,'' Phys.\ Lett.\ B {\bf 308}, 51
(1993).
}
\lref\MartinecHT{ E.~J.~Martinec, G.~W.~Moore and N.~Seiberg, ``Boundary
operators in 2-D gravity,'' Phys.\ Lett.\ B {\bf 263}, 190 (1991).
}

\lref\DavidTX{ F.~David, ``Planar Diagrams, Two-Dimensional Lattice Gravity And
Surface Models,'' Nucl.\ Phys.\ B {\bf 257}, 45 (1985).
}

\lref\KazakovDS{ V.~A.~Kazakov, ``Bilocal Regularization Of Models Of Random
Surfaces,'' Phys.\ Lett.\ B {\bf 150}, 282 (1985).
}

\lref\MooreMG{ G.~W.~Moore, ``Geometry Of The String Equations,'' Commun.\
Math.\ Phys.\  {\bf 133}, 261 (1990).
}

\lref\MooreCN{ G.~W.~Moore, ``Matrix Models Of 2-D Gravity And Isomonodromic
Deformation,'' Prog.\ Theor.\ Phys.\ Suppl.\  {\bf 102}, 255 (1990).
}

\lref\KazakovEA{ V.~A.~Kazakov, A.~A.~Migdal and I.~K.~Kostov, ``Critical
Properties Of Randomly Triangulated Planar Random Surfaces,'' Phys.\ Lett.\ B
{\bf 157}, 295 (1985).
}

\lref\AmbjornAZ{ J.~Ambjorn, B.~Durhuus and J.~Frohlich, ``Diseases Of
Triangulated Random Surface Models, And Possible Cures,'' Nucl.\ Phys.\ B {\bf
257}, 433 (1985).
}

\lref\DouglasVE{ M.~R.~Douglas and S.~H.~Shenker, ``Strings In Less Than
One-Dimension,'' Nucl.\ Phys.\ B {\bf 335}, 635 (1990).
}

\lref\GrossVS{ D.~J.~Gross and A.~A.~Migdal, ``Nonperturbative Two-Dimensional
Quantum Gravity,'' Phys.\ Rev.\ Lett.\  {\bf 64}, 127 (1990).
}

\lref\BrezinRB{ E.~Brezin and V.~A.~Kazakov, ``Exactly Solvable Field Theories
Of Closed Strings,'' Phys.\ Lett.\ B {\bf 236}, 144 (1990).
}

\lref\DouglasDD{ M.~R.~Douglas, ``Strings In Less Than One-Dimension And The
Generalized K-D-V Hierarchies,'' Phys.\ Lett.\ B {\bf 238}, 176 (1990).
}

\lref\KostovHN{ I.~K.~Kostov, ``Multiloop correlators for closed strings with
discrete target space,'' Phys.\ Lett.\ B {\bf 266}, 42 (1991).
}

\lref\KostovGS{ I.~K.~Kostov, ``Loop space Hamiltonian for $c \le 1$ open
strings,'' Phys.\ Lett.\ B {\bf 349}, 284 (1995) [arXiv:hep-th/9501135].
}

\lref\KostovUQ{ I.~K.~Kostov, ``Boundary correlators in 2D quantum gravity:
Liouville versus discrete approach,'' Nucl.\ Phys.\ B {\bf 658}, 397 (2003)
[arXiv:hep-th/0212194].
}

\lref\DaulBG{ J.~M.~Daul, V.~A.~Kazakov and I.~K.~Kostov, ``Rational theories
of 2-D gravity from the two matrix model,'' Nucl.\ Phys.\ B {\bf 409}, 311
(1993) [arXiv:hep-th/9303093].
}

\lref\GinspargIS{ P.~Ginsparg and G.~W.~Moore, ``Lectures On 2-D Gravity And
2-D String Theory,'' arXiv:hep-th/9304011.
}

\lref\DiFrancescoNW{ P.~Di Francesco, P.~Ginsparg and J.~Zinn-Justin, ``2-D
Gravity and random matrices,'' Phys.\ Rept.\ {\bf 254}, 1 (1995)
[arXiv:hep-th/9306153].
}

\lref\McGreevyKB{ J.~McGreevy and H.~Verlinde, ``Strings from tachyons: The $c
= 1$ matrix reloaded,'' arXiv:hep-th/0304224.
}

\lref\MartinecKA{ E.~J.~Martinec, ``The annular report on non-critical string
theory,'' arXiv:hep-th/0305148.
}

\lref\KlebanovKM{ I.~R.~Klebanov, J.~Maldacena and N.~Seiberg, ``D-brane decay
in two-dimensional string theory,'' JHEP {\bf 0307}, 045 (2003)
[arXiv:hep-th/0305159].
}

\lref\McGreevyEP{ J.~McGreevy, J.~Teschner and H.~Verlinde, ``Classical and
quantum D-branes in 2D string theory,'' arXiv:hep-th/0305194.
}

\lref\MooreAG{ G.~W.~Moore and N.~Seiberg, ``From loops to fields in 2-D
quantum gravity,'' Int.\ J.\ Mod.\ Phys.\ A {\bf 7}, 2601 (1992).
}

\lref\AlexandrovNN{ S.~Y.~Alexandrov, V.~A.~Kazakov and D.~Kutasov,
``Non-perturbative effects in matrix models and D-branes,'' JHEP {\bf 0309},
057 (2003) [arXiv:hep-th/0306177].
}

\lref\KostovUH{ I.~K.~Kostov, B.~Ponsot and D.~Serban, ``Boundary Liouville
theory and 2D quantum gravity,'' arXiv:hep-th/0307189.
}
\lref\TakayanagiSM{ T.~Takayanagi and N.~Toumbas, ``A matrix model dual of type
0B string theory in two dimensions,'' JHEP {\bf 0307}, 064 (2003)
[arXiv:hep-th/0307083].
}
\lref\TeschnerYF{ J.~Teschner, ``On the Liouville three point function,''
Phys.\ Lett.\ B {\bf 363}, 65 (1995) [arXiv:hep-th/9507109].
}
\lref\AganagicQJ{ M.~Aganagic, R.~Dijkgraaf, A.~Klemm, M.~Marino and C.~Vafa,
``Topological Strings and Integrable Hierarchies,'' arXiv:hep-th/0312085.
}
\lref\BanksDF{ T.~Banks, M.~R.~Douglas, N.~Seiberg and S.~H.~Shenker,
``Microscopic And Macroscopic Loops In Nonperturbative Two-Dimensional
Gravity,'' Phys.\ Lett.\ B {\bf 238}, 279 (1990).
}
\lref\CardyIR{ J.~L.~Cardy, ``Boundary Conditions, Fusion Rules And The
Verlinde Formula,'' Nucl.\ Phys.\ B {\bf 324}, 581 (1989).
}
\lref\HosomichiXC{ K.~Hosomichi, ``Bulk-boundary propagator in Liouville theory
on a disc,'' JHEP {\bf 0111}, 044 (2001) [arXiv:hep-th/0108093].
}
\lref\PonsotSS{ B.~Ponsot, ``Liouville theory on the pseudosphere:
Bulk-boundary structure constant,'' arXiv:hep-th/0309211.
}
\lref\ItohIX{ K.~Itoh and N.~Ohta, ``Spectrum of two-dimensional
(super)gravity,'' Prog.\ Theor.\ Phys.\ Suppl.\  {\bf 110}, 97 (1992)
[arXiv:hep-th/9201034].
}
\lref\LianCB{ B.~H.~Lian, ``Semiinfinite Homology And 2-D Quantum Gravity,''
UMI-92-21392, Ph.D. Thesis, Yale University (1991).
}
\lref\LianAJ{ B.~H.~Lian and G.~J.~Zuckerman, ``Semiinfinite Homology And 2-D
Gravity. 1,'' Commun.\ Math.\ Phys.\  {\bf 145}, 561 (1992).
}

\def\mod{{\rm mod}}

\def\CP{{\cal P}}

\def\IL{\relax{\rm I\kern-.18em L}}
\def\IH{\relax{\rm I\kern-.18em H}}
\def\IR{\relax{\rm I\kern-.18em R}}
\def\IC{\relax\hbox{$\inbar\kern-.3em{\rm C}$}}
\def\IZ{\relax\ifmmode\mathchoice
{\hbox{\cmss Z\kern-.4em Z}}{\hbox{\cmss Z\kern-.4em Z}}
{\lower.9pt\hbox{\cmsss Z\kern-.4em Z}} {\lower1.2pt\hbox{\cmsss Z\kern-.4em
Z}}\else{\cmss Z\kern-.4em Z}\fi}
\def\CM {{\cal M}}

\def\CP {{\cal P }}
\def\CL {{\cal L}}

\def\CO {{\cal O}}

\def\CB {{\cal B}}

\def\CM {{\cal M}}

\def\CO {{\cal O}}

\def\CP {{\cal P }}

\def\Tr{{\rm Tr}}

\font\manual=manfnt \def\dbend{\lower3.5pt\hbox{\manual\char127}}

\def\IZ{\relax\ifmmode\mathchoice
{\hbox{\cmss Z\kern-.4em Z}}{\hbox{\cmss Z\kern-.4em Z}}
{\lower.9pt\hbox{\cmsss Z\kern-.4em Z}} {\lower1.2pt\hbox{\cmsss Z\kern-.4em
Z}}\else{\cmss Z\kern-.4em Z}\fi}

\def\lfm#1{\medskip\noindent\item{#1}}

\def\bar{\overline}

\def\rt2{\sqrt{2}}
\def\irt2{{1\over\sqrt{2}}}

\def\hat{\widehat}
\def\slashchar#1{\setbox0=\hbox{$#1$}           
   \dimen0=\wd0                                 
   \setbox1=\hbox{/} \dimen1=\wd1               
   \ifdim\dimen0>\dimen1                        
      \rlap{\hbox to \dimen0{\hfil/\hfil}}      
      #1                                        
   \else                                        
      \rlap{\hbox to \dimen1{\hfil$#1$\hfil}}   
      /                                         
   \fi}
\writedefs

%
%
%
\newbox\tmpbox\setbox\tmpbox\hbox{\abstractfont }
\Title{\vbox{\baselineskip12pt\hbox to\wd\tmpbox{\hss
}}\hbox{PUPT-2102}}
{\vbox{\centerline{Branes, Rings and Matrix
Models in }\smallskip\centerline{Minimal (Super)string Theory}}}
\smallskip
\centerline{ Nathan Seiberg$^1$ and David Shih$^2$}
\smallskip
\bigskip
\centerline{$^1$School of Natural Sciences, Institute for Advanced Study,
Princeton, NJ 08540 USA}
\medskip
\centerline{$^2$Department of Physics, Princeton University, Princeton, NJ
08544 USA}
\bigskip
\vskip 1cm
 \noindent
We study both bosonic and supersymmetric $(p,q)$ minimal models
coupled to Liouville theory using the ground ring and the various
branes of the theory. From the FZZT brane partition function,
there emerges a unified, geometric description of all these
theories in terms of an auxiliary Riemann surface $\CM_{p,q}$ and
the corresponding matrix model. In terms of this geometric
description, both the FZZT and ZZ branes correspond to line
integrals of a certain one-form on $\CM_{p,q}$. Moreover, we argue
that there are a finite number of distinct $(m,n)$ ZZ branes, and
we show that these ZZ branes are located at the singularities of
$\CM_{p,q}$. Finally, we discuss the possibility that the bosonic
and supersymmetric theories with $(p,q)$ odd and relatively prime
are identical, as is suggested by the unified treatment of these
models.

\Date{December, 2003}

\newsec{Introduction and conclusions}

In this work we will explore minimal string theories.  These are
simple examples of string theory with a small number of
observables. Their simplicity makes them soluble and therefore
they are interesting laboratories for string dynamics.

The worldsheet description of these minimal string theories is based on the
$(p,q)$ minimal conformal field theories coupled to two-dimensional gravity
(Liouville theory), or the $(p,q)$ minimal superconformal field theories
coupled to two-dimensional supergravity (super-Liouville theory). Even though
these two worldsheet descriptions appear different, we find that actually they
are quite similar.  In fact, our final answer depends essentially only on the
values of $(p,q)$. This suggests a uniform presentation of all these theories
which encompasses the two different worldsheet frameworks and extends them.

These theories were first solved using their description in terms of matrix
models \refs{\DavidTX\KazakovDS\KazakovEA\AmbjornAZ\DouglasVE\GrossVS
\BrezinRB-\DouglasDD} (for reviews, see e.g.
\refs{\GinspargIS,\DiFrancescoNW}). The matrix model realizes the important
theme of open/closed string duality in the study of string theory. Recent
advances in the study of Liouville theory
\refs{\DornSV\TeschnerYF-\ZamolodchikovAA} and its D-branes
\refs{\FateevIK\TeschnerRV-\ZamolodchikovAH} has led to progress by
\refs{\McGreevyKB\MartinecKA\KlebanovKM\McGreevyEP\AlexandrovNN
\KostovUH\DouglasUP-\KlebanovWG} and others in making the connection between
the matrix model and the worldsheet description more explicit. Much of the
matrix model treatment of these theories has involved a deformation by the
lowest dimension operator. In order to match with the worldsheet description
based on Liouville theory, we should instead tune the background such that only
the cosmological constant $\mu$ is turned on.  In \MooreIR\ such backgrounds
were referred to as conformal backgrounds.

In the first part of this work, we will examine the simplest examples of
minimal string theory, which are based on the bosonic $(p,q)$ minimal models
coupled to two-dimensional gravity. These theories, which we review in section
2, exist for all relatively prime integers $(p,q)$. It is important that these
theories have only a finite number of standard physical vertex operators
$\CT_{r,s}$ of ghost number one; we will refer to these as tachyons. They are
constructed by ``dressing" the minimal model primaries $\CO_{r,s}$ with
Liouville exponentials, subject to the condition that the combinations have
conformal dimension one. They are labelled by integers $r=1,\dots,p-1$ and
$s=1,\dots,q-1$, and they are subject to the identification $\CT_{r,s} \sim
\CT_{p-r,q-s}$, which can be implemented by restricting to $qr>ps$.

In addition to the tachyons, there are infinitely many physical operators at
other values of the ghost number \LianGK.  Of special importance are the
operators at ghost number zero. These form a ring \WittenZD\ under
multiplication by the operator product expansion modulo BRST commutators. We
will argue that the ring is generated by two operators $\hat{\CO}_{1,2}$ and
$\hat{\CO}_{2,1}$. In terms of the operators
\eqn\grrescdef{
    \hat{x}={1\over 2}{\hat{\CO}_{1,2}}\ ,\quad \hat{y}={1\over
    2}{\hat{\CO}_{2,1}}\ ,
}
the other elements in the ring are
\eqn\ringfidi{
    \hat{\CO}_{r,s} = U_{s-1}(\hat{x})U_{r-1} (\hat{y})
 }
where the $U$ are the Chebyshev polynomials of the second kind $U_{r-1}(\cos
\theta )={\sin(r\theta)\over \sin
 \theta}$
and for simplicity we are setting here the cosmological constant $\mu=1$.  The
ring has only $(p-1)(q-1)$ elements. The restriction on the values of $(r,s)$
is implemented by the ring relations
\eqn\ringrelfi{
     U_{q-1} (\hat{x})=0\ ,\qquad U_{p-1} (\hat{y})=0\ .
}
The tachyons $\CT_{r,s}$ form a module of the ring, which is simply understood
by writing $\CT_{r,s}=\hat{\CO}_{r,s} \CT_{1,1}$. The restriction in the
tachyon module $rq>ps$ implies that the module is not a faithful representation
of the ring. Instead, there is an additional relation in the ring when acting
on the tachyon module:
\eqn\tachyonreli{
    \big(U_{q-2} (\hat{x})-U_{p-2} (\hat{y})\big)\CT_{r,s}=0\ .
}
The ground ring will enable us to compute some simple correlation functions
such as
\eqn\corfui{\langle \CT_{r_1,s_1} \CT_{r_2,s_2} \CT_{r_3,s_3}
    \rangle = N_{(r_1,s_1)(r_2,s_2)(r_3,s_3)} \langle  \CT_{1,1}
    \CT_{1,1} \CT_{1,1}\rangle
}
with $N_{(r_1,s_1)(r_2,s_2)(r_3,s_3)}$ the integer fusion rules of the minimal
model.

In sections 3 and 4, we study the two types of branes in these theories, which
are referred to as the FZZT and ZZ branes
\refs{\FateevIK\TeschnerRV-\ZamolodchikovAH}. The former were previously
explored in the context of matrix models as operators that create macroscopic
loops \refs{\BanksDF,\MooreIR,\MooreAG,\KostovHN}. The Liouville expressions
for these loops \refs{\FateevIK,\TeschnerRV} lead to more insight. The general
Liouville expressions simplify in our case for two reasons. First, the
Liouville coupling constant $b^2$ is rational, $b^2={p \over q}$; and second,
we are not interested in the generic Liouville operator but only in those which
participate in the physical (i.e.\ BRST invariant) operators of the minimal
string theory. It turns out that these branes are labelled by a continuous
parameter $x$ which can be identified with the boundary cosmological constant
$\mu_B$. This parameter can be analytically continued, but it does not take
values on the complex plane. Rather, it is defined on a Riemann surface
$\CM_{p,q}$ which is given by the equation
 \eqn\CMeqi{
    F(x,y)\equiv T_q(x)-T_p(y)=0
 }
with $T_p(\cos\theta)=\cos(p\,\theta)$ a Chebyshev polynomial of the first
kind. This Riemann surface has genus zero, but it has ${(p-1)(q-1)/2}$
singularities that can be thought of as pinched $A$-cycles of a higher-genus
surface. In addition to \CMeqi, the singularities must satisfy
 \eqn\CMeqsini{\eqalign{
    &\partial_x F(x,y)= q\, U_{q-1}(x)=0\cr
    &\partial_y F(x,y)= p\, U_{p-1}(y)=0\ .\cr
 }}
Rewriting the conditions \CMeqi\ and \CMeqsini\ as
 \eqn\CMeqsini{\eqalign{
    &U_{q-1}(x)=U_{p-1}(y)=0\cr
    &U_{q-2}(x)-U_{p-2}(y)=0\ ,
 }}
we immediately recognize the first line as the ring relations \ringrelfi, and
the second line as the relation in the tachyon module \tachyonreli.

Deformations of the curve \CMeqi\ correspond to the physical operators of the
minimal string theory. Among them we find all the expected bulk physical
operators, namely the tachyons, the ring elements and the physical operators at
negative ghost number. There are also deformations of the curve that do not
correspond to bulk physical operators. We interpret these to be open string
operators.

A useful object is the one form $y\, dx$.  Its integral on $\CM_{p,q}$ from a
fixed reference point (which we can take to be at $x \to \infty$) to the point
$\mu_B$ is the FZZT disk amplitude with boundary cosmological constant $\mu_B$:
\eqn\FZZTinti{
    Z(\mu_B)=\int^{\mu_B} y\, dx\ .
}
We can also consider closed contour integrals of this one-form through the
pinched cycles of $\CM_{p,q}$. This leads to disk amplitudes for the ZZ branes:
\eqn\ZZinti{
  Z_{m,n}=\oint_{B_{m,n}} y\, dx\ .
 }
This integral can also be written as the difference between FZZT
branes on the two sides of the singularity \KlebanovWG.  Such a
relation between the two types of branes follows from the work of
\refs{\ZamolodchikovAH, \HosomichiXC, \PonsotSS} and was made most
explicit in \MartinecKA.

The relations \CMeqsini\ which are satisfied at the singularities, together
with the result \ZZinti, suggest the interpretation of the $(m,n)$ ZZ brane
states as eigenstates of the ring generators $\hat{x}$ and $\hat{y}$, with
eigenvalues $x_{mn}$ and $y_{mn}$ corresponding to the singularities of
$\CM_{p,q}$:
\eqn\ZZeigenval{\eqalign{
    &\hat{x}| m,n\rangle_{\rm ZZ}=x_{mn}| m,n\rangle_{\rm ZZ}\ , \qquad
    \hat{y}| m,n\rangle_{\rm ZZ}=y_{mn}| m,n\rangle_{\rm ZZ}\ .\cr
}}
We also find that these eigenvalues completely specify the BRST cohomology
class of the ZZ brane. In other words, branes located at the same singularity
of $\CM_{p,q}$ differ by a BRST exact state, while branes located at different
singularities are distinct physical states. Moreover, branes that do not
correspond to singularities of our surface are themselves BRST exact. Thus
there are as many distinct ZZ branes as there are singularities of $\CM_{p,q}$
in $(p,q)$ minimal string theory.

A natural question is the physical interpretation of the uniformization
parameter $\theta$ of our surface, defined by $x=\cos\theta$.  The answer is
given in appendix A, where we discuss the (nonlocal) Backlund transformation
that maps the Liouville field $\phi$ to a free field $\tilde\phi$. We will see
that $\theta$ can be identified with the Backlund field, which satisfies
Dirichlet boundary conditions.

In the second part of our paper we will study the minimal superstring theories,
which are obtained by coupling $(p,q)$ superminimal models to supergravity.
These theories fall in two classes.  The odd models exist for $p$ and $q$ odd
and relatively prime. They (spontaneously) break worldsheet supersymmetry.  The
even models exist for $p$ and $q$ even, $p/2$ and $q/2$ relatively prime and
$(p-q)/2$ odd. Our discussion parallels that in the bosonic string and the
results are very similar. There are however a few new elements.

The first difference from the bosonic system is the option of
using the 0A or 0B GSO projection. Most of our discussion will
focus on the 0B theory. In either theory, there is a global ${\Bbb
Z}_2$ symmetry $(-1)^{F_L}$, where $F_L$ is the left-moving
spacetime fermion number. This symmetry multiplies all the RR
operators by $-1$ and is broken when background RR fields are
turned on. Orbifolding the 0B theory by $(-1)^{F_L}$ leads to the
0A theory and vice versa.

A second important difference relative to the bosonic system is that here the
cosmological constant $\mu$ can be either positive or negative, and the results
depend on its sign
\eqn\zetadef{
\zeta={\rm sign}(\mu)\ .
}
This sign can be changed by performing a ${\Bbb Z}_2$
R-transformation in the super-Liouville part of the theory. It
acts there as $(-1)^{f_L}$ with $f_L$ the left moving worldsheet
fermion number. Since this is an R-transformation, it does not
commute with the worldsheet supercharge. In order to commute with
the BRST charge such an operation must act on the total
supercharge including the matter part.  This operation is usually
not a symmetry of the theory.  In particular it reverses the sign
of the GSO projection in the Ramond sector. Therefore, our answers
will in general depend on $\zeta$.

In section 5 we discuss the tachyons, the ground ring and the correlation
functions.  The results are essentially identical to those in the bosonic
string except that we should use the appropriate values of $(p,q)$, and the
ring relation in the tachyon module \tachyonreli\ depends on $\zeta$
 \eqn\tachyonrelis{
    \big(U_{q-2}(\hat x)- \zeta U_{p-2} (\hat y)\big)\CT_{r,s}=0\ .
 }
In sections 6 and 7 we consider the supersymmetric FZZT and ZZ
branes. Here we find a third element which is not present in the
bosonic system. Now there are two kinds of branes labelled by a
parameter $\eta=\pm 1$; this parameter determines the combination
of left and right moving supercharges that annihilates the brane
boundary state: $(G+i\eta \tilde G)|B\rangle=0$.  We must also
include another label $\xi=\pm 1$ which multiplies the Ramond
component of the boundary state. It is associated with the ${\Bbb
Z}_2$ symmetry $(-1)^{F_L}$. Changing the sign of $\xi$ maps a
brane to its antibrane.

Most of these new elements do not affect the odd models.  The answers are
independent of $\zeta$ and $\eta$, and we again find the Riemann surface
$\CM_{p,q}$ given by the curve $T_q(x)-T_p(y)=0$. The characterization of the
FZZT and ZZ branes as contour integrals of $y\,dx$ is identical to that in the
bosonic string.

The even models are richer.  Here we find two Riemann surfaces $\CM^{\hat
\eta}_{p,q}$ depending on the sign of
 \eqn\etahatdi{
    \hat \eta = \zeta \eta\ .
 }
These surfaces are described by the curves
 \eqn\Fetaha{
 F_{\hat \eta}(x,y) = T_q(x)-\hat{\eta}\,T_p(y)=0\ .
 }
The discussion of the surface $\CM^-_{p,q}$ is similar to that of the bosonic
models. The surface $\CM^{+}_{p,q}$ is more special, because it splits into two
separate subsurfaces $T_{p\over 2}(y)=\xi T_{q\over 2}(x)$ which touch each
other at singular points. The two subsurfaces are interpreted as associated
with the choice of the Ramond ``charge'' $\xi=\pm 1$. The two kinds of FZZT
branes which are labelled by $\xi$ correspond to line integrals from infinity
in the two sub-surfaces.  ZZ branes are again given by contour integrals. These
can be either closed contours which pass through the pinched singularities in
each subsurface, or they can be associated with contour integrals from infinity
in one subsurface through a singularity which connects the two subsurfaces to
infinity in the other subsurface.

It is surprising that our results depend essentially only on $p$
and $q$.  The main difference between the bosonic models and the
supersymmetric models is in the allowed values of $(p,q)$.  Odd
$p$ and $q$ which are relatively prime occur in both the bosonic
and the supersymmetric models. This suggests that these two models
might in fact be the same.  This suggestion is further motivated
by the fact that the two theories have the same KPZ scalings
\refs{\KlebanovWG,\JohnsonHY}. Indeed, all our results for these
models (ground ring, sphere three point function, FZZT and ZZ
branes) are virtually identical. We will discuss the evidence for
the equivalence of these models in section 8.

Our work goes a long way to deriving the matrix model starting from the
worldsheet formulation of the theory. We will discuss the comparison with the
matrix model in section 9. Our Riemann surface $\CM_{p,q}$ occurs naturally in
the matrix model and determines the eigenvalue distribution. Of all the
possible matrix model descriptions of the minimal string theories, the closest
to our approach are Kostov's loop gas model
\refs{\KostovHN\KostovXT\KostovCG\KostovGS-\KostovUQ} and the two matrix model
\DaulBG, in which expressions related to ours were derived. For the theories
with $p=2$, we also have a description in terms of a one-matrix model. Some
more detailed aspects of the comparison to the one-matrix model are worked out
in appendix B.

The matrix model also allows us to explore other values of $(p,q)$ which are
not on the list of (super)minimal models.  For example, the theories with
$(p,q)=(2,2k+2)$ were interpreted in \KlebanovWG\ as a minimal string theory
with background RR fields. It is natural to conjecture that all values of
$(p,q)$ correspond to some minimal string theory or deformations thereof. This
generalizes the worldsheet constructions based on (super)minimal models and
(super)Liouville theory.

The Riemann surface which is central in our discussion is closely related to
the target space of the eigenvalues of the matrix model.  However, it should be
stressed that neither the eigenvalue direction nor the Riemann surface are the
target space of the minimal string theory itself.  Instead, the target space is
the Liouville direction $\phi$, which is related to the eigenvalue space
through a nonlocal transform \MooreAG.  This distinction between $\phi$ and the
coordinates of the Riemann surface is underscored by the fact that $x$ and $y$
on the Riemann surface are related to composite operators in the worldsheet
theory (the ground ring generators), rather than to the worldsheet operator
$\phi$.

Our discussion is limited to the planar limit, where the worldsheet topology is
a sphere or a disk (with punctures). It would be interesting to extend it to
the full quantum string theory. It is likely that the work of
\refs{\MooreMG,\MooreCN} is a useful starting point of this discussion.

A crucial open problem is the nonperturbative stability of these
theories. For instance, some of the bosonic $(p,q)$ theories are
known to be stable while others are known to be unstable. In fact,
as we will discuss in section 9, the matrix model of all these
models have a small instability toward the tunnelling of a small
number of eigenvalues. The nonperturbative status of models based
on generic values of $(p,q)$ remains to be understood.

After the completion of this work an interesting paper \AganagicQJ\ came out
which overlaps with some of our results and suggests an extension to the
quantum string theory.

\newsec{Bosonic minimal string theory}

\subsec{Preliminaries}

We start by summarizing some relevant aspects of the $(p,q)$ minimal models and
Liouville theory that we will need for our analysis, at the same time
establishing our notations and conventions.  For the bosonic theories, we will
take $\alpha'=1$. The $(p,q)$ minimal models exist for all $p,q \ge 2$ coprime.
Our convention throughout will be $p<q$. The central charge of these theories
is given by
\eqn\cpq{
    c=1-{6(p-q)^2\over p\,q}
}
The $(p,q)$ minimal model has a total of $N_{p,q}=(p-1)(q-1)/2$ primary
operators $\CO_{r,s}$ labelled by two integers $r$ and $s$, with $r=1,\dots,
p-1$ and $s=1,\dots, q-1$. They satisfy the reflection relation
$\CO_{p-r,q-s}\equiv \CO_{r, s}$ and have dimensions
\eqn\dimpq{
\Delta(\CO_{r,s})=\bar{\Delta}(\CO_{r,s})={(r q-s p)^2-(p-q)^2 \over 4p\,q}\ .
}
In general the operator $\CO_{r,s}$ contains two primitive null vectors among
its conformal descendants at levels $rs$ and $(p-r)(q-s)$. Because they are
degenerate, primary operators must satisfy the fusion rules
 \eqn\fusionpq{\eqalign{
    &\CO_{r_1,s_1}\CO_{r_2,s_2}=\sum [\CO_{r,s}]\cr
    &\qquad r=|r_1-r_2|+1,|r_1-r_2|+3,\dots,\cr
    &\qquad\qquad\min(r_1+r_2-1,2p-1-r_1-r_2)\cr
    &\qquad s=|s_1-s_2|+1,|s_1-s_2|+3,\dots,\cr
    &\qquad\qquad\min(s_1+s_2-1,2q-1-s_1-s_2)\cr
 }}
The fusion of any operator with $\CO_{1,2}$ and $\CO_{2,1}$ is especially
simple:
\eqn\fusionsimppq{\eqalign{
    & \CO_{1,2} \CO_{r,s}=[\CO_{r,s+1}]+[\CO_{r,s-1}]\cr
    & \CO_{2,1} \CO_{r,s}=[\CO_{r+1,s}]+[\CO_{r-1,s}]\ .\cr
 }}
Thus in a sense, $\CO_{1,2}$ and $\CO_{2,1}$ generate all the primary operators
of the minimal model.  In particular, $\CO_{1,2 } $ generates all primaries of
the form $\CO_{1,s}$ and $\CO_{2,1}$ generates all primaries of the form
$\CO_{r,1}$. The fusion of two such operators is very simple:
$\CO_{1,s}\CO_{r,1}=[\CO_{r,s}]$.

The central charge of Liouville field theory is
 \eqn\cLiouville{
    c=1+6Q^2=1+6\left(b+{1\over b}\right)^2
 }
where $Q=b+{1\over b}$ is the background charge. The basic primary
operators of the Liouville theory are the vertex operators
$V_\alpha=e^{2\alpha\phi}$ with dimension
 \eqn\dimLiouville{
    \Delta(\alpha)=\bar{\Delta}(\alpha)=\alpha(Q-\alpha)\ .
 }

Of special interest are the primaries in degenerate Virasoro
representations
\eqn\degenL{
    V_{\alpha_{r,s}}=e^{2\alpha_{r,s}\phi},\quad
    2\alpha_{r,s}={1\over b} (1-r) + b (1-s)\ .
}
For generic $b$, these primaries have exactly one singular vector
at level $rs$. Therefore, their irreducible character is given by:
\eqn\gench{
    \chi_{r,s}(q) = {q^{\Delta(\alpha_{r,s})-(c-1)/24}\over \eta(q)}(1-q^{rs})
}
The degenerate primaries also satisfy the analogue of the minimal
model fusion rule \fusionsimppq\ \refs{\DornSV, \ZamolodchikovAA}:
 \eqn\opefusionLa{\eqalign{
    &V_{-{b\over 2}}V_\alpha=[V_{\alpha-{b\over 2}}]+C(\alpha)[V_{\alpha+{b\over
    2}}]\cr
    &C(\alpha)=-\mu{\pi\gamma(2b\alpha-1-b^2)\over
    \gamma(-b^2)\gamma(2b\alpha)}\cr
 }}
with a similar expression for $V_{-{1\over 2b}}$:
\eqn\opefusionLb{\eqalign{
    &V_{-{1\over 2b}}V_\alpha=[V_{\alpha-{1\over 2b}}]+\tilde{C}(\alpha)[V_{\alpha+{1\over
    2b}}]\cr
    &\tilde{C}(\alpha)=-\tilde{\mu}{\pi\gamma(2\alpha/b-1-1/b^2)\over
    \gamma(-1/b^2)\gamma(2\alpha/b)}\ .\cr
 }}
Here $\gamma(x)=\Gamma(x)/\Gamma(1-x)$. Notice that the second OPE may be
obtained from the first by taking $b\rightarrow 1/b$ and also $\mu\rightarrow
\tilde{\mu}$. The quantity $\tilde\mu$ is the {\it dual cosmological constant},
which is related to $\mu$ via
$\pi\tilde{\mu}\gamma(1/b^2)=(\pi\mu\gamma(b^2))^{1/b^2}$. From this point
onwards, we will find it convenient to rescale $\mu$ and $\tilde{\mu}$ so that
 \eqn\dualmu{
    \tilde{\mu}=\mu^{1/b^2}\ .
 }
This will simply many of our later expressions.

In order to construct a minimal string theory, we must couple the $(p,q)$
minimal model to Liouville. Demanding the correct total central charge implies
that the Liouville theory must have
\eqn\bpqdd{
 b=\sqrt{p\over q}\ .
}
We will see throughout this work that taking $b^2$ to be rational
and restricting only to the BRST cohomology of the full string
theory gives rise to many simplifications, and also to a number of
subtleties. One such a simplification is that not all Liouville
primaries labelled by $\alpha$ correspond to physical (i.e.\ BRST
invariant) vertex operators of minimal string theory. For
instance, physical vertex operators may be formed by first
``dressing" an operator $\CO_{r,s}$ from the matter theory with a
Liouville primary $V_{\beta_{r,s}}$ such that the combination has
dimension $(1,1)$, and then multiplying them with the ghosts
$c\bar{c}$. We will refer to such operators as ``tachyons.''
Requiring the sum of \dimpq\ and \dimLiouville\ to be 1 gives the
formula for the Liouville dressing of $\CO_{r,s}$:
 \eqn\dressL{\eqalign{
 &\CT_{r,s}= c \bar c\CO_{r,s} V_{\beta_{r,s}}\cr
 & 2\beta_{r,s}= {p+q-rq+s p\over \sqrt{p\, q}},\quad r q-s p\ge
 0\ .
 }}
Note that in solving the quadratic equation for $\beta_{r,s}$, we have chosen
the branch of the square root so that $\beta_{r,s}<Q/2$ \SeibergEB.

An important subtlety that arises at rational $b^2$ has to do with the
irreducible character of the degenerate primaries. The formula \gench\ that we
gave above for generic $b$ must now be modified for a number of
reasons.\foot{We thank B. Lian and G. Zuckerman for helpful discussions about
the structure of these representations.}

\lfm {1.} Different $(r,s)$ can now lead to the same degenerate
primary, and therefore labelling the representations with $(r,s)$
is redundant. It will sometimes be convenient to remove this
redundancy by defining
\eqn\Ntmn{
 N(t,m,n)\equiv |\,tpq+mq+np\,|
 }
and labelling each representation by $(t,m,n)$ satisfying
\eqn\tmn{
 \qquad 0< m \le p\ , \qquad 0 < n \le q \ , \qquad t\ge 0
}
such that the degenerate primary has dimension
\eqn\labeldegen{
\Delta(t,m,n)={(p+q)^2 -N(t,m,n)^2\over 4pq}
}
Then each $(t,m,n)$ satisfying \tmn\ corresponds to a unique degenerate
representation and vice versa.

\lfm {2.} The Verma module of a degenerate primary can now have
more than one singular vector. For instance, we can write
\eqn\Ntmnk{
N(t,m,n)=((t-j)p+m)q+(j q+n)p
}
for $j=0,\dots,t$, and by continuity in $b$ we expect the Verma
module of $(t,m,n)$ to contain singular vectors at levels
$((t-j)p+m)(j q+n)$, with dimensions
\eqn\dimsing{
\Delta ={(p+q)^2 -N(t-2j,m,-n)^2\over 4pq}
}

\lfm {3.} A related subtlety is that a singular vector can itself be
degenerate. For example, the singular vectors discussed above are degenerate as
long as $t-2j \ne 0$. In general, this will lead to a complicated structure of
nested Verma submodules contained within the original Verma module of the
degenerate primary. A similar structure is seen, for instance, in the minimal
models at $c<1$. An important distinction is that here there are only finitely
many singular vectors.

\medskip

Taking into account these subtleties leads to a formula for the character
slightly more complicated than the naive formula \gench. The answer is
\refs{\LianAJ,\LianCB}:
\eqn\modchi{\eqalign{
    \hat\chi_{t,m,n}(q)={1\over
    \eta(q)}\sum_{j=0}^{t}\left(q^{-N(t-2j,m,n)^2/4pq}-q^{-N(t-2j,m,-n)^2/4pq}\right)
}}
Notice that we can also write this as a sum over naive characters
\gench:
\eqn\modchinc{
    \hat\chi_{t,m,n}(q)=\sum_{j=0}^{[{t\over2}]}\chi_{(t-2j)p+m,n}-
                  \sum_{j=0}^{[{t-1\over2}]}\chi_{(t-2j)p-m,n}
}
For $t=0$, the formula reduces to \gench, i.e.\
$\hat\chi_{t=0,m,n}=\chi_{m,n}$. These results will be important when we come
to discuss the ZZ boundary states of minimal string theory in section 3.

\subsec{The ground ring of minimal string theory}

The other important collection of BRST invariant operators in minimal string
theory is the ground ring of the theory. The ground ring consists of all
dimension 0, ghost number 0, primary operators in the BRST cohomology of the
theory \LianGK, and it was first studied for the $c=1$ bosonic string in
\WittenZD. Ring multiplication is provided by the OPE modulo BRST commutators.
As in the matter theory, elements $\hat{\CO}_{r,s}$ of the ground ring are
labelled by two integers $r$ and $s$, $r=1,\dots, p-1$ and $s=1,\dots, q-1$. In
contrast to the matter primaries, however, $\hat{\CO}_{r,s}$ and
$\hat{\CO}_{p-r,q-s}$ are distinct operators. Thus the ground ring has
$(p-1)(q-1)$ elements, twice as many as the matter theory.

The construction of the ground ring starts by considering the combination
$\CO_{r,s}V_{\alpha_{r,s}}$ of a matter primary and a corresponding degenerate
Liouville primary. Using \dimpq\ and \dimLiouville, it is easy to see that this
combination has dimension $1-rs$. Acting on $\CO_{r,s}V_{\alpha_{r,s}}$ with a
certain combination of level $rs-1$ raising operators then gives the ground
ring operator
\eqn\ringoddd{
    \hat{\CO}_{r,s}=\CL_{r,s}\cdot
 \CO_{r,s}V_{\alpha_{r,s}},\qquad 2\alpha_{r,s}={p+q-r q-s p\over
 \sqrt{pq}}\ .
}
{}From the construction, it is clear that $\hat{\CO}_{r,s}$ has
Liouville momentum $2\alpha_{r,s}$. When $b$ and $1/b$ are
incommensurate (which is the case for the $(p,q)$ minimal models),
\ringoddd\ then implies that there is a unique ground ring
operator at a given Liouville momentum $2\alpha_{r,s}$ for
$r=1,\dots, p-1$ and $s=1,\dots, q-1$. It follows that the
multiplication table for the ground ring can be derived from
kinematics alone. When $\mu=0$, Liouville momentum is conserved in
the OPE, and therefore one must have
\eqn\grmultpq{
    \hat{\CO}_{r,s}=\hat{\CO}_{1,2}^{s-1}\hat{\CO}_{2,1}^{r-1}\ .
}
Thus the ground ring is generated by two elements, $\hat{\CO}_{1,2}$ and
$\hat{\CO}_{2,1}$. Moreover, the fact that the ring has finitely many elements
leads to non-trivial relations for the ring generators:
\eqn\grrelpq{\eqalign{
    &\hat{\CO}_{1,2}^{q-1}=0   \cr
    &\hat{\CO}_{2,1}^{p-1}=0\ .\cr
}}
Both \grmultpq\ and \grrelpq\ (and all the ground ring equations that follow)
are understood to be true modulo BRST commutators.

Using the $\mu$-deformed Liouville OPEs \opefusionLa,
\opefusionLb\ it is easy to see how ring multiplication and the
ring relations are altered at $\mu\ne 0$. It will be convenient to
define the dimensionless combinations
 \eqn\grrescdefII{
    \hat{x}={1\over 2\sqrt{\mu}}{\hat{\CO}_{1,2}},
    \qquad \hat{y}={1\over 2\sqrt{\tilde \mu}}\hat{\CO}_{2,1}\ .
 }
Let us start by analyzing the operators $\hat{\CO}_{r,1}$. It is
clear that $\hat{\CO}_{r,1}=\CP_{r-1}(\hat y)$ is a polynomial of
degree $r-1$ in the generator $\hat y$. These polynomials are
constrained by the fusion rules
 \eqn\ringfur{ \eqalign{
    &\CP_{r-1}(\hat y) \CP_{l-1}(\hat y) =
    \sum a_{r,l,k} \CP_{k-1}(\hat y)\cr
    &\qquad k=|r-l|+1,|r-l|+3,\dots,\cr
    &\qquad\qquad\min(r+l-1,2p-1-r-l) \cr
 }}
The restrictions on the sum in \ringfur\ determine most of the coefficients in
$\CP_{r-1}$ even without using the known values of the operator product
coefficients (\opefusionLa\ and the similar coefficient in the minimal
model).\foot{For example, all the coefficients in $\CP_r$ and all the
coefficients $a_{r,l,k}$ in \ringfur\ are determined in terms of $a_{1,r,r-1}$
which appears in $\CP_1(\hat y) \CP_r(\hat y)= \CP_{r+1}(\hat y) + a_{1,r,r-1}
\CP_{r+1}(\hat y)$.  The $p$-dependence of the truncation of the sum in
\ringfur\ leads to the ring relation $\CP_{p-1}(\hat y)=0$, which generalizes
\grrelpq\ to nonzero $\mu$, and leads to relations among the coefficients
$a_{1,r,r-1}$.} The remaining coefficients can be computed using the OPE in the
minimal model and Liouville, but we will not do that here. Instead we will
simply state the answer. We claim that
\eqn\ringexmu{
    \hat {\CO}_{r,1}= \mu^{{q(r-1)\over2p}} U_{r-1} (\hat y)
}
where the $U$ are Chebyshev polynomials of the second kind. We will postpone
the full justification of our claim until we come to discuss the ZZ brane
one-point functions in section 3.1. Also, the computation of the tachyon
three-point functions below will provide a non-trivial check of \ringexmu.

For now, let us simply show that our ansatz is consistent with \ringfur. This
is a result of the following trigonometric identity for the multiplication of
the Chebyshev polynomials $U$:
 \eqn\cheIIi{\eqalign{
     &U_{r-1}(\hat y) U_{l-1}(\hat y) = \sum_{k} U_{k-1}(\hat y)\cr
     &\qquad k=|r-l|+1, |r-l|+3,\dots, r+l-1\ .
 }}
The fact that the Chebyshev polynomials can be expressed as $SU(2)$ characters
\eqn\cheII{
    U_{r-1}(\cos \theta )={\sin(r\theta)\over \sin
    \theta}=\Tr_{j={r-1 \over 2}} e^{2i\theta J_3}\ .
}
underlies the identity \cheIIi. This identity is almost of the form \ringfur.
The $p$-dependence in \ringfur\ is implemented by the ring relation
\eqn\ringrelp{
    U_{p-1} (\hat y)=0\ .
}
(This is a standard fact in the representation theory of $\widehat {SU(2)}$.)
This shows that the expression \ringexmu\ with the relation \ringrelp\
satisfies \ringfur\ with all nonzero $a_{r,l,k}$ equal to one.

It is trivial to extend these results to the operators $\hat{\CO}_{1,s}$ which
are generated by $\hat x$.  Finally, using $\hat{\CO}_{r,s} = \hat{\CO}_{r,1}
\hat{\CO}_{1,s}$ we derive the expressions for the ring elements
\eqn\ringfid{
    \hat{\CO}_{r,s} = \mu^{q(r-1)+p(s-1)\over 2p}
    U_{s-1}(\hat x) U_{r-1} (\hat y)
}
and the relations
\eqn\ringrelf{\eqalign{
 &U_{q-1} (\hat x)=0 \cr
 &U_{p-1} (\hat y)=0\ .\cr
}}

Having found the ring multiplication, we can use it to analyze the tachyon
operators \dressL. Ghost number conservation implies that the tachyons are a
module of the ring \KutasovQX.  Using our explicit realization in terms of the
generators \ringfid\ it is clear that
\eqn\tachmod{
    \CT_{r,s}=\mu^{1-s}\hat{\CO}_{r,s} \CT_{1,1}=
    \mu^{q(r-1)-p(s-1)\over 2p}
    U_{s-1}\big(\hat x\big) U_{r-1}\big(\hat y\big)
    \CT_{1,1}\ .
}
Using this expression it is easy to act on $\CT_{r,s}$ with any ring operator.
This can be done by writing the ring operator and the tachyon in terms of the
generators as in \ringfid\ and \tachmod\ and then simply multiplying the
polynomials in the generators subject to the ring relations \ringrelf.

It is clear, however, that this cannot be the whole story.  There are
$(p-1)(q-1)$ different ring elements $\hat{\CO}_{r,s}$, but there are only
$(p-1)(q-1)/2$ tachyons $\CT_{r,s}$ because they are subject to the
identification
\eqn\tachident{
    \CT_{p-r,q-s}=\mu^{p s-q r\over p}\CT_{r,s}\ .
}
This means that in addition to the two ring relations
\ringrelf\ there must be more relations which are satisfied only in the tachyon
module; i.e.\ it is not a faithful representation of the ring. It turns out
that one should impose
\eqn\tachyonrel{
    \big(U_{q-2} (\hat x)-U_{p-2} (\hat y)\big)\CT_{r,s}=0\ .
}
It is easy to show, using trigonometric/Chebyshev identities, that
this relation guarantees the necessary identifications
\DiFrancescoZK\ (see also \refs{\ImbimboXT,\ImbimboFW}). Note that
equation \tachyonrel\ can also be written using the ring relations
\ringrelf\ in terms of the Chebyshev polynomials of the first kind
$T_p(x)=\cos(p\,\theta)$ as
\eqn\tachyonrela{
 \big(T_q (\hat x)-
       T_p (\hat y)\big)\CT_{r,s}=0\ .
}
This expression will be useful in later sections. We should also point out that
the effect of the relations \ringrelf\ and \tachyonrel\ on ring multiplication
is to truncate it to precisely the fusion rules \fusionpq\ of the minimal
model.

Using this understanding we can constrain the correlation functions of these
operators.  The simplest correlation functions involve three tachyons and any
number of ring elements on the sphere. Because of the conformal Killing vectors
on the sphere, this calculation does not involve any moduli integration.  It is
given by
\eqn\corfu{\eqalign{
    \langle \CT_{r_1,s_1}
    \CT_{r_2,s_2} \CT_{r_3,s_3} \prod_{i \ge 4} \hat{\CO}_{r_i,s_i} \rangle
    &=
    \mu^{3-s_1-s_2-s_3}\langle  \CT_{1,1} \CT_{1,1} \CT_{1,1}\prod_{i \ge 1} \hat{\CO}_{r_i,s_i}\rangle\ .\cr
}}
The product of ring elements can be recursively simplified using the ring
relations \ringrelf\ and \tachyonrel. This leads to a linear combination of
ring elements, of which only $\hat{\CO}_{1,1}$ survives in the expectation
value \corfu. As an example, consider the three-point function:
 \eqn\corfuthree{\eqalign{
     \langle \CT_{r_1,s_1} \CT_{r_2,s_2} \CT_{r_3,s_3} \rangle &=
     \mu^{3-s_1-s_2-s_3}\langle  \CT_{1,1} \CT_{1,1} \CT_{1,1}
     \hat{\CO}_{r_1,s_1}\hat{\CO}_{r_2,s_2}\hat{\CO}_{r_3,s_3} \rangle \cr
    &=N_{(r_1,s_1)(r_2,s_2)(r_3,s_3)} \mu^{\kappa}
 }}
where $N_{(r_1,s_1)(r_2,s_2)(r_3,s_3)}\in \{0,1\}$ represent the integer fusion
rules of the conformal field theory and $\kappa$ is the KPZ exponent of the
correlation function:
\eqn\kappadef{
\kappa = {Q-\sum_{i} \beta_{r_i,s_i}\over b}\ .
}
Note that in calculating the three-point function, we have normalized
$\langle\CT_{1,1}^3\rangle = \mu^{{Q\over b}-3}$.

The surprising result that the three-point functions are given simply by the
minimal model fusion rules was noticed many years ago using different methods
\refs{\DiFrancescoJD\GoulianQR-\DiFrancescoUD}. The agreement between our
calculation and the results in \refs{\DiFrancescoJD\GoulianQR-\DiFrancescoUD}\
serves as a check of our ansatz \ringfid\ for the $\mu$-deformed ring
multiplication. (We will also give an independent derivation of \ringfid\ in
section 3.1 using the ZZ branes.) From our current perspective, the simplicity
of the tachyon three-point functions follows from the simplicity of the
expressions for the ring elements \ringfid\ and the relations \ringrelf,
\tachyonrel. At a superficial level we can view the minimal string theory as a
topological field theory based on the chiral algebra $Virasoro/Virasoro$. Its
physical operators are the primaries of the conformal field theory and its
three point functions are the fusion rule coefficients.

For correlation functions involving fewer than three tachyons we simply insert
the necessary power of the cosmological constant $\CT_{1,1}$ in order to have
three tachyons. Then we integrate with respect to $\mu$ to find the desired
two- or one-point functions.

It would be nice to generalize this method to four and higher-point correlation
functions. For such correlators, there are potential complications involving
contact terms between the integrated vertex operators.

\newsec{FZZT and ZZ branes of minimal string theory}

\subsec{Boundary states and one-point functions}

In this section, we will study in detail the FZZT and ZZ branes of Liouville
theory coupled to the bosonic $(p,q)$ minimal models. Let us first review what
is known about the FZZT and ZZ boundary states in these theories. To make the
former, we must tensor together an FZZT boundary state in Liouville and a Cardy
state from the matter. This gives the boundary state \refs{\FateevIK,
\TeschnerRV, \CardyIR}:
\eqn\FZZTbs{
    |\sigma; k,l\rangle = \sum_{k',l'} \int_{0}^{\infty}dP\, \cos
    (2\pi P \sigma) {\Psi^{*}(P)S(k,l;k',l')\over
    \sqrt{S(1,1;k',l')}}|P\rangle\rangle_L|k',l'\rangle\rangle_M\ .
}
Here $(k,l)$ labels the matter Cardy state associated to the minimal model
primary $\CO_{k,l}$ and $|p\rangle\rangle_L$ and $|k',l'\rangle\rangle_M$ are
Liouville and matter Ishibashi states, respectively. The Liouville and matter
wavefunctions are $\cos(2\pi P\, \sigma)\Psi(P)$ and $S(k,l;k',l')$, where
\eqn\FZZTwvfn{\eqalign{
    &\Psi(P)=\mu^{-{iP\over b}}{\Gamma\left(1+{2iP\over b}\right)\Gamma(1+2iPb)\over i\pi P}\cr
    &S(k,l;k',l')=(-1)^{k l'+k' l}\sin({\pi p l l'\over q})\sin({\pi q
    k k'\over p})\ .
}}
The matter wavefunction is essentially the modular $S$-matrix of the minimal
model. Note that we have separated out the $\sigma$ dependent part of the
Liouville wavefunction for later convenience. Finally, the parameter $\sigma$
is related to the boundary cosmological constant $\mu_B$ via\foot{We have
rescaled the usual definition of $\mu_B$, together with the rescaling of $\mu$
that we mentioned above \dualmu. Our conventions for $\mu$ and $\mu_B$ relative
to, e.g.\ \FateevIK, are $\mu_{\rm here}=\pi\mu_{\rm there}\gamma(b^2)$ and
$(\mu_B)_{\rm here}=(\mu_B)_{\rm there}\sqrt{\pi\gamma(b^2)\sin\pi b^2}$.}
\eqn\fzzsmub{
    {\mu_B\over \sqrt{\mu}}=\cosh\pi b \sigma\ .
}
By thinking of the FZZT states as Liouville analogues of Cardy
states, one also finds that the state labelled by $\sigma$ is
associated to the non-degenerate Liouville primary with
$2\alpha=Q+i\sigma$ \ZamolodchikovAH.

Using the expression \FZZTbs\ for the FZZT boundary state, we can easily
calculate the one-point functions of physical operators on the disk with the
FZZT boundary condition. Let us start with the physical tachyon operators
$\CT_{r,s}=\CO_{r,s}e^{2\beta_{r,s}\phi}$ with $\beta_{r,s}$ defined in
\dressL. Using $P=i(Q/2-\beta_{r,s})$ in \FZZTbs, we find
\eqn\fzztoptach{\eqalign{
    \langle \CT_{r,s}|\,\sigma; k,l\rangle &=
    A_{r,s}(-1)^{k s+l r}
    \cosh\left({\pi(qr-ps)\sigma\over \sqrt{pq}}\right)
    \sin({\pi q k r\over p})\sin({\pi p l s\over q})
}}
where the $(\sigma,k,l)$ independent normalization factor $A_{r,s}$ will be
irrelevant for our purposes.

We can similarly compute the one-point functions for the ground ring elements
$\hat{\CO}_{r,s}$. As discussed in section 2, these take the general form
\eqn\grpqgen{
    \hat{\CO}_{r,s}=\CL_{r,s}\cdot\CO_{r,s}e^{2\alpha_{r,s}\phi}
}
where $\CL_{r,s}$ denotes a certain combination of Virasoro raising operators
of total level $rs-1$. These only serve to contribute an overall $(\sigma,k,l)$
independent factor to the one-point functions. Since
$\alpha_{r,s}=\beta_{r,-s}$, the ground ring ring one-point functions are
identical to the tachyon one-point functions \fzztoptach\ up to normalization.

Finally, let us consider the physical operators at negative ghost number
\LianGK.\foot{In some of the literature, physical operators at positive ghost
number are also discussed. However, these violate the Liouville bound
$\alpha<Q/2$ \SeibergEB. Thus they are not distinct from the negative ghost
number operators, but are related by the reflection $\alpha\rightarrow
Q-\alpha$.} These are essentially copies of the ground ring, and their
construction is analogous to \grpqgen. Their Liouville momentum is also given
by $\beta_{r,s}$, but with $s$ taking the values $s < -q$ and $s\ne
0\,\mod\,q$. Thus their one-point functions will also be given by \fzztoptach\
up to normalization, with the appropriate values of $s$.

The tachyons, the ground ring, and the copies of the ground ring at negative
ghost number are the complete set of physical operators in the minimal string
theory. Their one-point functions \fzztoptach\ have several interesting
properties as functions of $(\sigma,k,l)$. First, they satisfy an identity
relating states with arbitrary matter label to states with matter label
$(k,l)=(1,1)$
\eqn\fzztopfrelate{\eqalign{
\langle \CO | \,\sigma; &k,l\rangle = \sum_{m',n'} \langle \CO |
\,\sigma+{i(m'q+n'p)\over \sqrt{pq}};1,1\rangle\cr
}}
with $(m',n')$ ranging over the values
\eqn\mpnpval{\eqalign{
    m'&=-(k-1),-(k-1)+2,\dots,k-1\cr
    n'&=-(l-1),-(l-1)+2,\dots,l-1\ .
}}
Here $\CO$ stands for an arbitrary physical operator. This is
evidence that in the full string theory, where the boundary states
are representatives of the BRST cohomology, the following is true
\eqn\fzztstrelate{\eqalign{
| \,\sigma; &k,l\rangle = \sum_{m',n'} | \,\sigma+{i(m'q+n'p)\over
\sqrt{pq}};1,1\rangle\cr
 }}
modulo BRST exact states. We should emphasize that \fzztstrelate, which relates
branes with different matter states, is an inherently quantum mechanical
result. This relation is difficult to understand semiclassically, where branes
with different matter states appear distinct. But there is no contradiction,
because \fzztstrelate\ involves a shift of $\sigma$ by an imaginary quantity,
which amounts to analytic continuation of $\mu_B$ from the semiclassical region
where it is real and positive.

According to \fzztstrelate, the FZZT branes with $(k,l)=(1,1)$ form a complete
basis of all the FZZT branes of the theory. The branes with other matter states
should be thought of as multi-brane states formed out of these elementary FZZT
branes. This allows us to simplify our discussion henceforth by restricting our
attention, without loss of generality, to the elementary FZZT (and ZZ) branes
with $(k,l)=(1,1)$. We will also simplify the notation by dropping the label
$(1,1)$ from the boundary states; this label will be implicit throughout.

A second interesting property of the one-point functions is that they are
clearly invariant under the transformations
\eqn\sigmainvt{
\sigma \rightarrow -\sigma,\quad \sigma\pm 2i\sqrt{pq}
}
Again, this is evidence that the states labelled by $\sigma$ should be
identified under the transformations \sigmainvt\ modulo BRST exact states.
Thus, labelling the states by $\sigma \in \Bbb C$ infinitely overcounts the
number of distinct states. Therefore, it makes more sense to define
\eqn\zdef{
z=\cosh {\pi \sigma\over \sqrt{pq}}
}
and to label the states by $z$,
\eqn\fzztrelabelz{\eqalign{
    &|\,\sigma\rangle \rightarrow |\,z\rangle\cr
}}
such that two states $|\,z\rangle$ and $|\,z'\rangle$ are equal if
and only if $z=z'$. In section 4, we will interpret geometrically
the parameter $z$ and the infinite overcounting by $\sigma$.

Now let us discuss the ZZ boundary states. As was the case for the FZZT states,
these are formed by tensoring together a Liouville ZZ boundary state and a
matter Cardy state (in this case the $(1,1)$ matter state). However, here there
are subtleties arising from the fact that $b^2$ is rational: the Liouville ZZ
states are in one-to-one correspondence with the degenerate representations of
Liouville theory, which, as we discussed in section 2, have rather different
properties at generic $b$ and at $b^2$ rational. In either case, the
prescription \refs{\ZamolodchikovAH, \MartinecKA} for constructing the ZZ
boundary states is to take the formula for the irreducible character (\modchi\
for rational $b^2$) and replace each term ${1\over\eta(q)}q^{-N^2/4pq}$ with an
FZZT boundary state with $\sigma= iN$. Thus \modchi\ becomes
\eqn\ZZbs{\eqalign{
    |\,t,m,n\rangle &=
    \sum_{j=0}^{t}\left(\big|\,z=\cos{\pi N(t-2j,m,n)\over pq}\big\rangle-
    \big|\,z=\cos{\pi N(t-2j,m,-n)\over pq}\big\rangle\right)\cr
    &=
    (t+1)\left(\big|\,z=(-1)^t\cos{\pi (m q+np)\over pq}\big\rangle-
    \big|\,z=(-1)^t\cos{\pi (m q-np)\over pq}\big\rangle\right)\cr
}}
In the second equation we have substituted \Ntmn\ and simplified
the arguments of the cosines -- surprisingly, they become
independent of $j$. We recognize the quantity in parentheses to be
a ZZ state with $t=0$; thus we conclude that
\eqn\ZZreduce{
    |\,t,m,n\rangle =\cases{
    +(t+1)|\,t=0,m,n\rangle & \quad $t$ even\cr
    -(t+1)|\,t=0,m,q-n\rangle & \quad $t$ odd\cr}
}
It is also straightforward to show using \ZZbs\ that
\eqn\ZZreflect{
|\,t,m,n\rangle = |\,t,p-m,q-n\rangle
}
and that
\eqn\ZZvanishes{
|\,t,m,n\rangle = 0\quad{\rm when}\,\,\, m=p\,\,\,{\rm
or}\,\,\,n=q
}
One should keep in mind that \ZZreduce--\ZZvanishes\ are meant to
be true modulo BRST null states.

As implied by the comment below \modchinc, the states with $t=0$ appearing in
\ZZreduce\ are identical to the ZZ boundary states for generic $b$, which can
be written as differences of just two FZZT states \MartinecKA:
\eqn\ZZbsnaive{\eqalign{
    &|\,t=0,m,n\rangle = |\,z=\cos {\pi\sigma(m,n)\over\sqrt{pq}}\rangle -
    |\,z=\cos {\pi\sigma(m,-n)\over\sqrt{pq}}\rangle\cr
    &\quad =2\sum_{k',l'} \int_{0}^{\infty}dP\,
    \sinh ({2\pi m P\over b})\sinh (2\pi n P b){\Psi^{*}(P) \sqrt{S(1,1;k',l')}
    }|P\rangle\rangle_L|k',l'\rangle\rangle_M\cr
}}
with
\eqn\sigmamn{
    \sigma(m,n)=i\left({m\over b}+n b\right)\ .
}
These expressions will be useful below. The boundary cosmological
constant corresponding to $\sigma(m,n)$ is
\eqn\zzmuB{
    \mu_B(m,n)=\sqrt{\mu}\,(-1)^{m} \cos \pi n b^2\ .
}
Thus the two subtracted FZZT states in \ZZbsnaive\ have the same
boundary cosmological constant. In the next section, we will
interpret geometrically this fact, together with the formula
\ZZreduce\ for the general ZZ boundary state.

Using the identifications \ZZreduce--\ZZvanishes, we can reduce
any ZZ brane down to a linear combination of $(t=0,m,n)$ branes
with $1\le m\le p-1$, $1\le n\le q-1$ and $mq-np> 0$. We will call
these $(p-1)(q-1)/2$ branes the {\it principal ZZ branes}. It is
easy to see from \ZZbsnaive\ that the one-point functions of
physical operators are sufficient to distinguish the principal ZZ
branes from one another. Thus the principal ZZ branes form a
complete and linearly independent basis of physical states with
the ZZ-type boundary conditions.

We will conclude this section by discussing an interesting feature
of the principal ZZ branes. The ground ring one-point functions in
the principal ZZ brane states can be normalized so that (we drop
the label $t=0$ from these states from this point onwards)
 \eqn\zzoneptgr{
    \langle \hat{\CO}_{r,s}| m,n\rangle =
    U_{s-1}\left((-1)^m\cos {\pi np\over q}\right)
    U_{r-1}\left((-1)^n\cos {\pi mq\over p}\right)
        \langle 0| m,n\rangle\ ,
}
where $\langle 0|m,n\rangle$ denotes the ZZ partition function
(i.e.\ the one-point function of the identity operator). This is
consistent with the ring multiplication rule
\eqn\grmultrs{
    \hat{\CO}_{r,s}=U_{s-1}(\hat x) U_{r-1}(\hat y)
}
assuming that the principal ZZ branes are eigenstates of the ring generators:
\eqn\zzoneptgrgen{\eqalign{
    \hat{x}| m,n\rangle&=
    x_{mn}| m,n\rangle\cr
    \hat{y}| m,n\rangle&=
    y_{mn}| m,n\rangle\cr
}}
with eigenvalues
\eqn\zzoneptgrev{
    x_{mn}=(-1)^m\cos {\pi np \over q},\qquad y_{mn}=(-1)^n\cos {\pi mq\over
    p}\ .
}
Assuming the principal ZZ branes are eigenstates of the ring
elements, the expression \zzoneptgr\ constitutes an independent
derivation of our ansatz \ringfid\ for the $\mu$-deformed ring
multiplication. This derivation allows us to avoid the explicit
computation of minimal model and Liouville OPEs that would have
otherwise been necessary to obtain \ringfid.

Let us make a few more comments on the result \zzoneptgrgen.
 \lfm{1.} It is clear that the general FZZT boundary
state labelled by $\sigma$ will not be an eigenstate of the ring generators:
this property is special to the ZZ boundary states.
 \lfm {2.} Once we have normalized the ring elements to bring their one-point
functions to the form \zzoneptgr, the ZZ branes with other matter
labels $(k,l)$ will not be eigenstates of the ring elements. Of
course, we could have normalized the ring elements with respect to
a different $(k,l)$; then \zzoneptgr\ would have applied to this
matter label. But in view of the decomposition \fzztstrelate, it
is natural to assume that the branes with matter label $(1,1)$ are
eigenstates of the ring.
 \lfm{3.} Finally, notice that $x_{mn}$ is essentially the value of the
boundary cosmological constant \zzmuB\ associated with the $(m,n)$ ZZ brane. In
the next section, we will see what $y_{mn}$ corresponds to.

\newsec{Geometric interpretation of minimal string theory}

\subsec{The surface $\CM_{p,q}$ and its analytic structure}

In this section, we will provide a geometric interpretation of minimal string
theory. We will see how the structure of an auxiliary Riemann surface can
explain many of the features of the FZZT and ZZ branes that we found above. Of
primary importance will be the partition function $Z$ of the FZZT boundary
state. This partition function depends on the bulk and boundary cosmological
constants, $\mu$ and $\mu_B$. Differentiating with respect to $\mu$ gives the
expectation value of the bulk cosmological constant operator with FZZT boundary
conditions:
\eqn\muevfzzt{
\partial_{\mu}Z\big|_{\mu_B}=\langle c\bar{c}\,V_b \rangle\big|_{\rm FZZT}\ .
}
Using the formulas \FZZTbs\ and \FZZTwvfn\ for the FZZT boundary state with
$P=i(Q/2-b)$, we obtain
\eqn\muevfzztbec{
    \partial_{\mu}Z\big|_{\mu_B}=
    {1\over 2(b^2-1)}(\sqrt{\mu})^{1/b^2-1}\cosh\left(b-{1\over
    b}\right)\pi\sigma\ ,
}
where we have fixed the normalization of $Z$ for later convenience. We have
also suppressed the dependence on the matter state, since this will be $\mu_B$
and $\mu$ independent. Integrating \muevfzztbec\ with respect to $\mu$ then
gives
\eqn\Zfzzt{
Z={b^2\over b^4-1}(\sqrt{\mu})^{1/b^2+1}\left(b^2 \cosh\pi b \sigma\, \cosh{\pi
\sigma\over b}-\sinh\pi b \sigma\,\sinh{\pi \sigma\over b}\right)\ .
}
Finally, we can differentiate with respect to $\mu_B$ at fixed $\mu$ to obtain
the relatively simple formula:
\eqn\ymfzzt{
    \partial_{\mu_B}Z\big|_{\mu}=(\sqrt{\mu})^{1/b^2}\cosh{\pi \sigma\over b}\
    .
}
The normalization of $Z$ was chosen so that the coefficient of this expression
would be unity.

We can similarly consider the dual FZZT brane, which is given in terms of the
dual bulk and boundary cosmological constants $\tilde\mu$ and $\tilde{\mu}_B$.
The former was defined in \dualmu, while the latter is given by
\eqn\dualmu{
    {\tilde{\mu}_B\over \sqrt{\tilde{\mu}}}=\cosh{\pi \sigma\over b}\ ,
}
where again we have rescaled the usual definition of $\tilde{\mu}_B$ by a
convenient factor. It was observed in
\refs{\ZamolodchikovAA\FateevIK\TeschnerRV-\ZamolodchikovAH} that the Liouville
observables are invariant under $b\rightarrow 1/b$ provided one takes $\mu,
\mu_B \rightarrow \tilde\mu, \tilde\mu_B$ as well. Thus the dual brane provides
a physically equivalent description of the FZZT boundary conditions. Mapping
$b\rightarrow 1/b$ and applying the transformation \dualmu\ to the FZZT
partition function \Zfzzt, we find the dual partition function
\eqn\Zfzztdual{
    \tilde{Z} =
    {b^2\over 1-b^4}(\sqrt{\tilde{\mu}})^{b^2+1}\left({1\over b^2}
    \cosh{\pi b \sigma}\, \cosh{\pi \sigma\over b}-\sinh\pi b \sigma\,
    \sinh{\pi\sigma\over b}\right)\ .
}
Note that $\tilde{Z}\ne Z$, although the formula \FZZTwvfn\ for the one-point
functions is self-dual under $b\rightarrow 1/b$. The two loops are physically
equivalent by construction: physical observables can be calculated equivalently
with either the FZZT brane or its dual. Differentiating \Zfzztdual\ with
respect to $\tilde{\mu}_B$ (while holding $\tilde{\mu}$ fixed) leads to
\eqn\ymfzztdual{
    \partial_{\tilde{\mu}_B}\tilde{Z}\big|_{\tilde\mu}=(\sqrt{\tilde{\mu}})^{b^2}\cosh{\pi b
    \sigma}\ .
}

So far, the discussion has been for general $b$. Now let us consider what
happens at the special, rational values of $b^2=p/q$ that describe the $(p,q)$
minimal string theories. Let us define the dimensionless variables
\eqn\defvar{\eqalign{
    &x={\mu_B\over\sqrt{\mu}}\ , \qquad y ={\partial_{\mu_B}Z\over\sqrt{\tilde{\mu}}}\ ,\cr
    &\tilde{x}={\tilde{\mu}_B\over\sqrt{\tilde{\mu}}}\ ,
    \qquad \tilde{y} ={\partial_{\tilde{\mu}_B}\tilde{Z}\over\sqrt{\mu}}\ .\cr
}}
Then we can rewrite the equations \ymfzzt\ and \ymfzztdual\ as polynomial
equations in these quantities:
\eqn\twomatrespoly{\eqalign{
    F(x,y)= T_q(x)-T_p(y) &=0 \cr
    \tilde{F}(\tilde{x},\tilde{y})= T_p(\tilde{x})-T_q(\tilde{y}) &=0\ .\cr
}}
Therefore, $x$ and $y$ have a natural analytic continuation to a Riemann
surface $\CM_{p,q}$ described by the curve $F(x,y)=0$. Moreover, since
$\tilde{F}(x,y)=F(y,x)$, the dual FZZT brane gives rise to the {\it same}
Riemann surface. In fact, it is clear from \ymfzzt\ and \ymfzztdual\ that
$\tilde{x}=y$ and $\tilde{y}=x$.

The existence of the auxiliary Riemann surface $\CM_{p,q}$
suggests that we recast our discussion in a more geometric
language. Consider first the FZZT branes. The definition \defvar\
of $y$ and $\tilde{y}$ implies that we can think of the FZZT
partition function and its dual as integrals of the one-forms
$y\,dx$ and $x\,dy$:
\eqn\fzztint{
    Z(\mu_B)=\mu^{{p+q\over2p}}\int_{\CP}^{x(\mu_B)}y\,dx\ ,\quad\quad
    \tilde{Z}(\tilde{\mu}_B)=\tilde{\mu}^{{p+q\over2q}}\int_{\CP}^{y(\tilde{\mu}_B)}x\,dy
}
for some arbitrary fixed point $\CP\in \CM_{p,q}$. The fact that $Z\ne
\tilde{Z}$ is simply the statement that $y\,dx$ and $x\,dy$ are distinct
one-forms on $\CM_{p,q}$. It is true, however, that they are related by an
exact form:
\eqn\fzztexform{
y\,dx+x\,dy = d(x y)\ .
}
It is not surprising then that the brane and the dual brane are physically
equivalent descriptions of the FZZT boundary conditions.

Integrating \fzztexform\ leads to another interesting result. We can think of
$\int y\,dx$ and $\int x\,dy$ as effective potentials $V_{eff}(x)$ and
$\tilde{V}_{eff}(y)$. Then \fzztexform\ implies that these effective potentials
are related by a Legendre transform:
\eqn\legendreVeff{
V_{eff}(x) = x y - \tilde{V}_{eff}(y)\ .
}
It seems then that $x$ and $y$ play the role of coordinate and conjugate
momentum on our Riemann surface. We will comment on this interpretation some
more when we discuss the matrix model description in section 9.

We can also give a geometric interpretation to the parameter $z$
defined in \zdef. Notice that $z$ is related to the coordinates of
our surface via
\eqn\uniformII{
x=T_p(z),\quad y=T_q(z)
}
Thus $z$ can be thought of as a uniformizing parameter of $\CM_{p,q}$ that
covers it exactly once. The parameter $\sigma$ also gives a uniformization of
the surface, but it covers the surface infinitely many times. This is
consistent with what we saw in section 3, that the parametrization of the FZZT
branes by $\sigma$ is redundant, while the $z$ parametrization is unambiguous.
We conclude that points on the surface $\CM_{p,q}$ are in one to one
correspondence with the distinct branes.

In terms of the parameter $z$, the surface $\CM_{p,q}$ appears to have genus
zero, since $z$ takes values on the whole complex plane. However, there are
special distinct values of $z$ that correspond to the same point $(x,y)$ in
$\CM_{p,q}$. Such points are singularities of $\CM_{p,q}$, and they can be
thought of as the pinched $A$-cycles of a higher genus surface. It is easy to
see that the singularities correspond to the following points in $\CM_{p,q}$:
\eqn\xysing{\eqalign{
    &(x_{mn},y_{mn})=\left((-1)^{m}\cos {\pi np\over q}\ ,\,\,
                    (-1)^n\cos{\pi mq\over p}\right),\cr
}}
which come from pairs $z^{+}_{mn}$ and $z^{-}_{mn}$, with
\eqn\zsing{
z^{\pm}_{mn}=\cos {\pi (mq\pm np)\over pq}
}
with $m$ and $n$ ranging over
\eqn\mnrange{
    m=1,\dots,p-1,\quad
    n=1,\dots,q-1,\quad mq-np>0
}
Therefore there are exactly $(p-1)(q-1)/2$ singularities of $\CM_{p,q}$.

These singularities can also be found directly from the curve \twomatrespoly\
by solving the equations
\eqn\singcond{
    F(x,y)=0\ , \quad \partial_x F(x,y)=\partial_y F(x,y)=0\ .
}
It is straightforward to check that this leads to the same conclusion as
\xysing.

With this understanding of the singularities of $\CM_{p,q}$, the geometric
interpretation of the ZZ branes is clear. Recall that the ZZ branes were found
to be eigenstates of the ground ring generators, with eigenvalues given in
\zzoneptgrgen. It is easy to see that these eigenvalues lie on the curve
$F(x,y)=0$; therefore the ground ring generators measure the location of the
principal ZZ branes on $\CM_{p,q}$. Moreover, comparing \zzoneptgrgen\ with the
singularities \xysing, we find that they are the same. Therefore, the principal
ZZ branes are located at the singularities of $\CM_{p,q}$!

In fact, since each principal brane is located at a different
singularity, and there are exactly as many such branes as there
are singularities, every singularity corresponds to a unique
principal ZZ brane and vice versa. All of the other $(t,m,n)$ ZZ
branes are related to multiples of principal branes by a BRST
exact state, and thus the location of a ZZ brane on $\CM_{p,q}$
determines its BRST cohomology class. Notice that branes that do
not correspond to singularities, namely those with $m=0\,\mod\,p$
or $n=0\,\mod\,q$, are themselves BRST null.

\medskip \ifig\figI{The surface $\CM_{p,q}$,
    along with examples of FZZT and ZZ brane contours, shown here for $(p,q)=(2,5)$.
    Here $\CP$ is an
    arbitrary, fixed point in $\CM_{p,q}$, while $z$ is the uniformizing parameter of $\CM_{p,q}$
    that corresponds through \uniformII\ to the boundary
    cosmological constant and the dual boundary cosmological constant of the associated FZZT brane.
    The points $z_{1,1}$ and $z_{1,2}$
    are the pinched $A$-cycles of the surface, and they correspond
    to the $(1,1)$ and $(1,2)$ principal ZZ branes. The $(m,n)$ ZZ contour is clearly
    a difference of two FZZT contours beginning at $\CP$ and ending at $z_{mn}$.}
    {\epsfxsize=0.6\hsize\epsfbox{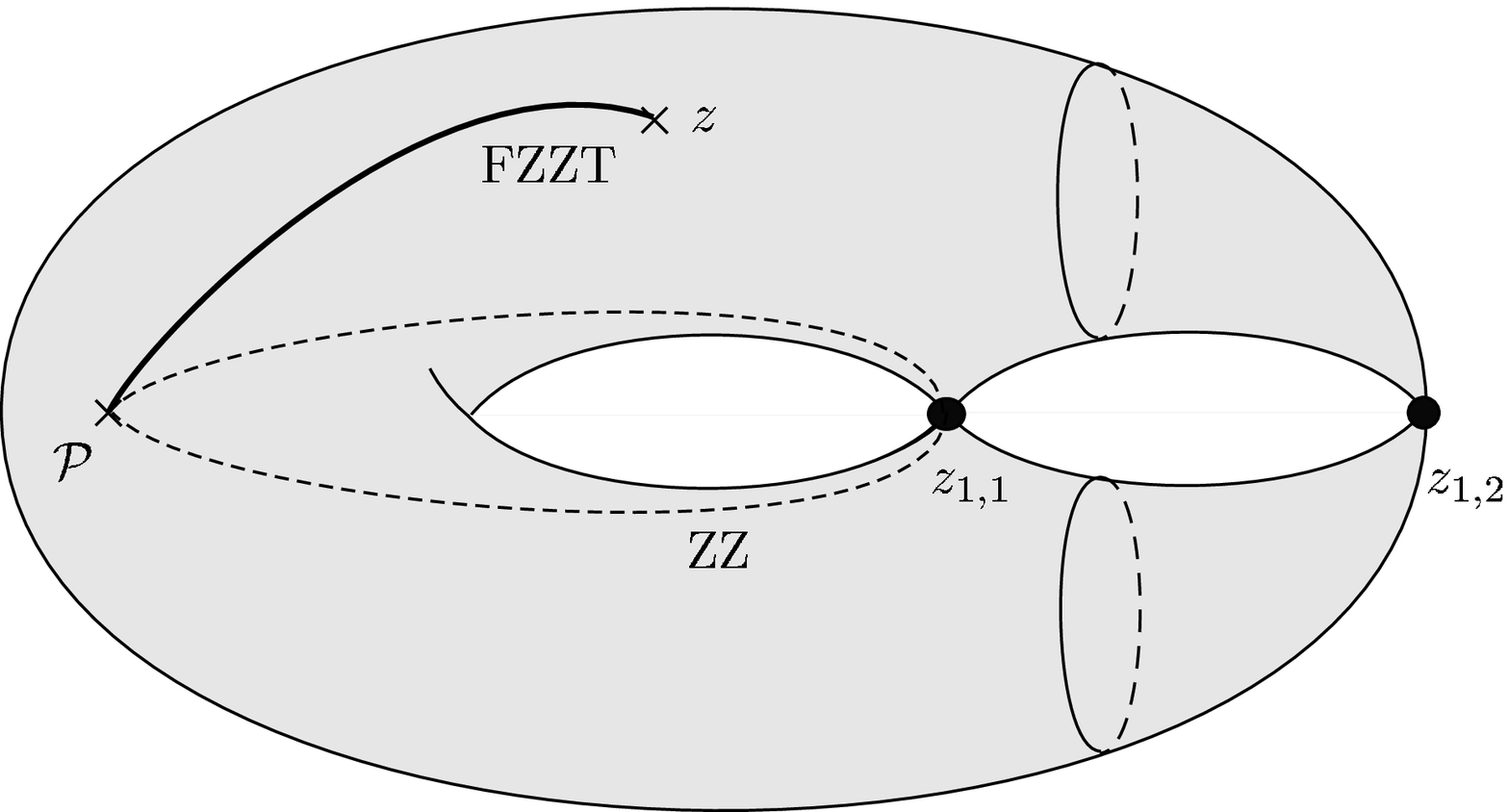}}

The relation \ZZbsnaive\ between the FZZT and the principal ZZ
branes also has a simple geometric interpretation. The fact that
the FZZT branes are equivalent to line integrals of $y\,dx$ on
$\CM_{p,q}$ implies that the ZZ branes can be thought of as {\it
periods} of $y\,dx$ around cycles $B_{m,n}$ of $\CM_{p,q}$ that
run between the singularity $(x_{mn}, y_{mn})$ and an open cycle
of $\CM_{p,q}$:
\eqn\zzint{
    Z_{m,n}=\mu^{p+q\over 2p}\oint_{B_{m,n}}y\,dx\ .
}
This generalizes an observation in \KlebanovWG. An illustration of
the surface $\CM_{p,q}$ and the various contour integrals is shown
in figure 1. Note that since the ZZ branes are defined using
closed contours, they will be the same whether we use the FZZT
brane or its dual. Thus, the (physically equivalent) dual brane
does not lead to an overcounting of the ZZ brane spectrum.
Actually, the relation between FZZT and principal ZZ boundary
states implies a stronger statement than \zzint:
\eqn\zzintbdst{
    | m,n\rangle_{\rm ZZ} = \oint_{B_{m,n}}\partial_{x}|  z(x)\rangle_{\rm
    FZZT}\,\,dx\ .
}
That is, we can still think of the ZZ branes as periods on
$\CM_{p,q}$ of the FZZT branes, even at the more general level of
boundary states.

Our geometrical interpretation implies that all of the distinct principal ZZ
branes exist and play a role in minimal string theory. We can also ask for the
interpretation of the general $(t,m,n)$ ZZ brane (let us keep the matter state
$(1,1)$ for simplicity). From \ZZreduce, the answer is clear. The $(t,m,n)$ ZZ
brane is simply $(t+1)$ copies of a closely related principal ZZ brane, and
therefore it corresponds to a contour integral of $y\,dx$ that winds $(t+1)$
times around the $B$-cycle of the principal ZZ brane.

We should point out that a proposal for the interpretation of the $(1,n)$
branes in the $c=1$ matrix model was recently advanced \AlexandrovUN. It would
be interesting to understand its relation to our work.

\subsec{Deformations of $\CM_{p,q}$}

So far, the discussion has been entirely in the ``conformal background" where
$\mu\ge 0$ and we have a simple world-sheet description of the theory. We have
seen that in this case, the surface $\CM_{p,q}$ is given by a very simple curve
$T_{q}(x)-T_{p}(y)=0$. Here we will consider deformations of this curve, and
for simplicity we will require that they simply shift the
$N_{p,q}={(p-1)(q-1)/2}$ singularities of $\CM_{p,q}$:
\eqn\shiftsing{
    (x_{i},y_{i})\rightarrow (x_{i}+\delta x_{i},y_{i}+\delta y_{i})\ ,
}
without changing their total number. This leads to constraints on the form of
the deformations. To see what these constraints are, it is sufficient to work
to linear order in the deformation. Before the deformation, the singularities
satisfy the following equations:
\eqn\singbef{\eqalign{
    &T_{q}(x_{i})-T_{p}(y_{i})=0\cr
    &T_{q}'(x_{i})=T_{p}'(y_{i})=0\ .\cr
}}
After the deformation, the first equation becomes
\eqn\singaft{\eqalign{
    0&=T_{q}(x_{i}+\delta x_{i})-T_{p}(y_{i}+\delta y_{i})+\delta F(x_{i},y_{i})\cr
    &=T_{q}(x_{i})-T_{p}(y_{i})+
    T_{q}'(x_{i})\delta x_{i}-T_{p}'(y_{i})\delta y_{i}+\delta F(x_{i},y_{i})\ .\cr
}}
Using \singbef, we see that the deformation must vanish at the singularities.
This gives $N_{p,q}$ constraints on the polynomial $\delta F(x,y)$. The
deformation of the second and third equations of \singbef\ gives no further
constraints; instead, solving them yields formulas for $\delta x_{i}$ and
$\delta y_{i}$.

Intuitively, we expect such deformations of $\CM_{p,q}$ to
correspond in the minimal string theory to perturbations of the
background by physical vertex operators:
\eqn\defpq{\eqalign{
    &\delta S=t_{r,s}V_{r,s}\cr
    &r=1,\dots,p-1,\quad s\le q-1,\quad s\ne 0\,\mod\,q, \quad q r-p s> 0\ .\cr
}}
These vertex operators exist at all ghost numbers $\le 1$ \LianGK. They have
KPZ scaling
\eqn\cftpqscal{
    t_{r,s} \sim \mu^{p+q-q r+p s\over 2p}
}
and their Liouville momenta are given by \dressL, with the range of $s$
extended as in \defpq.

In order to relate these perturbations to the deformations of
$\CM_{p,q}$, notice that to leading order in $t_{r,s}$, the change
in the FZZT partition function under the perturbation
\defpq\ is simply the one-point function $t_{r,s}\langle
V_{r,s}\rangle$ in the FZZT boundary state. Using \FZZTbs, we find
that
\eqn\fzztoptach{
    \delta Z=t_{r,s}\langle
    V_{r,s}\rangle=t_{r,s}\mu^{{qr-ps\over 2p}}\cosh \left({q r-p
    s\over p}\right)\pi b \sigma\ .
}
(The matter wavefunction will be irrelevant for this calculation, as will be
other overall normalization factors.) Since $y=\mu^{-{p+q\over
2p}}\partial_{x}Z$, we find that if we hold $x$ fixed, then $y$ is deformed by
\eqn\udef{
    \delta y \sim \tilde{t}_{r,s}{\sinh \left({q r-ps\over p}\right)\pi b
    \sigma\over \sinh\pi b\sigma}\ ,
}
where we have defined the dimensionless parameter
$\tilde{t}_{r,s}=t_{r,s}\mu^{{qr-ps-p-q\over 2p}}$. In terms of $x$ and $y$,
the deformation of the curve is then
\eqn\ymdefudef{
    \delta_{r,s} F(x,y)=p\,U_{p-1}(y)\delta y \sim
    \tilde{t}_{r,s}\bigg(U_{q-1}(x)T_{s}(x)U_{r-1}(y)-U_{p-1}(y)T_{r}(y)U_{s-1}(x)\bigg)\
    .
}
Here we have extended the definition of the Chebyshev polynomials to negative
$s$ in a natural way: $U_{-s-1}(x)=-U_{s-1}(x)$. In particular, $U_{-1}(x)=0$.
Finally, using the original curve $F(x,y)=0$, we can obtain a more compact
formula for the deformations of $\CM_{p,q}$ corresponding to the bulk physical
operators \defpq:
\eqn\ymdefbasis{\eqalign{
    &\delta_{r,s}F(x,y)=\tilde{t}_{r,s}\bigg(
    U_{q-s-1}(x)U_{r-1}(y)-U_{p-r-1}(y)U_{s-1}(x)\bigg)\cr
    &r=1,\dots,p-1\ ,\quad s\le q-1\ ,\quad s\ne 0\,\mod\,q\ , \quad q r-p s> 0\ .
}}
These deformations of $\CM_{p,q}$ indeed vanish on the singularities, where
$U_{q-1}(x)=U_{p-1}(u)=0$. Therefore, all of the perturbations of the string
theory background by bulk physical vertex operators correspond to
singularity-preserving deformations of $\CM_{p,q}$.

It is interesting to ask whether the converse is also true, i.e.\ whether every
possible singularity-preserving deformation of $\CM_{p,q}$ is given by
\ymdefbasis. A careful analysis reveals that a complete basis of such
deformations is indeed given by the same formula \ymdefbasis\ for
$\delta_{r,s}F$, but with $r$ and $s$ taking values in a slightly larger range:
\eqn\rsrange{
r=1,\dots,p\ ,\quad s\le q-1\ , \quad q r-p s > 0\ .
}
Thus, there are deformations of $\CM_{p,q}$ that do not correspond to
perturbations by bulk physical vertex operators. These have $r=p$ or
$s=0\,\mod\,q$. It is easy to see that the former category can be thought of as
reparametrizations of $y$ alone:
\eqn\reparamu{
y\rightarrow y+\tilde{t}_{p,s}U_{q-s-1}(x)\ , \quad  s\le q-1\ .
}
Deformations with $s=0\,\mod\,q$ are more complicated, but they also correspond
to polynomial reparametrizations of $x$ alone.

The worldsheet interpretation of these extra deformations with $r=p$ or
$s=0\,\mod\,q$ is not always clear. Recall however that the surface $\CM_{p,q}$
originally arose from the disk amplitude of the FZZT brane. It is reasonable
then to expect the deformations associated with such reparametrizations to be
associated with open strings on the FZZT brane. For example, $U_{q-1}(x)$ is
the boundary length operator as in \MartinecHT. A more detailed correspondence
would require a better understanding of open minimal string theory. While this
is certainly an interesting problem, it is beyond the scope of this paper.

In order to examine the relevance of the perturbations we assign
weight $(p,q)$ to $(x,y)$ so that $F(x,y)$ is a quasi-homogeneous
polynomial of degree $pq$, the degree of $\delta_{r,s}F(x,y)$ is
always $q(r-1)+p(q-s-1)$. This is less than the degree of $F(x,y)$
only for
\eqn\onlyfor{
r q-s p < p+q\ .
}
Such deformations of $\CM_{p,q}$ become important at small $x$ and $y$. In the
string theory, they correspond to perturbations by tachyons $\CT_{r,s}$ with
positive KPZ scaling. These are, of course, precisely the perturbations that
become increasingly relevant in the IR. This agrees well with the intuition
that flowing to the IR in the string theory corresponds to taking $x$ and $y\to
0$ in $\CM_{p,q}$.

So far we have limited ourselves to deformations which do not open
the singularities to smooth $A$ cycles.  It is then natural to ask
what deformations which smooth the singularities correspond to.
Given that the $(m,n)$ ZZ brane is associated with the $(m,n)$
singularity, it is reasonable to expect that a background with
$(m,n)$ ZZ branes is described by a Riemann surface with the cycle
$A_{m,n}$ smoothed out.  Since a small number of D-branes does not
affect classical string theory, we expect that the number of
D-branes $N_{m,n}$ needed to change the Riemann surface to be of
order $1/g_s$.  This number can be measured by computing the
period of the one form $ydx$ around the $A$ cycle
 \eqn\Nmn{\oint_{A_{m,n}} y dx = g_s N_{m,n}\ .}
The cycle $A_{m,n}$ is conjugate to $B_{m,n}$ because they have
intersection number one. Therefore, the ZZ brane creation operator
\zzintbdst\ which is an integral around $B_{m,n}$ changes
$N_{m,n}$ which is conjugate to it by one unit. This
interpretation of smoothing the $(m,n)$ singularity is similar to
the picture which has emerged in recent studies of four
dimensional gauge theories and matrix models
\refs{\DijkgraafDH,\CachazoRY}.

\newsec{Minimal superstring theory}

\subsec{Preliminaries}

We now turn to the study of minimal superstring theory. We start by reviewing
the $(p,q)$ superminimal models. For these theories, it will be more convenient
to work in $\alpha'=2$ units. As in the bosonic theories, one must have $p, q
\ge 2$. Moreover, one must have either $(p,q)$ odd and coprime, or else $(p,q)$
even, $(p/2,q/2)$ coprime, and $(p-q)/2$ odd (the last condition follows from
modular invariance \refs{\DiFrancescoXZ,\KlebanovWG}). The superminimal models
have central charge
\eqn\chatspq{
    \hat{c}=1-{2(p-q)^2\over p\,q}
}
As in the bosonic theories, primary operators $\CO_{r,s}$ are labelled by
integers $r$ and $s$ with $r=1, \dots, p-1$ and $s=1, \dots, q-1$ and
$\CO_{p-r,q-s}\equiv \CO_{r,s}$. The crucial difference here however is that we
must distinguish between NS and the R sector operators. The NS (R) operators
have $r-s$ even (odd). As a result, the operator dimensions are given by a
slightly more complicated expression:
 \eqn\dimspq{\eqalign{
    \Delta(\CO_{r,s})=\overline \Delta(\CO_{r,s}) &={(r q-s p)^2-(p-q)^2\over
     8pq}+{1-(-1)^{r-s}\over 32}\ .
 }}
Of particular interest is the operator $\CO_{{p\over 2},{q\over 2}}$ which
exists only in the $(p,q)$ even theories.  It corresponds to the supersymmetric
Ramond ground state.  Since it is absent in the $(p,q)$ odd theories, they
break supersymmetry.

The operators of the $(p,q)$ superminimal models obey fusion rules identical to
those of the bosonic models \fusionpq. Note that the ``generators" $\CO_{1,2}$
and $\CO_{2,1}$ of the $(p,q)$ superminimal models are in the R sector, and
thus the fusion of these operators with a general R (NS) primary results in a
set of NS (R) operators.

Now consider super-Liouville theory with central charge
 \eqn\chatsL{
    \hat{c}=1+2Q^2=1+2\left(b+{1\over b}\right)^2\ .
 }
The basic vertex operators of super-Liouville are the NS operators
$N_\alpha=e^{\alpha\phi}$ and the R operators
$R_\alpha^{\pm}=\sigma^{\pm}e^{\alpha\phi}$. These have dimensions
\eqn\dimsL{\eqalign{
    &\Delta(N_\alpha)=\bar{\Delta}(N_\alpha)={1\over
    2}\alpha(Q-\alpha)\cr
    &\Delta(R_\alpha^\pm)=\bar{\Delta}(R_\alpha^\pm)={1\over
    2}\alpha(Q-\alpha)+{1\over 16}\ .\cr
 }}
Here $\sigma^{\pm}$ denotes the dimension $1/16$ spin fields of
the super-Liouville theory.  If we study super-Liouville as an
isolated quantum field theory, only one of these fields exists
following the GSO projection, say $R_\alpha^-$. However, if we
combine it with another ``matter'' theory, we sometimes need both
of them before performing a GSO projection on the combined theory.

The supersymmetric Ramond ground state has $\alpha={Q\over 2}$ and hence
$\Delta={\hat c \over 16}$. It does not correspond to two degenerate fields
$R^\pm$ which are related by the action of the supercharge. Instead, from
solving the minisuperspace equations we find two possible wave functions
$\psi^+ = e^{\mu e^{b\phi}}$ with $\psi^-=0$, or $\psi^- = e^{-\mu e^{b\phi}}$
with $\psi^+=0$ \DouglasUP.  Imposing that the wave function goes to zero as
$\phi \to +\infty$, we see that depending on the sign of $\mu$
 \eqn\hatsdef{
 \zeta = {\rm sign}(\mu)\ ,
 }
we have only one wave function $\psi^{-\zeta}=e^{-|\mu|
e^{b\phi}}$ and $\psi^{\zeta}=0$. Therefore, in pure
super-Liouville theory, where we keep only $R_{\alpha={Q\over
2}}^-$, the Ramond ground state exists only for positive $\mu$.
Below we will review how this conclusion changes when we add the
matter theory.

The degenerate primaries of super-Liouville are also similar to those of
ordinary Liouville. These are given by
\eqn\degensuperL{\eqalign{
    &N_{\alpha_{r,s}}=e^{\alpha_{r,s}\phi}\ , \quad r-s\in 2\Bbb Z\cr
    &R_{\alpha_{r,s}}^{\pm}=\sigma^{\pm}e^{\alpha_{r,s}\phi}\ ,
    \quad r-s\in 2\Bbb Z+1\cr
    &2\alpha_{r,s}={1\over b}(1-r)+b(1-s)\ .\cr
 }}
The analogue of the bosonic fusion rules \opefusionLa, \opefusionLb\ are for
super-Liouville theory \refs{\RashkovJX\PoghosianDW\FukudaBV-\AhnEV}:
\eqn\opefusionsL{\eqalign{
    R_{-{b\over2}}R_\alpha&=[N_{\alpha-{b\over 2}}]+C_{-}^{(R)}
    (\alpha)[N_{\alpha+{b\over 2}}]\cr
    R_{-{b\over2}}N_\alpha&=[R_{\alpha-{b\over 2}}]+C_{-}^{(NS)}
    (\alpha)[R_{\alpha+{b\over
    2}}]\cr
    C_{-}^{(R)}(\alpha)&={1\over 4}\mu b^2 {\gamma(\alpha b-{b^2\over2})\over
    \gamma\left({1-b^2\over 2}\right)\gamma\left(\alpha b+{1\over2}\right)}\cr
    C_{-}^{(NS)}(\alpha)&={1\over 4}\mu b^2 {\gamma(\alpha b-{b^2\over2}-{1\over2})\over
    \gamma\left({1-b^2\over 2}\right)\gamma\left(\alpha b\right)}\cr
}}
with similar expressions for $R_{-{1\over2b}}$ with $b\rightarrow 1/b$ and
$\mu\rightarrow \tilde{\mu}$, where $\tilde{\mu}$ is the dual cosmological
constant $\pi \tilde{\mu}\gamma({Q\over 2b})=(\pi \mu \gamma({b Q\over
2}))^{1/b^2} $. As in the bosonic string, we will rescale $\mu$ and
$\tilde{\mu}$ so that they are more simply related
\eqn\dualmusuper{
    \tilde{\mu}=\mu^{1/b^2}\ .
}

We now couple the super-minimal matter theory to the super-Liouville theory to
form the minimal superstring. Imposing that the combined system has the correct
central charge fixes the Liouville parameter
 \eqn\bvalues{b=\sqrt{p\over q}}
The tachyon operators $\CT_{r,s}$ are obtained by dressing the matter operators
$\CO_{r,s}$ with
 \eqn\dresssuperL{\eqalign{
    &N_{\beta_{r,s}}=e^{\beta_{r,s}\phi}\ ,\quad r-s \in 2\Bbb Z \cr
    &R_{\beta_{r,s}}^\pm=\sigma^{\pm} e^{\beta_{r,s}\phi}\ , \quad
    r-s
    \in 2{\Bbb Z}+1\cr
    &2\beta_{r,s}={p+q-r q+s p\over \sqrt{p\,q}}\ , \quad r q-s p\ge 0
 }}
and the appropriate superghosts.

The matter Ramond operators are labelled by their fermion number
$\CO_{r,s}^\pm$. (In the superminimal model without gravity we keep only one of
them, say $\CO_{r,s}^+$.) This fermion number should be correlated with the
fermion number in super-Liouville. However, there is an important subtlety
which should be explained here.  So far we discussed two fermion number
operators $(-1)^{f^{L,M}}$, one in Liouville and one in the matter sector of
the theory. Their action on the lowest energy states in a Ramond representation
is proportional to $i G_{0}^{L,M} \tilde G_0^{L,M} $.  When we combine the
Liouville and the matter states it is natural to define left and right moving
fermion numbers $(-1)^{f_L}$ and $(-1)^{f_R}$ such that their action on the
lowest energy states in a Ramond representation is proportional to $ i G^L_0
G^M_0$, and $ i \tilde G^L_0 \tilde G^M_0$ respectively.  Then it is
conventional to define the total fermion number as $(-1)^f=(-1)^{f_L+f_R}$.
Therefore,
 \eqn\totalferm{(-1)^f=(-1)^{f_L+f_R}=(-1)^{f^{L}+f^{M}+1}\ ,
 }
i.e.\ the total (left plus right) worldsheet fermion number
differs from the sum of the fermion numbers in the Liouville and
the matter parts of the theory.

Consider now the vertex operators in the $(-1/2,-1/2)$ ghost
picture. In the 0B theory we project on $(-1)^f=1$.  Therefore,
following \totalferm\ we project on $(-1)^{f^{L}+f^{M}}=-1$.  The
candidate operators are $\CO_{r,s}^\pm R_{\beta_{r,s}}^\mp$ (we
suppress the ghosts). Only one linear combination of them is
physical \KlebanovWG.  The situation is more interesting when we
try to dress the supersymmetric matter groundstate
$(r,s)=({p\over2},{q\over2})$. Now there is only one matter
operator,\foot{This fact is not true in other systems like the
$\hat c=1$ theory.} say $\CO^+_{{p\over2}, {q\over2}}$. Hence it
should be dressed with $R_{Q\over2}^-$. But as we said above, this
operator exists only for positive $\mu$. Therefore, in the 0B
minimal superstring theories with $(p,q)$ even, the Ramond ground
state exists only for $\mu>0$. For $(p,q)$ odd, there is of course
no Ramond ground state to begin with.

The situation is somewhat different in the $(-1/2,-3/2)$ or
$(-3/2,-1/2)$ ghost picture, where because of \totalferm, we
project on $(-1)^{f^{L}+f^{M}}=+1$. For generic $(r,s)$ we can use
inverse picture changing to find a single vertex operator which is
a linear combination of $\CO_{r,s}^\pm R_{\beta_{r,s}}^\pm$. The
orthogonal linear combination is a gauge mode which is annihilated
when we try to picture change to the $(-1/2,-1/2)$ picture. There
are new subtleties for the Ramond ground state
$(r,s)=({p\over2},{q\over2})$. The operator
$\CO_{{p\over2},{q\over2}}^+ R_{Q\over2}^+$ exists only for
negative $\mu$. If we try to picture change it to the
$(-1/2,-1/2)$ picture we find zero. But there could be another
operator $\CO^+_{{p\over2},{q\over2}}\tilde R_{Q\over2}^+$, which
exists only for positive $\mu$ and is related by picture changing
to the Ramond ground state
$\CO^+_{{p\over2},{q\over2}}R_{Q\over2}^-$ we found in the
$(-1/2,-1/2)$ picture. The wave function of $\tilde R^+_{Q\over2}$
is obtained by solving the equation
 \eqn\wvfnRgs{
 (\partial_\phi - b \mu e^{b \phi}) \tilde \psi^+=
 e^{-\mu e^{b\phi}}
 }
subject to the boundary condition that $\tilde \psi^+ (\phi \to
+\infty)=0$. The answer is given by an incomplete Gamma function
 \eqn\asympt{
 \tilde \psi^+=  -{ e^{-\mu e^{b\phi} } \over b} \int
 _0^\infty {e^{-2t} \over t+ \mu e^{b\phi} } dt=\cases{
 \phi + ... & $\phi \to -\infty $ \cr
  -{1\over 2b\mu} e ^{-b\phi}e^{-\mu e^{b\phi}}(1 + ...)
 & $ \phi \to +\infty $\ . \cr}
 }
The asymptotic form as $\phi \to -\infty$ leads to the form of the
operator $\tilde R_{Q\over2}^+= \sigma^+\phi e^{{Q\over2} \phi}$.
Because of the factor of $\phi$, this is not a standard Liouville
operator and its analysis is subtle.  It is possible, however, to
picture change it to the $(-1/2,-1/2)$ picture, where it is
simple.

In the spacetime description of these theories, each operator is
the ``on-shell mode" of a field which depends on the Liouville
coordinate $\phi$. Ramond vertex operators in the $(-1/2,-3/2)$ or
$(-3/2,-1/2)$ picture describe the RR scalar $C$, while operators
in the $(-1/2,-1/2)$ picture describe its gradient
$\partial_{\phi}C$ (more precisely, $(\partial_\phi - b \mu e^{b
\phi}) C$ \KlebanovWG). Thus in the spacetime description, the
first candidate Ramond ground state
$\CO^+_{{p\over2},{q\over2}}R_{Q\over2}^+$, which exists only for
$\mu<0$ and is zero in the $(-1/2,-1/2)$ picture, describes the
constant mode of $C$. It decouples from all correlation functions
of local vertex operators but is important in the coupling to
D-branes. On the other hand, the second candidate operator for the
Ramond ground state $\CO^+_{{p\over2},{q\over2}}\tilde
R_{Q\over2}^+$, which exists only for $\mu>0$, corresponds
(asymptotically as $\phi \to -\infty$) to changes of
$\partial_\phi C$, i.e.\ it describes changes in RR flux.
Therefore its wave function is linear in $\phi$ in the
$(-1/2,-3/2)$ picture, and it is constant in the $(-1/2,-1/2)$
picture. To summarize, in the bulk we have for $\mu>0$ a Ramond
ground state operator that describes changes in RR flux, but no
such operator for $\mu<0$. But in the presence of D-branes, there
exists for $\mu<0$ a Ramond ground state operator that couples to
D-brane charge. In other words, for $\mu>0$ we can have RR flux
but no charged D-branes, while the opposite is true for $\mu<0$.

All of these 0B theories have a global ${\Bbb Z}_2$ symmetry which
acts as $-1$ ($+1$) on all R (NS) operators.  We refer to this
operation as $(-1)^{F_L}$ where $F_L$ is the left-moving spacetime
fermion number. Orbifolding by this symmetry yields the 0A
theories. In the 0A theories, all of the physical operators in the
Ramond sector are projected out. The only physical operators from
the twisted sector are constructed out of the Ramond ground state
$(r={p\over 2},s={q\over 2})$. They exist only when such an
operator could not be dressed in the 0B theory. Thus in 0A, we can
have RR flux but no charged D-branes for $\mu<0$ and the opposite
for $\mu>0$.

Finally we should discuss the operation of $(-1)^{f_L}$ with $f_L$ the
left-moving worldsheet fermion number.  This operation is an $R$-transformation
because it does not commute with the supercharge.  However, it is not a
symmetry of the theory for two reasons. First, the cosmological constant term
in the worldsheet action explicitly breaks the would-be symmetry: acting with
$(-1)^{f_L}$ sends $\mu \to -\mu$. The second reason this symmetry is broken is
that the transformation by $(-1)^{f_L}$ reverses the way the GSO projection is
applied in the Ramond sector. We saw that in the 0B theory, one linear
combination of the matter operator $\CO_{r,s}^\pm R_{\beta_{r,s}}^\mp$ was
physical. After the transformation by $(-1)^{f_L}$, the orthogonal linear
combination will be physical. As usual, the situation is more complicated for
the Ramond ground state. The $(p,q)$ odd models do not have a Ramond ground
state; therefore the spectrum of physical states before and after this
operation are identical. Hence, we expect that in these models, the theory with
$\mu$ is dual to the theory with $-\mu$. Moreover, at $\mu=0$ the operation of
$(-1)^{f_L}$ becomes a symmetry of the theory.\foot{A similar situation exists
in the $\hat c=1$ theory where the Ramond ground state appears twice with two
different fermion numbers \DouglasUP.}  On the other hand, the $(p,q)$ even
models have a Ramond ground state, so the spectrum at $\mu$ differs from the
spectrum at $-\mu$. An interesting special case is the pure supergravity theory
$(p=2,q=4)$ where the 0B theory at $\mu$ is the same as the 0A theory at $-\mu$
\KlebanovWG. For the $(p,q)$ even models, the transformation $(-1)^{f_L}$ is
generally not useful.

\subsec{The ground ring of minimal superstring theory}

As in the bosonic string, the minimal superstring theories have a ground ring
consisting of all dimension 0, ghost number 0 operators in the BRST cohomology
of the theory. Since this ground ring is nearly identical to that of the
bosonic string, our discussion will be brief. We will start with the 0B models.
Just as in the bosonic string, the ground ring has $(p-1)(q-1)$ elements. The
operator $\hat{\CO}_{r,s}$ has Liouville momentum $\alpha_{r,s}$ given by
\degensuperL, and it is constructed by acting on the product $\CO_{r,s}
V_{\alpha_{r,s}}$ with some combination of raising operators
\refs{\PandaGE\BouwknegtAM\ItohIX-\ItohIY}. Operators with $r-s$ even (odd) are
in the NS (R) sectors. We point out that since we use the Liouville momenta
$\alpha_{r,s}$ of \degensuperL\ rather than $\beta_{r,s}$ of \dresssuperL,
there is no subtlety associated with the dressing of the Ramond ground state.

When $\mu=0$, Liouville momentum is conserved in the OPE, and thus one expects
on kinematical grounds that just as in the bosonic models, the ground ring is
generated by the R sector operators $\CO_{1,2}$ and $\CO_{2,1}$:
\eqn\grmultsuperpq{
    \hat{\CO}_{r,s}=\hat{\CO}_{1,2}^{s-1}\hat{\CO}_{2,1}^{r-1}
}
with the ring relations
\eqn\grrelsuperpq{
\hat{\CO}_{1,2}^{q-1}=\hat{\CO}_{2,1}^{p-1}=0\ .
 }
For $\mu\ne 0$, Liouville momentum is no longer conserved, and instead one has
super-Liouville fusion rules \opefusionsL\ very similar to those of the bosonic
theory.  We expect the expressions for the ring elements \grmultsuperpq\ and
the relations \grrelsuperpq\ to be modified in the same way as in the bosonic
string.  Once again it will be convenient to define dimensionless generators
$\hat x$ and $\hat y$ as in \grrescdefII. Then \grmultsuperpq\ becomes
\eqn\grmultsuperpqm{
    \hat{\CO}_{r,s}=\mu^{q(r-1)+p(s-1)\over 2p} U_{s-1}(\hat x) U_{r-1}(\hat y)\
    ,
}
while the relations \grrelsuperpq\ become
\eqn\grrelsuperpqm{
    U_{q-1}(\hat x)=U_{p-1}(\hat y)=0\ .
}

Let us now discuss the relations in the tachyon module. As in the bosonic
theory we have
\eqn\tachyonops{
    \CT_{r,s}=\mu^{1-s}\hat{\CO}_{r,s} \CT_{1,1}=
    \mu^{q(r-1)-p(s-1)\over 2p}
    U_{s-1}\big(\hat x\big)U_{r-1}\big(\hat y\big)
    \CT_{1,1}\ .
}
The Ramond ground state in the even $(p,q)$ models leads to a new complication.
As we have seen, the tachyon $\CT_{{p\over 2},{q\over 2}}$ exists in the 0B
theory only for positive $\mu$. (For the discussion of correlation functions of
physical vertex operators without D-branes we can neglect the similar operator
in the $(-1/2,-3/2)$ ghost picture which exists only for negative $\mu$.)
Therefore, for positive $\mu$ we have tachyons with $rq-sp \ge 0$, while for
negative $\mu$ we have only the tachyons with $rq-sp > 0$.  This truncation can
be achieved by imposing
 \eqn\tachcond{\CT_{p-r,q-s}=\zeta \mu^{ps-qr\over p}\CT_{r,s}\ .}
As in the bosonic string, it is enough to require that
\eqn\tachrelen{
\left(U_{q-2}(\hat x)-\zeta U_{p-2}(\hat y) \right)\CT_{1,1}=0\ .
}
This will guarantee all of the relations \tachcond.

The discussion of the ground ring and the tachyon module has so far been
entirely for the 0B models. The ground ring for 0A follows trivially from that
of the 0B: since we need not worry about the Ramond ground state, we simply
project out the R sector of the 0B ground ring to obtain the ground ring of 0A.
Thus the ground ring of 0A is generated by the NS operators $\hat{\CO}_{1,3}$,
$\hat{\CO}_{2,2}$ and $\hat{\CO}_{3,1}$.

It is straightforward to extend the calculation of tachyon correlation
functions in the bosonic string (section 2.2) to the superstring. Since the
results are nearly identical to the bosonic string, we will not discuss them
here.

\newsec{FZZT and ZZ branes of minimal superstring theory}

\subsec{Boundary states}

Here we will extend the discussion of the FZZT and the ZZ branes of the
previous sections to the $(p,q)$ superminimal models coupled to gravity. The
analysis will become more complicated, owing to the presence of the NS and R
sectors, the two different GSO projections, and the option of having negative
$\mu$. Nevertheless, we will find that the supersymmetric and bosonic theories
share many features in common.

As in the bosonic string, the boundary states are labelled by a Liouville
parameter $\sigma$ and matter labels $(k,l)$.  In addition to these labels
there are also a supercharge parameter $\eta$ related to the linear combination
of left and right moving supercharges $G_r+ i\eta \tilde G_{-r}$ which
annihilate the state, and the R-R ``charge'' $\xi=\pm 1$. Finally, the boundary
states also depend on $\zeta = {\rm sign} (\mu)$.

Consider first the branes in the positive $\mu$ ($\zeta=+1$) theory. In the
expressions that follow, it will be implicit that the matter representation
$(k,l)$ and the parameter $\eta$ are correlated by the condition
\eqn\kletacond{
(-1)^{k+l}=\eta\ .
}
This follows from the boundary conditions on the supercharge, which imply that
the Cardy states with $\eta=1$ ($\eta=-1$) in are in the NS (R) sector. Using
the results of \refs{\FukudaBV,\AhnEV}, we write the boundary states as:
 \eqn\superFZZT{\eqalign{
    |\sigma,(k,l); \xi, \eta=+1\rangle\big|_{\zeta=+1}=
    \int_{0}^{\infty}dP\, \bigg(&\cos(\pi P \sigma)
    A_{NS}(P)|P,(k,l);\eta=+1\rangle\rangle_{NS}\cr
        +\xi &\cos(\pi P \sigma)
    A_R(P)|P,(k,l);\eta=+1\rangle\rangle_R
    \bigg)\cr
    |\sigma,(k,l);\xi, \eta=-1\rangle\big|_{\zeta=+1}=
    \int_{0}^{\infty}dP\,  \bigg( &\cos(\pi
    P\sigma)A_{NS}(P)|P,(k,l);\eta=-1\rangle\rangle_{NS}\cr
    -i\xi&\sin (\pi P\sigma)A_R(P)|P,(k,l);\eta=-1
    \rangle\rangle_R \bigg)\ ,\cr
 }}
where the Liouville wavefunctions are given by
 \eqn\sFZZTprop{\eqalign{
    &A_{NS}(P)=\left({|\mu|\over 4}\right)^{iP/b}{\Gamma(1-i P b)
    \Gamma(1-{iP\over b})\over  \sqrt{2}\pi    P}\cr
    &A_{R}(P)=\left({|\mu|\over4}\right)^{iP/b}{1\over \pi b^2}
    \Gamma\left({1\over 2}-i P
    b\right)\Gamma    \left({1\over 2}-{iP\over b}\right)\ ,\cr
 }}
and the boundary cosmological constant depends on the various parameters as
 \eqn\sFZZTmuBp{
    {\mu_B\over \sqrt{|\mu|}}=\cases{
        \xi\cosh({\pi b\sigma\over 2}) & \quad ${\eta}=+1$\cr
        \xi\sinh({\pi b \sigma\over 2}) & \quad ${\eta}=-1$\ .\cr
 }}
The factor of $\xi$ in the definition of $\mu_B$ was included for convenience,
so as to make explicit the $(-1)^{F_L}$ symmetry of the theory. The action of
$(-1)^{F_L}$ maps $\xi\rightarrow -\xi$ since it acts as $-1$ ($+1$) on all R
(NS) states. It also maps $\mu_B\rightarrow -\mu_B$, since $\mu_B$ is the
coefficient of a boundary fermion which transforms like a bulk spin field.

The states $|P,(k,l);\eta\rangle\rangle_{NS,R}$ appearing on the
RHS of \superFZZT\ are shorthand for the following linear
combination the matter ($M$), Liouville ($L$) and superghost ($G$)
Ishibashi states (i.e.\ strictly, they are not Ishibashi states):
 \eqn\ishsic{\eqalign{
    &|P,(k,l); \eta\rangle\rangle_{NS} =
    \sum_{k'+l'\,\,{\rm even}}
    \psi_{(\eta,\zeta=+1)}(k,l;k',l')
    |NS;(k',l');\eta\rangle\rangle_M
    |NS;P;\eta\rangle\rangle_L
    |NS; \eta\rangle\rangle_G\cr
    &|P,(k,l); \eta\rangle\rangle_{R} =
    \sum_{k'+l'\,\,{\rm odd}}
    \psi_{(\eta,\zeta=+1)}(k,l;k',l')
    |R;(k',l');{\eta}\rangle\rangle_M
    | R;P;\eta\rangle\rangle_L
    |R; \eta\rangle\rangle_G\ .\cr
}}
A few comments are in order.
 \lfm{1.} For the superghost states it is convenient to
work in the $(-1,-1)$ picture in the NS sector and in the
$(-1/2,-3/2)$ or $(-3/2,-1/2)$ in the R sector.  In these pictures
we see the elementary spacetime fields and the inner products of
the states are simplest.
 \lfm{2.} The individual Ishibashi states in each sum must have the same
label $\eta=\pm 1$. This guarantees that the linear combinations $G_r + i\eta
\tilde G_{-r}$ annihilate the state, where $G$ and $\tilde G$ are the total
left and right-moving supercharges of the system.  Actually, this would have
allowed also a linear combination of Liouville and matter states with opposite
$\eta$. Such a linear combination is incompatible with the $P$ dependence in
the Cardy state, which is different for the two signs of $\eta$.
 \lfm{3.}  The restriction in
the sums to $k'+l'$ even or $k'+l'$ odd guarantees that we include
only NS or R matter Ishibashi states. The ``matter wavefunctions''
$\psi_{(\eta,\zeta)}(k,l;k',l')$ are closely related to the
modular $S$-matrix elements of the combined superminimal model and
super-Liouville. Although they can be computed, we will not do
that here.
 \lfm{4.} It is important to note that the Cardy states
\superFZZT\ are not just products of Cardy states of the matter, Liouville and
ghost sectors, as they were in the bosonic string. This is because our system
is not simply the product of the superminimal model and super-Liouville theory.
Instead, the NS sector is constructed out of the NS sectors of the two theories
by projecting on $(-1)^{f}=1$, where $f=f_L+f_R$ is the total worldsheet
fermion number operator in the combined theory. Similarly, the R sector is made
out of the R sectors of the two theories, again by projecting on fixed
$(-1)^{f}$. In the $(-1/2,-3/2)$ or $(-3/2,-1/2)$ picture its value is $-1$ in
the 0B theory and $+1$ in the 0A theory.

\medskip

Let us discuss in some detail the construction of the individual
Ishibashi states $|NS,h,\eta\rangle \rangle_{L,M}$ and $
|R,h,\eta\rangle \rangle_{L,M}$ in \ishsic, where here we will use
$h$ to denote the conformal weight of the associated Liouville or
matter primary instead of the labels $P$ and $(k,l)$. In the NS
sector, the Ishibashi states are linear combinations of NS states
of fixed $(-1)^{f_L}$
 \eqn\NSishi{\eqalign{
     |NS,h,\eta\rangle \rangle_{L,M}=
     \left(1+\eta\,f(h,\hat c)\,   G^{L,M}_{-1/2}\tilde{G}^{L,M}
     _{-1/2}+\dots\right)|NS,h\rangle_{L,M}\ ,
}}
where $f(h,\hat c)$ is $\eta$-independent and is fixed by the
supersymmetry constraints, and $|NS,h\rangle_{L,M}$ denotes the
associated primary state. Note that this expression is analogous
to the fact that in the Ising model one uses the Ishibashi states
$|1 \rangle\rangle+ \eta |\psi \rangle\rangle$ in forming Cardy
states.

In the R sector, every generic (i.e.\ nonsupersymmetric)
representation leads to two Ishibashi states which are annihilated
by $G_{r}+i\eta \tilde{G}_{-r}$ for $\eta =\pm1$. Consider first
the matter or the Liouville part of the theory independently. Then
the Ishibashi state in each of them is
\eqn\Rishi{\eqalign{
    |R,h,\eta\rangle\rangle_{L,M}&=\left(1+a(h,\hat c)\, L^{L,M}_{-1}\tilde{L}^{L,M}_{-1}+
     \eta\,b(h,\hat c)\, G^{L,M}_{-1}\tilde{G}^{L,M}_{-1}\right)|R,h,\eta\rangle_{L,M}
     \cr
     &\qquad\qquad +\left(c(h,\hat c)\, G^{L,M}_{-1}\tilde{L}^{L,M}_{-1}+
     \eta\,d(h,\hat c)\, L^{L,M}_{-1}\tilde{G}^{L,M}_{-1}\right)
    |R,h,-\eta\rangle_{L,M}+ \dots\cr
}}
where again the coefficients appearing here are $\eta$-independent
and are determined by the supersymmetry constraints. These
Ishibashi states are eigenstates of the fermion number operator in
each sector $(-1)^{f^{L,M}}$
\eqn\Rfei{
    (-1)^{f^{L,M}}|R,h,\eta\rangle\rangle_{L,M}=- \eta
    |R,h,\eta\rangle\rangle_{L,M}\ .
}
Using \totalferm\ we conclude that
 \eqn\Rishifn{
    (-1)^{f}|R,h,\eta\rangle \rangle_{L}|R,h',\eta\rangle \rangle_{M}
    = -  |R,h,\eta\rangle \rangle_{L}|R,h',\eta\rangle \rangle_{M}
 }
with generic values of $h$ and $h'$.  Therefore, in the 0B theory
all the terms in the right hand side of \ishsic\ made out of these
representations satisfy the GSO projection, while no such state
survives the projection in the 0A theory.

The situation is somewhat different for the matter R ground state, which exists
in the even $(p,q)$ superminimal models and has conformal weight
$h_0={\hat{c}_M\over 16}$. The matter R ground state must have $(-1)^{f^M}=1$,
but it still leads to two Ishibashi states with $\eta=\pm 1$ and
$(-1)^{f^M}=1$. These take the form
\eqn\Rgsishi{
    |R,h_0,\eta\rangle\rangle_M =
    \left(1+a(h_0,\hat c)\, L_{-1}^M\tilde{L}^M_{-1}+\eta\,b(h_0,\hat c)\,  G_{-1}^M\tilde{G}^M_{-1}+\dots\right)
    |R,h_0\rangle_M\ ,
}
where the coefficients are the same as in \Rishi. (One can also
show that $c(h_0,\hat c)=d(h_0,\hat c)=0$.) Meanwhile, the
Liouville Ishibashi state still has $(-1)^{f^L}=-\eta$, and
therefore the contribution of the matter R ground state to
\ishsic\ is restricted by
\eqn\Rgscontrib{\eqalign{
&\psi_{(\eta=+1,\zeta=+1)}(k,l;{p\over 2},{q\over 2}) = 0\qquad (0{\rm B})\cr
&\psi_{(\eta=-1,\zeta=+1)}(k,l;{p\over 2},{q\over 2}) = 0\qquad (0{\rm A})\
.\cr
}}

Now let us examine the situation for negative $\mu$ ($\zeta=-1$). We start by
ignoring the matter and the ghosts.  The sign of $\mu$ can be changed in the
super-Liouville part of the theory by acting with the $Z_2$ R-transformation
$(-1)^{f_L}$. This is not a symmetry of the theory, as it changes the parameter
$\mu$. Also, this transformation changes the sign of the projection in the
Ramond sector. As for the boundary states, this transformation has the effect
of reversing the sign of $\eta$, since it sends $G\rightarrow -G$ without
changing $\tilde G$. Therefore, the boundary super-Liouville theory depends
only on $\hat \eta=\eta\zeta$.

Adding back in the matter and superghost sectors, we see that if we want to
perform such an R-transformation on the Liouville sector of the theory, we
should do it in all the sectors because the total supercharge $G$ is gauged.
This is complicated when the matter theory does not have such an R-symmetry.
Nevertheless, we can perform such a transformation and label the states by
their value of $\hat{\eta}$. Then our expression \superFZZT\ for the boundary
states becomes
 \eqn\superFZZTg{\eqalign{
    |\sigma,(k,l); \xi,\hat \eta=+1,\zeta\rangle=&
    \int_{0}^{\infty}dP\, \bigg(\cos(\pi P \sigma)
    A_{NS}(P)|P,(k,l);\hat\eta=+1,\zeta\rangle\rangle_{NS}\cr
    &+\xi \cos(\pi P \sigma)
    A_R(P)|P,(k,l);\hat \eta=+1,\zeta\rangle\rangle_R
    \bigg)\cr
    |\sigma,(k,l);\xi, \hat \eta=-1,\zeta\rangle= &
    \int_{0}^{\infty}dP\,  \bigg( \cos(\pi
    P\sigma)A_{NS}(P)|P,(k,l);\hat\eta=-1,\zeta\rangle\rangle_{NS}\cr
    &
    -i\xi\sin (\pi P\sigma)A_R(P)|P,(k,l);\hat\eta=-1,\zeta
    \rangle\rangle_R \bigg),\cr
 }}
where the Liouville wavefunctions are again given by \sFZZTprop\ and the
expression for the boundary cosmological constant \sFZZTmuBp\ is generalized to
\eqn\sFZZTmuBp{
    {\mu_B\over \sqrt{|\mu|}}=\cases{
        \xi\cosh({\pi b\sigma\over 2}) & \quad ${\hat\eta}=+1$\cr
        \xi\sinh({\pi b \sigma\over 2}) & \quad ${\hat\eta}=-1$\ . \cr}
}
Now the Cardy states with $\hat \eta=1$ ($\hat \eta=-1$) in \superFZZTg\ are in
the NS (R) sector, and therefore their matter representation $(k,l)$ should
satisfy
\eqn\kletacondg{
(-1)^{k+l}=\hat{\eta}\ .
}
The Ishibashi states \ishsic\ are generalized to
 \eqn\ishsicg{\eqalign{
    &|P,(k,l); \hat\eta,\zeta\rangle\rangle_{NS} =
    \sum_{k'+l'\,\,{\rm even}}
    \psi_{(\hat{\eta},\zeta)}(k,l;k',l')
    |NS;(k',l');\hat\eta\rangle\rangle_M  |NS;P;{\hat \eta}\rangle\rangle_L
    |NS; \hat\eta\rangle\rangle_G\cr
    &|P,(k,l); \hat\eta,\zeta\rangle\rangle_{R} =
    \sum_{k'+l'\,\,{\rm odd}}
    \psi_{(\hat{\eta},\zeta)}(k,l;k',l')
    |R;(k',l');{\hat \eta}\rangle\rangle_M
    |R;P;{\hat \eta}\rangle\rangle_L
    |R; \hat\eta\rangle\rangle_G\ .
 }}
Here the only dependence on $\zeta$ for fixed $\hat{\eta}$ is through the
``matter wavefunctions" $\psi_{(\hat{\eta},\zeta)}(k,l;k',l')$.  This is
important in the even $(p,q)$ models which include the Ramond ground state.
Here the GSO condition implies
\eqn\Rgscontribg{\eqalign{
    &\psi_{(\hat{\eta}=+\zeta,\zeta)}(k,l;{p\over 2},{q\over 2}) = 0 \qquad (0{\rm B})\cr
    &\psi_{(\hat{\eta}=-\zeta,\zeta)}(k,l;{p\over 2},{q\over 2}) = 0 \qquad (0{\rm A})\ .\cr
}}
That is, the Ramond ground state does not contribute to the sum over Ishibashi
states for the $\eta=+1$ ($-1$) brane in the 0B (0A) theory.

To illustrate the general discussion above, let us consider the simplest
example of pure supergravity $(p,q)=(2,4)$, which does not have matter at all.
(Alternatively, the matter includes the identity and the R-ground state.)
Imposing the 0B GSO condition,\foot{If instead we use the 0A GSO projection,
the results below are the same with $\zeta \to -\zeta$ and $\eta\to -\eta$
\KlebanovWG.} \superFZZTg\ becomes
 \eqn\superFZZTgp{\eqalign{
    &|\sigma; \xi, \eta=+1,\zeta=+1\rangle =\int_{0}^{\infty}dP\,
    \cos(\pi P \sigma)
    A_{NS}(P)| NS;P;{\hat \eta}\rangle\rangle_L
    |NS; \hat\eta\rangle\rangle_G \cr
    &|\sigma;\xi, \eta=-1,\zeta=+1\rangle=
     \int_{0}^{\infty}dP\,  \bigg( \cos(\pi
    P\sigma)A_{NS}(P)| NS;P;{\hat \eta}\rangle\rangle_L
    |NS; \hat\eta\rangle\rangle_G \cr
    &\qquad\qquad\qquad \qquad\qquad\qquad\quad
    -i\xi\sin (\pi P\sigma)A_R(P)|
    R;P;{\hat \eta}\rangle\rangle_L
    |R; \hat\eta\rangle\rangle_G\bigg)\cr
}}
for $\mu$ positive. In this phase, the vertex operator for the zero mode of $C$
does not exist, and therefore neither brane is charged. Now consider $\mu$
negative. We have
\eqn\superFZZTgm{\eqalign{
     &|\sigma; \xi, \eta=+1,\zeta=-1\rangle =\int_{0}^{\infty}dP\,
    \cos(\pi P \sigma)
    A_{NS}(P)| NS;P;{\hat \eta}\rangle\rangle_L
    |NS; \hat\eta\rangle\rangle_G
    \cr
    &|\sigma;\xi, \eta=-1,\zeta=-1\rangle=
     \int_{0}^{\infty}dP\,  \bigg( \cos(\pi
    P\sigma)A_{NS}(P)| NS;P;{\hat \eta}\rangle\rangle_L
    |NS; \hat\eta\rangle\rangle_G \cr
    &\qquad\qquad\qquad \qquad\qquad\qquad\quad\,\,
    +\xi \cos(\pi P \sigma)A_R(P)|R;P;{\hat \eta}\rangle\rangle_L
    |R; \hat\eta\rangle\rangle_G\bigg)\ .\cr
 }}
In this phase, the $\eta=+1$ brane is again uncharged, but now the $\eta=-1$
brane carries charge. This agrees with the discussion below \asympt, where we
saw that for $\mu<0$ we can have charged branes but not flux.

\subsec{FZZT one-point functions and the role of $\xi$}

Let us use the expression \superFZZTg\ for the FZZT boundary state to study the
one-point functions of physical operators on the disk with FZZT-type boundary
conditions. We expect to find similar results as in the bosonic string, where
not all FZZT states are distinct in the BRST cohomology. In particular, we
expect (see \fzztstrelate) that all of the FZZT branes with arbitrary matter
label can be reduced to elementary branes with fixed matter label, say $(1,1)$
for NS branes and $(1,2)$ or $(2,1)$ for R branes. To actually prove this at
the level of the one-point functions as we did for the bosonic string would
require knowledge of the matter wavefunctions, which we do not presently have.
Thus we will simply assume that it can be done, and from this point onwards we
will suppress the matter label and refer to the FZZT branes as $|\,\sigma;\xi,
\hat\eta,\zeta\rangle$.

Now consider the FZZT one-point functions of the tachyon operators $\CT_{r,s}$.
From \superFZZTg\ we find the $(\sigma,\xi)$-dependent part of these one-point
functions can be expressed compactly with the following:
\eqn\oneptrs{
    \langle \CT_{r,s}|\, \sigma;\xi,\hat\eta,\zeta\rangle \propto
        \xi^{r+s}\cosh \left({\pi(r q-s p)\sigma\over
        2\sqrt{pq}}+{i\pi(r+s)\hat\nu\over2}\right)
}
where we have defined
\eqn\hatnu{
    \hat\nu={1-\hat\eta\over 2} =\cases{
    0&  for $\hat \eta =+1$ \cr
    1&  for $\hat \eta =-1$ \cr}
}
The one-point functions of the other physical operators are also
given by \oneptrs, but with different values of $s$. Thus physical
one-point functions are invariant under the transformations
\eqn\sigmaxiinvt{
(\sigma,\ \xi)\rightarrow (-\sigma,\
\hat\eta\,\xi),\qquad (\sigma\pm 2i\sqrt{pq},\ (-1)^p\xi)
}
Note that the first transformation is true at the level of the
boundary state \superFZZTg. The second transformation is only true
for physical one-point functions, which is evidence that FZZT
states related by this transformation differ by a BRST exact
state. These transformations suggest that we define
\eqn\zdefetapm{
    z=\cases{
            \cosh\left({\pi\sigma\over 2\sqrt{pq}}-{i\pi\hat\nu\over2}\right) &\qquad $(p,q)$ odd\cr
            \cosh\left({\pi\sigma\over
            \sqrt{pq}}-{i\pi\hat\nu\over2}\right) &\qquad $(p,q)$ even \cr
            }
}
and label the FZZT brane by $z$:
\eqn\sfzztrelabel{\eqalign{
    &|\,\sigma; \xi,\hat\eta,\zeta\rangle \rightarrow
     |\,z;\xi,\hat\eta,\zeta\rangle
}}
so that at {\it fixed $\xi$}, two FZZT branes labelled by $z$ and
$z'$ are equal if and only if $z=z'$.

The parametrization in terms of $z$ eliminates some, but not all of the
redundancy of description implied by the transformations \sigmaxiinvt. The
remaining redundancy involves changing the sign of $\xi$. Indeed, we see that
when $\hat\eta=-1$ or $(p,q)$ is odd, a state with $(z,-\xi)$ is equivalent to
a state with $(-z,\xi)$. For these states, $\xi$ is a redundant label which can
be removed by analytic continuation in $z$. The only states for which $\xi$
cannot be eliminated in this way are those with $\hat{\eta}=+1$ when $(p,q)$ is
even.\foot{We could further reparametrize the FZZT branes so as to remove the
redundancy by $\xi$. However, as this would needlessly complicate the notation,
we prefer to continue labelling the FZZT branes with $\xi$, even in the cases
where it is unnecessary.} In section 7, we will see how these disparate facts
can be all understood geometrically and in a unified way, in terms of an
auxiliary Riemann surface.

Finally, we note that when $(p,q)$ is odd, the transformation
\eqn\sigmaxieta{
(\sigma,\ \hat\eta)\rightarrow (i\sqrt{pq}-\sigma,\ -\hat\eta)
}
leaves the $\sigma$-dependent part of the one-point functions \oneptrs\
unchanged. Moreover, the transformation also leaves $z$ unchanged. This
suggests that when $(p,q)$ is odd, the FZZT states with $\hat\eta=-1$ are
equivalent to the FZZT states with $\hat\eta=+1$ (but with different $\sigma$)
in the BRST cohomology. Let us assume that this is true, and focus our
attention only on the states with $\hat\eta=+1$ when $(p,q)$ is odd. We will
also motivate this simplification geometrically in section 7 and we will use it
in section 8.

\subsec{ZZ branes and their one-point functions}

The discussion of the ZZ boundary states is analogous to that of the FZZT
states, so we will simply write down the relevant expressions. Here we will not
attempt to analyze the subtleties of the degenerate super-Virasoro
representations that arise at rational $b^2$. For the bosonic string, we found
that these subtleties only pertained to $(m,n)$ ZZ branes with $m > p$ or $n>
q$. The ZZ boundary states with $m\le p$ and $n\le q$ were given by the formula
for generic $b$, and could be written as a difference of just two FZZT branes.
Therefore the subtleties at rational $b^2$ did not affect our final conclusion
in the bosonic string, which was that the set of all $(m,n)$ ZZ branes could be
reduced to a principal set with $m < p$ and $n<q$ and $mq-np>0$. As we expect a
similar conclusion in the superstring, let us restrict our attention from the
outset to ZZ boundary states with $m\le p$ and $n\le q$. Then by analogy with
the bosonic string, these should be given by the formula at generic $b$
\refs{\FukudaBV,\AhnEV}:
 \eqn\superZZ{\eqalign{
    &|(m,n),(k,l); \xi,\hat\eta=+1,\zeta\rangle=\cr
        &\quad 2\int_{0}^{\infty} dP \,
        \bigg( \sinh \big({\pi P\, m\over b}\big)\sinh \big(\pi P\, b\, n\big)
        A_{NS}(P)|P,(k,l);\eta,\zeta\rangle\rangle_{NS}\cr
        &\qquad\qquad\qquad
        +\xi\sinh \big({\pi m P\over b}+{i\pi n\over 2}\big)
        \sinh \big(\pi n P\,b-{i\pi n\over 2}\big)A_{R}(P)|P,(k,l);\eta,\zeta\rangle\rangle_R
        \bigg)\cr
    &|(m,n),(k,l);\xi,\hat\eta=-1,\zeta\rangle=\cr
        &\quad 2\int_{0}^{\infty} dP \,
        \bigg( \sinh \big({\pi P\, m\over b}\big)\sinh \big(\pi P\, b\, n\big)
        A_{NS}(P)|P,(k,l);\eta,\zeta\rangle\rangle_{NS}\cr
        &\qquad\qquad\qquad
        +\xi\cosh \big({\pi m P\over b}+{i\pi n\over 2}\big)
        \sinh \big(\pi n P\,b-{i\pi n\over 2}\big)A_{R}(P)|P,(k,l);\eta,\zeta\rangle\rangle_R
        \bigg)\ .\cr
}}
As for the FZZT branes, the parameter $\hat\eta$ determines
whether the Cardy state is NS or R. Thus we must require
\eqn\conmn{
    (-1)^{m+n}=(-1)^{k+l}=\hat{\eta}\ .
}
Also, we will assume as we did for the FZZT branes that the ZZ
branes with different matter labels can be reduced down to ZZ
branes with fixed matter label.

Expanding the products of $\cosh$ and $\sinh$ using standard trigonometric
identities and comparing with \superFZZTg, we see that the ZZ branes with $m\le
p$ and $n\le q$ can be written in terms of the FZZT branes as (we suppress the
labels $(\hat\eta,\zeta)$):
\eqn\sZZsFZZT{
    |\, m,n; \xi\rangle =
    |\,z=z(m,n);\xi\rangle-
    |\,z=z(m,-n);(-1)^n\xi\rangle
}
where $z(m,n)\equiv z(\sigma(m,n))$ was defined in \zdefetapm, and
\eqn\supersigmamn{
    \sigma(m,n)=i\left({m\over b}+n b\right)\ .
}
The boundary cosmological constant corresponding to $\sigma(m,n)$
is
 \eqn\smnsZZ{\eqalign{
   &\mu_B(m,n,\xi)=\mu_B(m,-n,(-1)^n\xi)=
    \cases{\xi\sqrt{|\mu|}\cos {\pi\over2}(m+n b^2)& $\hat{\eta}=+1$\cr
           i\,\xi\sqrt{|\mu|}\sin {\pi\over2}(m+n b^2) & $\hat{\eta}=-1$\ .\cr}
 }}
Note that the two FZZT branes in the right hand side of \sZZsFZZT\
always have the same value of $\mu_B$.

As in the bosonic string, we do not expect all of the ZZ branes to be distinct
in the full string theory; rather, we expect many to differ by BRST null
states. Indeed, the formula \sZZsFZZT\ for the ZZ branes immediately implies
the identification
\eqn\sZZidentify{
    | p-m,q-n ; (-1)^{q+m+n}\xi\rangle=
    |  m,n ;\xi\rangle
}
modulo BRST null states. Moreover, a straightforward computation using
\superZZ\ shows that the one-point functions of physical operators all vanish
when $m=p$ or $n=q$. This suggests that in the BRST cohomology:
\eqn\sZZidentnull{
    | m,n ;\xi\rangle=0,\quad\quad {\rm when}\,\,\,\, m=p\,\,\,\,{\rm
    or}\,\,\,\,n=q\ .
}
In the next section, we will also motivate these identifications
from a more geometrical point of view, as was done for the bosonic
string.

Using the identifications \sZZidentify\ and \sZZidentnull, we can
reduce the infinite set of $(m,n,\xi)$ ZZ branes down to what we
will call, as in the bosonic string, the principal ZZ branes. Let
us define $\CB=\{(m,n) |\,\, m=1,\dots,p-1,\,\, n=1,\dots,q-1\}$.
Then for $(p,q)$ even, the principal ZZ branes are
\eqn\diffsZZeven{\eqalign{
    &\hat{\eta}=+1:\,\, (m,n,\xi)\ ,\quad (m,n)\in \CB\ ,\quad m+n\,\,{\rm even}\ ,\quad mq-np\ge 0\ ,
        \quad \xi=\pm 1\cr
    &\hat{\eta}=-1:\,\, (m,n,\xi)\ ,\quad (m,n)\in \CB\ ,\quad m+n\,\,{\rm
    odd}\ ,\quad \xi=+1\ ,\cr
}}
giving a total of ${(p-1)(q-1)\pm1\over 2}$ principal ZZ branes for
$\hat{\eta}=\pm1$. On the other hand, for $(p,q)$ odd, the principal ZZ branes
are
\eqn\diffsZZodd{\eqalign{
    &(m,n,\xi)\ ,\quad  (m,n)\in \CB\ ,\quad m+n\,\,{\rm even}\ ,
    \quad \xi=+1\qquad ({\rm and}\,\,\,\hat\eta=+1)\cr
}}
giving a total of ${(p-1)(q-1)\over 2}$ principal ZZ branes for
$(p,q)$ odd. Notice that we have restricted ourselves to the ZZ
branes with $\hat\eta=+1$. Since the map \sigmaxieta\ $ (\sigma,\
\hat\eta)\rightarrow (i\sqrt{pq}-\sigma,\ -\hat\eta) $ leaves $z$
unchanged, it preserves the formula \sZZsFZZT\ and maps ZZ branes
with $\hat\eta$ to ZZ branes with $-\hat\eta$.

In the bosonic string, we saw that we could normalize the ground ring elements
so that the principal ZZ branes became eigenstates of the ground ring. We
expect similar phenomena to occur in the superstring. To make more precise
statements, we would need to have explicit formulas for the matter
wavefunctions. We will leave this for future work.

\newsec{Geometric interpretation of minimal superstring theory}

\subsec{The surfaces $\CM_{p,q}^{\pm}$ and their analytic structure}

As in the bosonic string, we can understand many of the features of the
boundary states using an auxiliary Riemann surface that emerges from the FZZT
partition function. We start as before with the disk one-point function of the
cosmological constant operator. This is given by:
\eqn\muevsfzzt{
    \partial_\mu Z\big|_{\mu_B}=\langle \psi \bar{\psi}e^{b\phi}\rangle=
    A(b)(\sqrt{|\,\mu|})^{1/b^2-1}\cosh\left(b-{1\over b}\right){\pi
    \sigma\over2}\ .
}
Notice the similarity to the bosonic one-point function \muevfzztbec. The
difference comes in the relation between $\sigma$ and $\mu_B$ for
$\hat{\eta}=-1$. For simplicity, we will limit our discussion in the
superstring to the FZZT brane and not its dual. The relationship between the
FZZT brane and its dual is exactly analogous to that in the bosonic string.

Integrating \muevsfzzt\ leads to
\eqn\Zsfzzt{
    Z= C(b)(\sqrt{|\mu|})^{1/b^2+1}\times
    \cases{
    b^2\cosh \left({\pi\sigma\over2b}\right)\cosh \left({\pi b \sigma\over 2}\right)-
    \sinh \left({\pi\sigma\over2b}\right)\sinh \left({ \pi b \sigma\over2}\right)
     &$\,\,   \hat\eta=+1$\cr
    b^2\sinh \left({\pi\sigma\over2b}\right)\sinh\left({\pi b \sigma\over 2}\right)-
    \cosh \left({\pi\sigma\over2b}\right) \cosh \left( { \pi b \sigma\over2}\right)
    &$\,\,   \hat\eta=-1$\ ,\cr
    }
}
and differentiating this with respect to $\mu_B$, we find
\eqn\ymsfzzt{
    \partial_{\mu_B}Z\big|_{\mu}
    =\cases{(\sqrt{|\mu|})^{1/b^2}\xi\cosh \left({\pi \sigma\over 2b}\right)
    &\quad $\hat{\eta}=+1$\cr
            (\sqrt{|\mu|})^{1/b^2}\xi\sinh \left({\pi \sigma\over 2b}\right)
            &\quad $\hat{\eta}=-1$\ .\cr}
}
In \muevsfzzt\ and \Zsfzzt\ above, $A(b)$ and $C(b)$ are overall normalization
factors that were chosen so as to make the normalization of \ymsfzzt\ unity. As
in the bosonic string, we suppress the contribution of the matter sector, since
this is $\mu$ and $\mu_B$ independent. The two types of FZZT brane lead to two
Riemann surfaces; we will call them $\CM_{p,q}^{\hat{\eta}}$. For reasons that
will shortly become clear, we will parametrize these surfaces with slightly
different dimensionless coordinates:
\eqn\coordxi{\eqalign{
    &\CM_{p,q}^{+}:\quad x={\mu_B\over\sqrt{\mu}}\ ,
    \quad y = {\partial_{\mu_B}Z\over\sqrt{\tilde\mu}}\ ,\cr
    &\CM_{p,q}^{-}:\quad x={i \mu_B\over\sqrt{\mu}}\ ,
    \quad y =  {i \partial_{\mu_B}Z\over\sqrt{\tilde\mu}}\ .\cr
}}
Then in terms of $x$ and $y$, the surfaces are described by the polynomial
equations:
\eqn\sfzztM{
    0=F(x,y)=\cases{T_q(x)-T_p(y)                 &\quad $\hat{\eta}=+1$\cr
                    (-1)^{p-q\over2}T_q(x)-T_p(y) &\quad $\hat{\eta}=-1$\ .\cr}
}
Note that we have assumed in writing \sfzztM\ that $(p,q)$ are
either both odd or both even; otherwise there could be additional
phases arising from the factor of $\xi$ in the definition of the
coordinates. It is interesting that for $(p,q)$ odd, the phase in
front of $T_q$ in the second line of \sfzztM\ can be absorbed into
the argument of the Chebyshev polynomial. Thus, for $(p,q)$ odd
the surfaces $\CM_{p,q}^+$ and $\CM_{p,q}^-$ are identical, and we
will limit our discussion to $\CM_{p,q}^+$ without loss of
generality. However, for $(p,q)$ even, since $(p-q)/2$ is odd the
phase cannot be absorbed, and therefore these surfaces are
different. This is consistent with the suggestion that for $(p,q)$
odd, boundary states with $\hat\eta=\pm 1$ become equivalent,
while for $(p,q)$ even, $\hat\eta$ labels distinct states.

As in the bosonic string, a natural uniformization of these
surfaces is provided by the parameter $z$ defined in \zdefetapm.
The coordinates of $\CM_{p,q}^{\hat\eta}$ are given in terms of
$z$ by
\eqn\uniformsuperII{
    (x,y)=\cases{
          \Big(\xi T_p(z),\xi T_q(z)\Big) &\qquad $(p,q)$ odd\cr
          \Big(\xi T_{p\over2}(z),\xi T_{q\over2}(z)\Big) &\qquad $(p,q)$ even and $\hat\eta=+1$\cr
          \Big(\xi z U_{{p\over2}-1}(\sqrt{1-z^2}),\xi z
          U_{{q\over2}-1}(\sqrt{1-z^2})\Big)&\qquad $(p,q)$ even and
          $\hat\eta=-1$\cr
          }
}
Therefore, for fixed $\xi$, $z\in \Bbb C$ covers
$\CM_{p,q}^{\hat\eta}$ exactly once, except for when $(p,q)$ is
even and $\hat\eta=-1$, where we must extend $z$ to a two-sheeted
cover of the complex plane in order to cover the surface once.
Notice also that $\xi$ can be absorbed into the definition of $z$
except when $(p,q)$ is even and $\hat\eta=+1$. (Recall that for
the $(p,q)$ even models, either $p/2$ or $q/2$ must be even.) This
gives a nice geometric interpretation to the results of section 6,
where we saw that only for $(p,q)$ even and $\hat{\eta}=1$ did the
parameter $\xi$ label two distinct FZZT boundary states. The fact
that $\xi$ cannot be eliminated in this case also has implications
for the analytic structure of the associated surface that we will
discuss below.

It is straightforward to work out the analytic structure of $\CM_{p,q}^+$ and
$\CM_{p,q}^-$. One can use either the condition $F=dF=0$, or equivalently one
can look for points $(x,y)$ that correspond to multiple values of $z$. Either
way, we must consider the following three cases:

\lfm{1.} {\it  $(p,q)$ odd and coprime} ($\hat\eta=+1$)

\item\indent For $(p,q)$ odd, $\CM_{p,q}^{+}$ is identical to the bosonic
surface $\CM_{p,q}$ (figure 1). Thus it has the ${(p-1)(q-1)/2}$ singularities
given by \xysing, which we include here for the sake of completeness:
\eqn\singpqodd{\eqalign{
    &(x,y)=\left(\cos {\pi (jq+kp)\over q}\ ,\,
                    \cos{\pi (jq+kp)\over p}\right)\ ,\cr
    &\qquad\qquad\qquad\qquad 1\le j \le p-1,\,\, 1\le k\le q-1,\,\,
    j\,q-k\,p>0
}}

\lfm{2.} {\it $(p,q)$ even, $({p\over2},{q\over2})$ coprime, ${p-q\over2}$ odd,
$\hat\eta=+1$}

\item\indent For $(p,q)$ even, the surface $\CM_{p,q}^{+}$ is actually quite
special. Since its curve can be written as $T_{q\over 2}(x)^2 = T_{p\over
2}(y)^2$, the surface splits into two separate branches described by $T_{q\over
2}(x) = \pm T_{p\over 2}(y)$, with each branch itself a Riemann surface. We
will denote these surfaces by $\left(\CM_{p,q}^{+}\right)_{\pm}$. Note that
since either ${p\over2}$ or ${q\over2}$ must be odd, the two branches of the
surface are related by the map $(x,y)\rightarrow (-x,-y)$. From the definition
\coordxi\ of $x$ and $y$, we see that this is just the action of
$\xi\rightarrow -\xi$. Thus we can think of the two branches as
$\left(\CM_{p,q}^{+}\right)_{\xi}$. We will elaborate on this identification of
the branches with $\xi$ shortly.

\item{}\indent One can easily check that $\CM_{p,q}^{+}$ has
${(p-1)(q-1)+1\over2}$ singularities described by
\eqn\singpqeven{\eqalign{
    (x,y)=\left(\cos {j\pi\over q},\, \cos {k \pi\over
    p}\right),\,\,
    j=1,&\dots,q-1,\,\, k=1\,\dots,p-1,\,\,
    k-j=0\,\mod\, 2.
}}
There are two types of singularity: those that are singularities of
$\left(\CM_{p,q}^{+}\right)_{\xi}$ individually, which we will call {\it
regular singularities}; and those that join the two branches, which we will
call {\it connecting singularities}. The distinction between the two is quite
simple algebraically: connecting singularities satisfy
$T_{q\over2}(x)=T_{p\over2}(y)=0$, while the regular singularities have
$T_{q\over 2}(x)$ and $T_{p\over2}(y)\ne 0$. Counting the number of
singularities of each type is also easy. There are ${p\,q\over 4}$ connecting
singularities; and ${(p-2)(q-2)\over 8}$ regular singularities on each branch
of $\CM_{p,q}^{+}$. An example of this kind of surface is shown in figure 2.

\medskip\ifig\figII{The surface $\CM^{+}_{p,q}$ for $(p,q)$ even, shown here for
$(p,q)=(4,6)$. The two subsurfaces of $\CM^{+}_{p,q}$ are labelled by $\xi=\pm
1$. The black points represent the connecting singularities that join the two
subsurfaces, while the gray points represent the regular singularities of each
subsurface.} {\epsfxsize=0.6\hsize\epsfbox{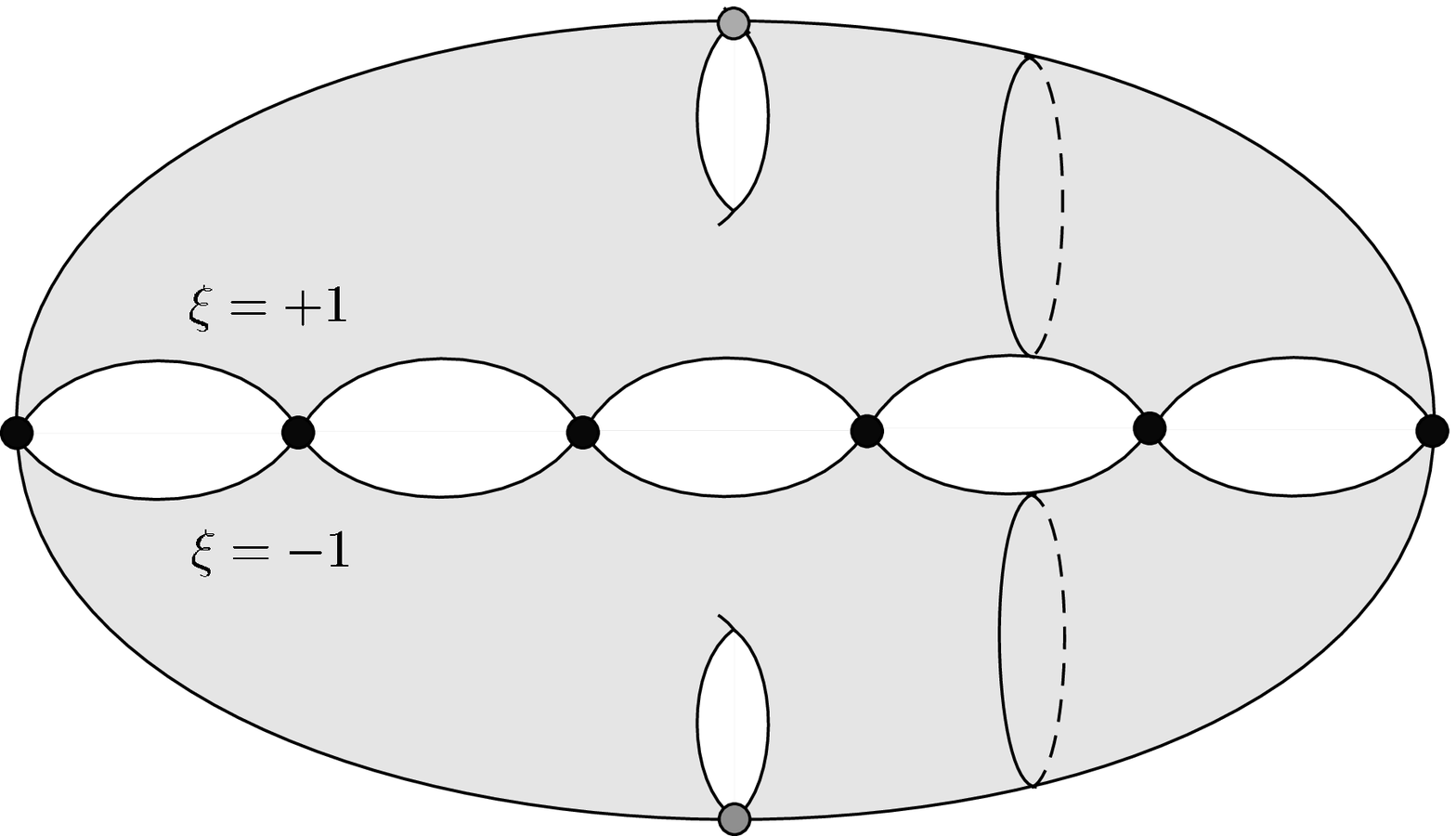}}

\lfm{3.} {\it $(p,q)$ even, $({p\over2},{q\over2})$ coprime, ${p-q\over2}$ odd,
$\hat\eta=-1$}

\item\indent Unlike the previous case, here $\CM_{p,q}^{-}$ does
not split into two branches, since the equation for this surface
takes the form $T_{p\over2}(y)^2+T_{q\over2}(x)^2=1$. One can show
that $\CM_{p,q}^{-}$ has ${(p-1)(q-1)-1\over 2}$ singularities
located at
\eqn\singm{\eqalign{
    (x,y)=\left(\cos {j\pi\over q},\, \cos {k \pi\over
    p}\right),\,\,
    j=1,&\dots,q-1,\,\, k=1\,\dots,p-1,\,\,
    k-j=1 \,\mod\, 2.
}}

\medskip

To summarize: in the four cases we considered above, the only
surface that split into two separate branches was $\CM_{p,q}^{+}$
with $(p,q)$ even. Moreover, we were able to identify these two
branches with the two signs of $\xi$.

Having worked out the singularities of our surfaces, let us now see that they
match the locations of the principal ZZ branes, as was the case in the bosonic
string. The principal ZZ branes have boundary cosmological constant given by
\smnsZZ. Using \ymsfzzt, we find that the $\hat\eta=+1$ branes are located at
the points
\eqn\sZZsingp{
    (x,y)=\left( \xi\cos{\pi\over 2q}\big(m q+n p\big),\,
    \xi\cos{\pi\over 2p}\big(m q +n p\big) \right)
}
in $\CM_{p,q}^{+}$, with $(m,n,\xi)$ taking values appropriate to the principal
$\hat\eta=+1$ ZZ branes as described in \diffsZZeven\ and \diffsZZodd.
Similarly, the $\hat\eta=-1$ branes are located at
\eqn\sZZsingm{
    (x,y)=\left( \xi\cos{\pi\over 2q}\big((m+1) q+n p\big),\,
    \xi\cos{\pi\over 2p}\big(m q +(n+1) p\big) \right)
}
in $\CM_{p,q}^{-}$, with $(m,n,\xi)$ again taking the appropriate values. We
claim that the locations of the principal ZZ branes are just a different
parametrization of the singularities of $\CM_{p,q}^{\pm}$. Let us briefly
sketch a proof. First, one can show that they indeed lie at the singularities
by checking that their locations $(x,y)$ satisfy the conditions $F(x,y)=0$ and
$dF(x,y)=0$. Since the locations of the principal ZZ branes are all distinct,
and there are exactly as many principal $\hat\eta\,$ ZZ branes as there are
singularities of $\CM_{p,q}^{\hat\eta}$, it follows that their locations match
the singularities exactly.

Combining the calculations of the one-point functions in the previous section
with the geometric results here, we can summarize our conclusions regarding the
ZZ branes very succintly. These conclusions are essentially the same as in the
bosonic string. The $\hat{\eta}=\pm 1$ principal ZZ branes are located at the
singularities of our surfaces $\CM_{p,q}^{\pm}$. ZZ branes located at the same
singularity of the surface differ by BRST null states, while ZZ branes located
at different singularities are distinct BRST cohomology classes. Finally, ZZ
branes that are not located at singularities of the surface are themselves BRST
null.

By analogy with the bosonic string, we also expect that the $\hat{\eta}$ FZZT
and ZZ branes can be represented as contour integrals of a one-form (or
boundary state) on the surfaces $\CM_{p,q}^{\hat\eta}$. For instance, in the
cases where $\xi$ is a redundant parameter, the $\hat{\eta}$ FZZT branes are
given by line integrals of the one-form $y\,dx$ on $\CM_{p,q}^{\hat\eta}$:
\eqn\fzztintsuper{
Z(\mu_B)=\mu^{p+q\over 2p}\int_{\CP}^{x(\mu_B)}y\,dx\ ,
}
where $\CP$ is an arbitrary fixed point in $\CM_{p,q}^{\hat\eta}$. For the case
of $(p,q)$ even and $\hat{\eta}=+1$ where $\xi$ actually labels two distinct
branes, we saw above that the corresponding surface splits into two separate
branches. These branches were also labelled by $\xi$, and they were connected
by singularities. Thus in this case alone can we define two inequivalent line
integrals on the surface; these clearly correspond to the two types of FZZT
brane:
\eqn\fzztintsuperxi{
Z(\mu_B,\xi)=\mu^{p+q\over 2p}\int_{\CP_{\xi}}^{x(\mu_B)}y\,dx\ .
}
Here $\CP_{\xi}\in \left(\CM_{p,q}^{+}\right)_{\xi}$ is an arbitrary fixed
point on the branch labelled by $\xi$.

We can consider the ZZ branes in a similar way. When $\xi$ is a redundant
parameter, the relation \sZZsFZZT\ between the ZZ and FZZT branes, together
with \fzztintsuper, implies that the ZZ branes can be written as closed contour
integrals of $y\,dx$. As in the bosonic string, we can promote this to a
relation between boundary states:
\eqn\intZZsuper{
    | m,n;\hat{\eta}\rangle=
    \oint_{B_{m,n}} \partial_x | z(x);\hat{\eta}\rangle\,\,dx\ ,
}
where the contour $B_{m,n}$ runs through the open cycle on $\CM_{p,q}^{\pm}$
and the singularity associated to the $(m,n)$ ZZ brane. For $\hat{\eta}=+1$ and
$(p,q)$ odd, we saw that $\xi$ was only redundant modulo BRST exact states.
Thus in this case, we must interpret \intZZsuper\ as a relation in the BRST
cohomology of the full theory.

Once again, the case of $(p,q)$ even and $\hat{\eta}=+1$ has more
structure. The relation \sZZsFZZT\ implies that for $n$ even, the
ZZ branes are differences of FZZT branes of the same $\xi$. Using
\fzztintsuperxi, we see that for $n$ even, we can write the ZZ
brane as a closed contour of the FZZT brane:
\eqn\sZZsFZZTMpe{
    | m,n ;\xi,\hat{\eta}=1\rangle\big|_{(m,n)\,\,{\rm even}} =
    \oint_{B_{m,n}^{\xi}}\partial_x |z(x);\xi,\hat{\eta}=1\rangle\,\,dx\ ,
 }
where the contour $B_{m,n}^{\xi}$ runs through the open
cycle on $\left(\CM_{p,q}^{+}\right)_{\xi}$ and the singularity
associated to the $(m,n,\xi)$ ZZ brane. On the other hand, for $n$
odd, the relation \sZZsFZZT\ says that the ZZ branes are
differences of FZZT branes of opposite $\xi$. Thus these ZZ branes
are not described by closed contours, but rather by line integrals
from one branch of $\CM_{p,q}^{+}$ to another:
\eqn\sZZsFZZTMpo{\eqalign{
    | m,n ;\xi,\hat{\eta}=1\rangle\big|_{(m,n)\,\,{\rm odd}} &=
    \int_{\CP_{\xi}}^{x_{m,n}}\partial_x
    | z(x),\xi,\hat{\eta}=1\rangle\,\,dx-
    \int_{\CP_{-\xi}}^{x_{m,n}}\partial_x|z(x),-\xi,\hat{\eta}=1\rangle\,\,dx\
    .
}}
Note that in order for these relations to be true, the properties of the
singularity associated to the $(m,n,\xi)$ ZZ brane must depend on $n$ in a very
non-trivial way: for $n$ even, the singularity must be a regular singularity of
the branch $\left(\CM_{p,q}^{+}\right)_{\xi}$, while for $n$ odd, it must be a
connecting singularity between the two branches. (See the discussion following
\singpqeven\ for a description of the two types of singularities.) Substituting
the locations of the ZZ branes into $T_{q\over2}(x)$ and $T_{p\over2}(y)$, we
immediately find that these polynomials are zero for the $n$ odd branes and
nonzero for the $n$ even branes. Therefore the $n$ odd (even) branes are indeed
located at the connecting (regular) singularities, which is a highly
non-trivial check of our relations \sZZsFZZTMpe\ and \sZZsFZZTMpo. Figure 3
illustrates the different types of FZZT and ZZ branes for this surface.

\medskip\ifig\figIII{The surface $\CM^{+}_{p,q}$ for $(p,q)$ even, along with
examples of FZZT and ZZ brane contours, shown here again for $(p,q)=(4,6)$.
There are now two types of FZZT brane labelled by $\xi$ (only one is drawn in
the figure), and these are given by contours (solid) from the base point
$\CP_{\xi}$ on the subsurface $(\CM_{p,q}^{+})_{\xi}$ to a point $z$ in
$\CM_{p,q}^{+}$. There are also two classes of ZZ branes: the neutral branes,
which are described by a closed contour (dashed) through a regular singularity;
and the charged branes, which are described by an open contour (dotted) that
runs from $\CP_{+}$ to $\CP_{-}$ through a connecting singularity.}
{\epsfxsize=0.6\hsize\epsfbox{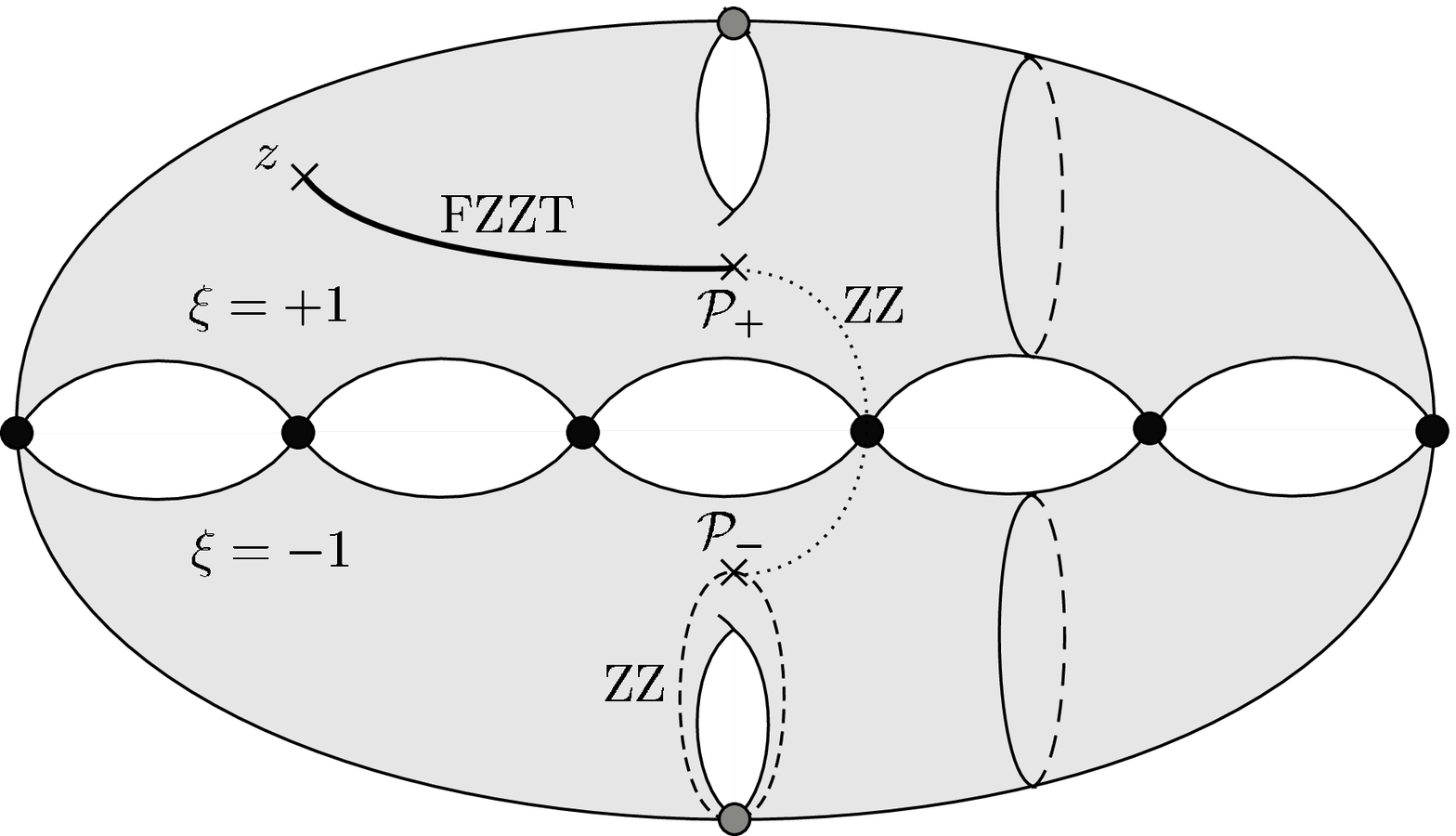}}

It follows immediately from these relations that the ZZ branes with $n$ even
are neutral branes, being differences of FZZT branes of the same charge.
Meanwhile the ZZ branes with $n$ odd are differences of two FZZT branes with
opposite values of $\xi$, so they are charged. We could have also seen this
directly from the behavior of Ramond part of the ZZ boundary state \superZZ\ at
zero momentum. Note that the connecting singularities come in pairs $(x,y)$ and
$(-x,-y)$. These correspond to a charged ZZ brane and its antibrane.

\subsec{Deformations of $\CM_{p,q}^{\pm}$}

Just as in the bosonic string, we can consider deformations of the surfaces
$\CM_{p,q}^{\pm}$ that preserve the number of singularities on each surface.
For $(p,q)$ odd, there is really only one surface, and it is identical to the
surface $\CM_{p,q}$ of the bosonic string. Therefore the discussion of
deformations in section 4.2 applies equally well here. This agrees with the
fact that the spectrum of physical operators in the $(p,q)$ odd bosonic and
supersymmetric minimal string theories are identical in every respect,
including their KPZ scalings. This also agrees with the fact that the $(p,q)$
odd minimal superstring theory is the same at $\mu>0$ and at $\mu<0$.

For $(p,q)$ even, there are two distinct surfaces $\CM_{p,q}^{\pm}$ described
by the equations $F_{\pm}(x,y)\equiv T_{p}(y)\mp T_q(x)=0$. The number of
singularities of the two surfaces is ${(p-1)(q-1)\pm 1 \over 2}$. By arguments
analogous to those in section 4.2, we find that a complete and independent
basis for the deformations is given by
\eqn\defbasispqeven{\eqalign{
    &\delta_{r,s}F_{\pm}(x,y)=\tilde{t}_{r,s}\bigg(
    U_{q-s-1}(x)U_{r-1}(y)\mp U_{p-r-1}(y)U_{s-1}(x)\bigg)\cr
    &1 \le r \le p\ , \quad s \le q-1\ , \quad  q r-p s \ge 0\ .
}}
Note the one essential difference between the deformations of
$\CM_{p,q}^{\pm}$: for $\CM_{p,q}^{+}$, the deformation with
$(r,s)=({p\over2},{q\over2})$ is not present, because the two terms in
\defbasispqeven\ are equal and cancel. In other words, there is no singularity
preserving deformation of $\CM_{p,q}^{+}$ that corresponds to the Ramond ground
state.

For the $\eta=-1$ brane, this is precisely what is needed. For $\mu>0$, the
brane is described by $\CM_{p,q}^{-}$, which does have a deformation
corresponding to the Ramond ground state. Meanwhile for $\mu<0$, the brane is
described by $\CM_{p,q}^{+}$, which does not have a deformation corresponding
to the Ramond ground state. This agrees with our study of the 0B spectrum in
section 5, where we saw that the Ramond ground state in the $(-1/2,-1/2)$
picture was present for $\mu>0$ and not present for $\mu<0$.

Consider now the $\eta=+1$ brane. For $\mu>0$ there is no problem, because
although the Ramond ground state exists in the string theory, its deformation
of the surface is zero. This occurs for two reasons: the Liouville part of the
tachyon one-point function vanishes for the Ramond ground state, and the matter
wavefunction also vanishes by \Rgscontribg. However, for $\mu<0$ the surface
$\CM_{p,q}^{-}$ has a deformation with $(r,s)=({p\over2},{q\over2})$, but the
Ramond ground state does not exist in the $(-1/2,-1/2)$ picture. We do not know
what this deformation corresponds to.

Finally, we can extend the bosonic string discussion of the effect
a background with many $(m,n)$ ZZ branes to the superstring.
Adding to the system $N_{m,n} \sim 1/g_s$ ZZ branes the
singularity associated with the pinched cycle $A_{m,n}$ is opened
up and as in \Nmn,
 \eqn\Nmns{\oint_{A_{m,n}} y dx = g_s N_{m,n}\ .}
When the $(m,n)$ D-brane is charged, this has the effect of adding
flux to the system which can be measured by an expression similar
to \Nmns.  Indeed, in \KlebanovWG\ the flux due to charged
D-branes was defined by such a contour integral, and its effect on
the dynamics was discussed.

\newsec{Comparing the $(p,q)$ odd supersymmetric and bosonic models}

In this section, we will compare the bosonic and 0B supersymmetric minimal
string theories with $(p,q)$ odd.

First, the two theories have the same spectrum of physical
operators \refs{\KlebanovWG, \JohnsonHY}. In particular, they both
have $(p-1)(q-1)/2$ tachyons, and their ground rings are
isomorphic. Moreover, since the $\mu$-deformed ground ring
multiplication is the same in both theories, the arguments of
section 2.2 and 5.2 imply that the tachyon $N$-point functions for
$N\le 3$ are the same in the two theories. Both will be given
essentially by the fusion rules in the minimal model, which are
the same as the fusion rules of the superminimal model. A more
interesting test of our conjecture would be to compare the general
($N>4$) correlation functions of the two theories. This might
probe structure of the theories that goes beyond their ground
rings.

Second, as we saw in section 5, the $(p,q)$ odd minimal superstring theories do
not have the Ramond ground state in their spectrum, for either sign of $\mu$.
Thus for these models, worldsheet supersymmetry is always broken. Moreover,
since the Ramond ground state measures RR charge, it is not possible to define
such a charge for the $(p,q)$ odd minimal superstring. These are of course both
necessary conditions if the supersymmetric and bosonic theories are to be the
same.

The boundary states of the bosonic and supersymmetric models offer
more opportunities for comparison. As we saw in section 7.1, they
are described by the same Riemann surface $\CM_{p,q}$. Thus it is
reasonable to expect the bosonic and supersymmetric theories to
have an identical set of FZZT and ZZ branes. We will not offer a
detailed comparison, but will instead focus on the simplest test:
counting the total number of such states. In the bosonic string,
we found that the set of all FZZT branes reduced to a
one-parameter family (parametrized by $\sigma$ or $z$) of FZZT
branes with matter state $(1,1)$. We also found exactly
$(p-1)(q-1)/2$ independent ZZ branes distinguished by their
Liouville label $(m,n)$.

Compare this with the boundary states in the superstring. These states are in
general labelled by $\xi=\pm 1$ and $\hat{\eta}=\pm 1$ in addition to their
Liouville and matter labels. For $(p,q)$ odd, we suggested in section 6.2 that
both $\xi$ and $\hat\eta$ are redundant parameters. These arguments were also
motivated geometrically in section 7. Then assuming that there is a reduction
of the FZZT branes with general matter state to the FZZT branes with $(1,1)$
matter state, as in the bosonic string, we find the same one-parameter family
of independent FZZT branes. We also find exactly $(p-1)(q-1)/2$ ZZ branes, as
in the bosonic string, again assuming the reduction in matter states. Clearly,
it would be nice to justify this assumption with an explicit computation. For
this, we would need explicit expressions for the matter wavefunctions.

We will see in the next section that the Riemann surface is precisely the
surface that defines the dual matrix model description. Therefore the bosonic
and supersymmetric $(p,q)$ models have the same matrix model, and we expect
them to agree to all orders in string perturbation theory.

We will conclude this section with a list of further tests of our
proposed duality that should be carried out. The 0B superstring
has a ${\Bbb Z}_2$ symmetry and orbifolding by it leads to the 0A
theory. It is not known whether an analogous construction exists
in the bosonic string. It also remains to understand what it means
to have negative $\mu$ in the bosonic string, since the
superstring clearly exists for both signs of $\mu$, and moreover
it is invariant under $\mu\rightarrow -\mu$. Finally, a more
detailed comparison of the boundary states should be undertaken,
using explicit formulas for the matter wavefunctions.

\newsec{Relation to matrix models}

Finally we come to discuss the connection to the dual matrix model. Clearly,
the matrix model description emerges from the Riemann surface $\CM_{p,q}$. The
FZZT brane corresponds to the macroscopic loop operator of the matrix model,
with the $x$ and $y$ the matrix eigenvalue and the resolvent, respectively. The
analytic structure of $\CM_{p,q}$ dictates the critical behavior of the matrix
model, which is the starting point of the double-scaling limit.

It is known that there are many equivalent matrix model descriptions of the
$(p,q)$ minimal string theories. The most natural descriptions for our purposes
are Kostov's loop gas formalism \refs{\KostovHN\KostovGS-\KostovUQ} and the
two-matrix model \DaulBG. We will focus mainly on the latter. However for
actual calculations, especially in the conformal background, the former is
often more useful. The two-matrix model consists of two random $N\times N$
hermitian matrices $X$ and $Y$, described by the partition function
\eqn\twopot{
    Z_{\rm matrix}=\int dX dY\,\exp\left(-{N\over g}\Tr\left(V_1(X)+V_2(Y)
    - X Y\right)\right)\ .
}
In the planar $N\rightarrow \infty$ limit, the eigenvalues of $X$ and $Y$ are
described by a continuous distribution which can be determined from the
structure of the surface $\CM_{p,q}$. For instance, when $p=2$, the surface can
be described as a two-sheeted cover of the complex plane. Then the eigenvalues
are localized to the branch cuts of the surface.

Important observables in the two-matrix model are the two resolvents, which are
derivatives of the loop operators:
\eqn\resolvents{
    R(x)\equiv W'(x)=\Tr{1 \over X-x}, \qquad \tilde{R}(y)\equiv \tilde{W}'(y)=\Tr{1\over
    Y-y}\ .
}
These resolvents were calculated in the equivalent loop-gas formalism
\refs{\KostovHN\KostovXT-\KostovCG} for the $(p,q)$ minimal string theories in
the conformal background, and at least for the bosonic models where $(p,q)$ are
relatively prime, they are given by the following expressions:
\eqn\twomatres{\eqalign{
    R(x)&=\left(x+\sqrt{x^2-1}\right)^{q/p}+\left(x-
    \sqrt{x^2-1}\right)^{q/p}\cr
    \tilde{R}(y)&=\left(y+\sqrt{y^2-1}\right)^{p/q}
    +\left(y-\sqrt{y^2-1}\right)^{p/q}\ .\cr
}}
One can show that the two resolvents are inverses of one another; in fact this
is a consequence of the saddle-point equations of the two-matrix model. Thus we
should identify $y=R(x)$ and $x=\tilde{R}(y)$. Moreover, if we write $x=\cosh
\theta$ and $y=\cosh \phi$, the equations above become simply the equation
$T_p(y)-T_q(x)=0$ of our surface $\CM_{p,q}$. Therefore, the eigenvalues $x$
and $y$ are the boundary cosmological constant and resolvent of the FZZT brane,
respectively. Since $y=R(x)$ and $x=\tilde{R}(y)$, we also identify the FZZT
brane and the dual brane with the two macroscopic loops built out of $X$ and
$Y$, respectively. The fact that there are only two macroscopic loops in the
two-matrix model also agrees well with our results in section 3, where we found
that there was only one type of FZZT brane (and its dual), labelled by matter
state $(1,1)$.

The picture is analogous for the superstring. For $(p,q)$ odd, the matrix model
is the same as in the bosonic string, and in particular the expressions for the
resolvents are unchanged. This agrees with the fact that we find only one
surface in the $(p,q)$ odd superstring, and moreover that it is the same
surface as in the bosonic string. For $(p,q)$ even, there are two surfaces, so
the two-matrix model description is more difficult. In particular, for a given
sign of $\mu$, we can make only one of the surfaces starting from the
resolvents \resolvents. Perhaps the other surface arises from a more
complicated resolvent.

Identifying the coordinates $(x,y)$ of $\CM_{p,q}$ with the eigenvalues of $X$
and $Y$ also suggests a way to go beyond tree-level in the minimal string
theory. In section 4.1, we discussed the sense in which $x$ and $y$ are
conjugate variables. This has a natural interpretation in the two-matrix model,
where in the double-scaling limit the two matrices become the
pseudo-differential operators $Q$ and $P$ which satisfy the equation
\DouglasDD:
 \eqn\douglaseq{ [Q,P]=\hbar\ .
}
This suggests that in order to quantize the minimal string theory, we should
promote $x$ and $y$ to operators and quantize the Riemann surface $\CM_{p,q}$
\refs{\MooreMG, \MooreCN}. This is a promising avenue of investigation that we
will leave for future work.

Another connection with the matrix model that we should mention is the relation
between matrix model operators and the operators of minimal string theory in
the conformal background ($\mu\ne 0$). The natural basis of operators in the
matrix model are simply products of $X$ and $Y$; while the tachyons, ground
ring elements, etc.\ are natural to use in minimal string theory. The relations
\ringfid\ and \tachmod\ in terms of Chebyshev polynomials (and their
counterparts in the superstring) tell us how to transform from one basis to
another. In \MooreIR, this change of basis was worked out in detail for the
$(p,q)=(2,2m-1)$ minimal string theories, and in appendix B we check explicitly
that our results are in complete agreement. Our results on the $\mu$-deformed
ring elements, tachyons and macroscopic loops generalize the work of \MooreIR\
to all $(p,q)$.

The most detailed picture we have is for theories with $p=2$ which can be
described by a one-matrix model. Here we have a unified description of the
bosonic models ($q$ odd), the supersymmetric models ($q=4k$) and its
generalizations ($q=4k+2$). For the bosonic models, the description is in terms
of a one-matrix model with one cut. The curve is
\eqn\onematbos{
2y^2 = T_q(x)+1={\big(T_{q+1\over2}(x)+T_{q-1\over2}(x)\big)^2\over x+1}\ ,
}
and the effective eigenvalue potential $V_{eff}(x)$ is obtained by integrating
$y$ with respect to $x$ \KlebanovWG:
\eqn\Veffbos{
    \sqrt{2}V_{eff}(x)={T_{q+3\over2}(x) + T_{q+1\over2}(x)\over (q+2)
    \sqrt{x+1}}-{T_{q-1\over2}(x) + T_{q-3\over2}(x)\over (q-2)
    \sqrt{x+1}}\ .
}
On the other hand, the supersymmetric 0B models are represented by a one-matrix
model, which has two cuts for $\mu>0$ and no cuts for $\mu<0$
\refs{\GrossHE\PeriwalGF-\CrnkovicWD,\KlebanovWG}. Here we have the curve
\eqn\onematetap{
    2y^2 = \hat{\eta}\, T_q(x)+1
}
for the $\hat \eta=\pm 1$ brane. For the $\hat{\eta}=-1$ brane, we find
\eqn\Veffgenm{
    V_{eff}(x)\big|_{\hat\eta=-1}={\big(2q\, T_{q\over2}(x)-x\,
    U_{{q\over2}-1}(x)\big)\sqrt{1-x^2}\over q^2-4 }\ ,
}
while for the $\hat\eta=+1$ brane the result is
\eqn\Veffgenp{
    V_{eff}(x)\big|_{\hat\eta=+1}=
    {T_{{q\over2}+1}(x)\over q+2}-{T_{{q\over2}-1}(x)\over q-2}\ .
}
For example in pure supergravity, $(p,q)=(2,4)$ and the curve for the $\hat
\eta =-1$ brane becomes $y^2=-4x^2 (x^2-1)$, while for $\hat \eta =1$ the curve
is  $y^2=(2x^2-1)^2$. In general, we find that the curve has no cuts for
$\hat\eta=+1$ (it is polynomial in $x$) and has two cuts starting at $x=\pm 1$
for $\hat\eta=-1$. Thus the resolvent of the 0B matrix model in the positive
(negative) $\mu$ phase corresponds to the $\hat{\eta}=-1$ ($+1$) brane.
Evidently, only the $\eta=-1$ brane can be interpreted as the resolvent in the
0B matrix model.

Using a change of variables, we can also obtain the resolvents of the 0A matrix
models for the $(2,2k)$ theories. The change of variables is motivated by the
fact that in 0A, the natural variable is not the eigenvalue $x_B$, but rather
$x_A=x_B^2$. If we think of the resolvents for 0A and 0B as a one-forms on the
Riemann surface, imposing that the one-forms transform into one another results
in $y_B\,dx_B=y_A\,dx_A=2y_A x_B\,dx_B$, so we find
\eqn\ABtransf{
x_A=x_B^2,\qquad y_A={y_B\over 2 x_B}\ .
}
Thus the resolvent of 0A can actually have a pole at $x_A=0$, a fact noticed in
\KlebanovWG. There it was convenient to define the curve of 0A in terms of
\eqn\Acurvedef{
\hat{y}_A\equiv 2x_A y_A= x_B y_B\ .
}
Using \ABtransf\ and \Acurvedef, we find the 0A curves:
\eqn\Acurvefind{
    2\hat{y}_A^2 = x_A \left(\hat{\eta}\, T_{q\over
    2}\big(2x_A-1\big)+1\right)\ .
 }
Again, for pure supergravity we recognize the resolvent of the 0A matrix model
in the two phases as the $\eta=-1$ brane.  For positive $\mu$ ($\hat \eta=-1$)
it satisfies $\hat{y}_A^2 =4 x_A^2 (1-x_A)$ and for negative $\mu$ ($\hat
\eta=+1$) it is $\hat{y}_A^2 = x_A (2x_A-1)^2$. Moreover, at least for pure
supergravity, the $\eta=\pm 1$ brane in 0A is the same as the $\eta=\mp 1$
brane in 0B. As we have already seen, the $\eta=+1$ 0B brane cannot be
described easily in the 0B language. Instead, it has a natural description as
the resolvent of the 0A theory.

The description in terms of the effective potential gives us a more physical
understanding of the ZZ branes in the one-matrix model. The singularities of
the surface correspond to points where $y(x)=T_q'(x)=0$, i.e.\ the regular
zeros of $y$. In terms of the effective eigenvalue potential $V_{eff}(x)$, this
is the statement that the singularities are local extrema of $V_{eff}(x)$.
Therefore the ZZ branes are matrix eigenvalues located at the zero-force points
of the effective potential. In order to create a ZZ brane, we must pull it out
of the Fermi sea, which corresponds to the branch cut of our surface. This is
the meaning of the ZZ brane as a contour integral: the integral pulls an
eigenvalue from the Fermi sea onto the singularity (zero-force point).

Note that the extrema of $V_{eff}(x)$ are not necessarily local
maxima. Let us focus on the bosonic models for concreteness, whose
effective potential is given by \Veffbos. It is easy to show that
the principal $(1,n)$ ZZ branes with $n$ even (odd) lie at local
minima (maxima) of $V_{eff}$, and moreover at these extrema,
$V_{eff}$ takes the values
\eqn\veffextrema{
V_{eff}(x_n)\sim (-1)^{n+1}\sin{2\pi n\over 2m-1}\ .
}
Recall that the ``Fermi level" of the perturbative vacuum is
located at $V_{eff}=0$. Then according to \veffextrema, all the
minima of $V_{eff}$ lie {\it below the Fermi sea}. Apparently, all
of these models are slightly unstable to the tunneling of a small
number of eigenvalues into the minima of $V_{eff}$.

This detailed discussion has so far been only for the one-matrix model. The
picture in the two matrix model is not nearly as complete. For example, it is
not clear how to obtain the 0A curves. Presumably, this can be done by
considering a two-matrix model of complex matrices. Also, a description in
terms of an effective potential is lacking, although some interesting proposals
were advanced in \KazakovYH. In any event, we still expect that in some sense
the singularities of $\CM_{p,q}$ are ``zero-force" points of some effective
potential.

The effect of adding order $1/g_s$ ZZ branes on the Riemann
surface has a natural interpretation in the matrix model.  As in
\refs{\DijkgraafDH,\CachazoRY}, the contour integral \Nmn\Nmns\
 \eqn\Nmnm{\oint_{A_{m,n}} y dx = g_s N_{m,n}\ .}
measures the number of eigenvalues around $(x_{mn}, y_{mn})$. We
again see that the matrix model eigenvalues can be thought of as
ZZ branes with the different $(m,n)$ ZZ brane differ by their
position in the surface.

\vskip 4cm

\noindent {\bf Acknowledgments:}

We would like to thank C.~Beasley, M.~Douglas, I.~Klebanov, D.~Kutasov,
J.~Maldacena, E.~Martinec, J.~McGreevy, G.~Moore, S.~Murthy, P.~Ouyang,
L.~Rastelli and E.~Witten for useful discussions. The research of NS is
supported in part by DOE grant DE-FG02-90ER40542. The research of DS is
supported in part by an NSF Graduate Research Fellowship and by NSF grant
PHY-0243680. Any opinions, findings, and conclusions or recommendations
expressed in this material are those of the author(s) and do not necessarily
reflect the views of the National Science Foundation.

\appendix{A}{The Backlund transformation and FZZT branes}

The purpose of this appendix is to give a semiclassical, intuitive
picture of the FZZT branes using the Backlund transformation.

\subsec{Bosonic Liouville theory}

We start with the bosonic Liouville theory in a two-dimensional,
Lorentzian signature spacetime (our conventions are chosen to
agree with \FateevIK):
 \eqn\boscon{
 \CL= {1\over 4\pi} \left( (\partial_\tau \phi)^2 -
 (\partial_\sigma  \phi)^2 - 4\pi\mu  e^{2b\phi}\right)\ .
 }
The most general classical solution to the equations of motion is given in
terms of two arbitrary functions $A^\pm(x^\pm)$
 \eqn\phicl{e^{2b\phi_{cl}} = {1 \over \pi b^2 \mu} {\partial_+A^+
 \partial_-A^-\over (A^+-A^-)^2}\ ,
 }
where $x^\pm=\tau\pm \sigma$ and $\partial_\pm={1\over 2}(\partial_\tau \pm
\partial_\sigma)$.  The Backlund field is a free field which is
defined in terms of $A^\pm$
 \eqn\backdef{\tilde\phi=  {1\over 2b}\log \left({\partial_+ A^+\over \partial_-
 A^-}\right)\ .
 }
Eliminating $A^\pm$ we find
 \eqn\phiphit{\eqalign{
 &\partial_\sigma \phi =\partial_\tau \tilde \phi - \sqrt{4\pi\mu}
 e^{b\phi}\cosh (b \tilde \phi) \cr
 &\partial_\tau \phi =\partial_\sigma \tilde \phi -\sqrt{4\pi\mu}
  e^{b\phi}\sinh( b \tilde \phi)\ ,
 }}
or equivalently
 \eqn\phiphitpm{\pm \partial_\pm \phi =\partial_\pm \tilde \phi
 - {\sqrt{4\pi\mu} \over 2} e^{b(\phi \pm \tilde \phi)\ .
 }}
The generating functional of this transformation is
 \eqn\gentran{\int d \sigma \left( \phi \partial_\sigma \tilde
 \phi -{\sqrt{4\pi\mu} \over b}e^{b \phi} \sinh (b \tilde
 \phi)\right)\ .
 }
Alternatively, we can start from the transformation \phiphit\ or \phiphitpm\
and check that they are compatible only if the appropriate equations of motion
are satisfied
 \eqn\compat{\partial_+\partial_-\phi =- \pi b \mu e^{2b\phi}\ ,
 \qquad\qquad \partial_+\partial_-\tilde \phi =0\ .
 }

It is straightforward to evaluate the energy momentum tensor
 \eqn\enet{T_{\pm\pm}= (\partial_\pm \phi)^2 - {1\over b}
 \partial_\pm^2 \phi = (\partial_\pm \tilde
 \phi)^2 \mp {1\over b} \partial_\pm^2\tilde\phi\ ,
 }
where the second term is the improvement term (note that it is different for
$\phi$ and $\tilde \phi$). Using the classical solution \phicl\ and the energy
momentum tensor \enet, one can check that
 \eqn\nullvea{\left[\partial_+^2 - b^2
 T_{++}(\phi_{cl})\right] {1\over \sqrt{\partial_+ A^+}}=
 \left[\partial_+^2 - b^2
 T_{++}(\phi_{cl})\right] {A^+\over \sqrt{\partial_+ A^+}}=0\ ,
 }
and therefore
 \eqn\nullve{\left[\partial_+^2 - b^2
 T_{++}(\phi_{cl})\right] e^{-b\phi_{cl}}=0\ .
 }
In the quantum theory this is the statement that $e^{-b\phi}$ has a null
descendant. This null vector is the basis for the exact solution of the quantum
theory \refs{\DornSV\TeschnerYF-\ZamolodchikovAA}.

The first term in the transformation \phiphitpm\ shows that it is like a
T-duality transformation.  Therefore, we expect that Dirichlet and Neumann
boundary conditions are exchanged.  Indeed, the FZZT brane which is associated
with the Neumann boundary conditions
 \eqn\fzztbc{
 \partial_\sigma\phi = -2\pi b\mu_B e^{b\phi}
 }
becomes a Dirichlet brane in terms of $\tilde \phi$ with
 \eqn\fzztbcb{
    {\mu_B\over\sqrt\mu}={1 \over b\sqrt{\pi}} \cosh (b\tilde \phi)\ .
 }
Note that \fzztbcb\ is the semiclassical ($b\to 0$) limit of \fzzsmub, before
the rescaling of $\mu$ and $\mu_B$. Therefore we identify the Backlund field with the
parameter $\pi\sigma$ of the FZZT brane.

The fact that Dirichlet and Neumann boundary conditions are exchanged by the
Backlund transformation is also clear from the form of the energy momentum
tensor $T_{\pm\pm}$ of \enet. Because of the improvement term Dirichlet
boundary conditions of $\phi$ are not conformally invariant. However, in terms
of $\tilde \phi$ the improvement terms in $T_{++}$ and $T_{--}$ have opposite
values.  Therefore, Neumann boundary conditions for $\tilde \phi$ are not
conformally invariant but Dirichlet boundary conditions are consistent.
Equivalently, a would-be localized D-brane at $\phi$ is expected to have mass
proportional to $e^{ \phi \over b}$. Therefore, it is unstable and is pushed to
$\phi \to +\infty$.  On the other hand a D-brane localized at $\tilde \phi$ is
stable.

\subsec{Minisuperspace wavefunctions}

In the minisuperspace approximation we focus on the zero mode of $\phi$.  We
can take as a complete basis of states the eigestates of $\phi$ or the states
with energy ${1\over 2} P^2 $.  Their inner products are
 \eqn\stbas{\eqalign{
 \langle \phi_1|\phi_2 \rangle &= \delta(\phi_1-\phi_2) \cr
 \langle P_1|P_2 \rangle &= \pi \delta(P_1-P_2)\cr
 \langle \phi|P \rangle &={2\left({\pi\mu \over b^2}\right)
 ^{-{iP\over b}} \over \Gamma\left(-{2i P \over b}\right)}
 K_{{2iP\over b}} \left({\sqrt {4\pi\mu} \over b} e^{b\phi}\right)\cr&=
 e^{2iP\phi}(1 + \dots) + \left({\pi\mu \over b^2}\right)^{-{2iP\over b}}
 { \Gamma\left({2i P \over b}\right) \over \Gamma\left(-{2i P \over b}
 \right) } e^{-2iP\phi} (1+\dots)\ .\cr
}}
The states $|P\rangle$ were normalized such that the incoming wave
$e^{2iP\phi}$ in $\langle \phi |P\rangle$ has weight one.  The coefficient in
front of $e^{-2iP\phi}$ is a pure phase and is the reflection amplitude.

The boundary states are eigenstates of $\tilde \phi$.  They satisfy
 \eqn\otherba{\eqalign{
 &\langle \phi|\tilde \phi \rangle = e^{-2\pi\mu_B e^{b\phi}},
 \qquad\qquad \mu_B={\sqrt\mu\over b\sqrt{\pi}} \cosh (b\tilde \phi) \cr
 &\langle \tilde \phi_1|\tilde \phi_2 \rangle = -{1\over b}
 \log\left[{\sqrt{4\pi \mu} \over b}\left(\cosh (b\tilde \phi_1 ) +
 \cosh(b\tilde \phi_2) \right)\right]+ const \cr
 &\qquad\qquad=-{1\over b}\log\left({2\sqrt{4\pi \mu} \over
 b}\cosh{b(\tilde \phi_1 + \tilde \phi_2) \over 2} \cosh{b
 (\tilde \phi_1 - \tilde \phi_2) \over 2} \right)+ const \cr
 &\langle \tilde \phi|p \rangle = {2\over b}
 \Gamma\left({2i P\over b}\right)\left({\pi\mu \over b^2}\right)
 ^{-i{P\over b}} \cos (P\tilde \phi)\ .
  }}
The additive constant in $\langle \tilde \phi_1|\tilde \phi_2 \rangle$ is a
nonuniversal infinite constant which is independent of $\tilde \phi_{1,2}$. The
wave function $\langle \phi|\tilde \phi \rangle$ can be derived using the
generating functional \gentran\ after exchanging the role of $\tau$ and $\sigma
$. Alternatively, we can derive it in a Euclidean worldsheet. Note that the
states $|\tilde \phi \rangle$ are not orthonormal. This means that the
canonical transformation from $\phi$ to $\tilde \phi$ is not unitary.  This
fact is also the reason for the somewhat unusual decomposition
 \eqn\decom{
 |\tilde \phi \rangle = {2\over b\pi } \int _0^\infty
 dP\ \cos (2P\tilde \phi) \Gamma\left(-{2i P\over
 b}\right)\left({\pi\mu \over b^2}\right)^{{iP\over b}} |P\rangle\ .
 }
The fact that is the semiclassical limit of the Liouville part of the FZZT
boundary state \FZZTbs\ confirms our identification of the Backlund field with
the parameter $\pi\sigma$ of the FZZT brane. It is interesting that if not for
the factor of $\Gamma\left(-{2i p\over b}\right)\left({\pi\mu \over
b^2}\right)^{i{p\over b}}$, this would have meant that $\tilde \phi$ is
conjugate to $p$ and $|\tilde \phi \rangle $ is a standard position eigenstate.
However, since this factor depends only on $p$ and not on $\tilde \phi$, that
conclusion is not completely wrong.

\subsec{Supersymmetric Liouville theory}

Now let us consider super-Liouville theory with a Euclidean worldsheet. We
follow the conventions of \DouglasUP, except that here we rescale $\mu\to
\mu/2$. The covariant derivatives, supercharges and algebra are
 \eqn\supco{\eqalign{
 & D={\partial \over \partial \theta} + \theta \partial
    \quad, \qquad
 \bar D={\partial \over \partial \bar\theta} + \bar\theta \bar \partial
    \quad,\qquad
 \{D,D\}=2  \partial
    \quad, \qquad
 \{\bar D,\bar D\}=2  \bar\partial \cr
 & Q={\partial \over \partial \theta} - \theta \partial
    \quad,\qquad
 \bar Q ={\partial \over \partial
 \bar\theta} - \bar\theta \bar\partial
    \quad ,\qquad
 \{Q,Q\}=-2  \partial
    \ \ ,\qquad
 \{\bar Q,\bar Q\}=-2  \bar \partial
 }}
with all other (anti)commutators vanishing. We define $z=x+iy$, $\bar z=x-iy$
and therefore $\partial=(\partial_x-i\partial_y)/2$,
$\bar\partial=(\partial_x+i\partial_y)/2$. Finally, the integration measure is
$\int\! d^2z d^2\theta = 2\int\! dxdyd\bar\theta d\theta$.  The action for
 \eqn\superf{
  \Phi = \phi + i\theta\psi + i \bar \theta
    \bar \psi + i\theta\bar \theta F
 }
is the super-Liouville action
 \eqn\slag{
  S = {1 \over 4\pi}\int\!d^2z d^2\theta\,
    \Bigl[ D\Phi \bar D\Phi + i\mu e^{b\Phi}\Bigr]\ .
  }

The Backlund transformation in super-Liouville theory
\refs{\ChaichianYZ,\DHokerZY} is
 \eqn\superback{\eqalign{
 &D\Phi = D\tilde \Phi + \xi b\sqrt{|\mu|} \Gamma
 e^{{b \over 2}(\Phi + \tilde \Phi)} \cr
  &\bar D\Phi = -\bar D\tilde \Phi - \xi\zeta b\sqrt{|\mu|}
  \Gamma e^{{b \over 2}(\Phi -\tilde \Phi)} \cr
  &D\Gamma = -i\xi\sqrt{|\mu|\over 4}
  e^{{b \over 2}(\Phi + \tilde\Phi)} \cr
   &\bar D\Gamma = -i\xi\zeta \sqrt{|\mu|\over 4}
   e^{{b \over 2}(\Phi - \tilde \Phi)} \cr
 &\zeta={\rm sign}(\mu)\ ,
 }}
where $\Gamma$ is a fermionic superfield ($\Gamma^2=0$), and
$\tilde \Phi$ is the Backlund field. (We have rescaled the
variables to agree with the conventions of \DouglasUP.) The
parameter $\xi=\pm 1$ implements the $(-1)^{F_L}$ symmetry, as
discussed in section 6.1. It is straightforward to check that the
integrability conditions for \superback\ are the equations of
motion
 \eqn\supereom{\eqalign{
 &\bar D D\Phi + {i\mu b \over2}e^{b\Phi}=0\cr
 &\bar D D\tilde \Phi =0\cr
 &\bar D D\Gamma -  {i\mu b^2 \over 8} e^{b\Phi}\Gamma =0\ ,
 }}
i.e.\ $\Phi$ satisfies the equation of motion of \slag, and $\tilde \Phi$ is
free.

The energy momentum tensor superfield is
 \eqn\superen{\eqalign{
 &T=D\Phi D^2\Phi -{1\over b} D^3\Phi = D\tilde \Phi D^2
 \tilde \Phi -{1\over b} D^3\tilde \Phi \cr
  &\bar T=\bar D\Phi \bar D^2\Phi -{1\over b} \bar
  D^3\Phi = \bar D\tilde \Phi\bar D^2
 \tilde \Phi + {1\over b} \bar D^3\tilde \Phi\ .
 }}
The second term is the improvement term. It has the same sign in $T$ and $\bar
T$ when expressed in terms of $\Phi$, but it has opposite signs in $T$ and
$\bar T$ when expressed in terms of $\tilde \Phi$.  The fermionic superfield
$\Gamma$ does not contribute to the energy momentum tensor.

The FZZT branes are D0-branes of $\tilde \Phi$; i.e.\ the Backlund field
satisfies Dirichlet boundary conditions
 \eqn\sdbr{
    D_t\tilde \Phi = (D + \eta \bar D)\tilde \Phi =0\ ,
 }
where $\eta=\pm 1$ denotes the preserved subspace of superspace: $\theta
=\eta\bar \theta$.  The opposite signs of the improvement term in $T$ and $\bar
T$ when expressed in terms of $\tilde \Phi$ \superen, make these boundary
conditions conformal (conversely, Neumann boundary conditions on $\tilde \Phi$
are not consistent). From \superback\ we find the boundary conditions of the
other fields
 \eqn\superbackb{\eqalign{
 &D_n\Phi=(D-\eta \bar D)\Phi = 2b\mu_B \Gamma e^{{b\over 2} \Phi}
 \cr
  &D_t\Gamma= (D + \eta \bar D)\Gamma = -i\mu_B  e^{{b \over
  2}\Phi}\ , \cr
}}
where
\eqn\smuBsemi{\eqalign{
 &\mu_B=
 \cases{\xi\sqrt{|\mu|} \cosh ({b\over 2}\tilde \Phi)& $\hat\eta=+1
 $\cr
  \xi\sqrt{|\mu|} \sinh ({b\over 2}\tilde  \Phi)& $ \hat\eta=-1 $ \cr
  }
  }}
with $\hat\eta=\zeta\eta$. This expression for $\mu_B$ is the semiclassical
limit of \sFZZTmuBp. Thus, as in the bosonic string, we are led to identify the
Backlund field with the parameter $\sigma$ of the FZZT brane. Note also that
the equations \superbackb\ are the boundary equations of motion when we add to
the action \slag\ the boundary term
 \eqn\openta{\eqalign{
  S_{\rm bry} &=  {1 \over 2\pi}\oint\! dxd\theta_t\,
 \Bigl(\Gamma D_t\Gamma + 2i\mu_B \Gamma e^{{b\over 2}\Phi}\Bigr) \cr
 &= {1 \over 2\pi}\oint\!dx\Bigl[- \gamma\partial_x\gamma- f^2
 -\mu_B\Bigl(b \gamma(\psi+\eta\bar\psi) e^{{b\over 2}\phi}
 + 2fe^{{b\over 2}\phi}\Bigr)\Bigr] \cr
  &= {1 \over 2\pi}\oint\!dx\Bigl[-\gamma\partial_x\gamma
  -\mu_B\, b\, \gamma(\psi+\eta\bar\psi) e^{{b\over 2}\phi}
   +\mu_B^2\, e^{b\phi}\Bigr]\ ,
 }}
where $\Gamma=\gamma + i\theta_t f$ is a fermionic superfield at the boundary.

\subsec{Minisuperspace wavefunctions}

We denote by $|\phi \pm\rangle$ and $|P\pm \rangle$ the states with the two
different fermion number in the R sector and by $|\phi 0 \rangle$ and $|P 0
\rangle$ the states in the NS sectors. Following \DouglasUP\ the wave functions
are expressed in terms of $z=|\mu|e^{b\phi}$. Let us assume first that $\mu$ is
positive ($\zeta={\rm sign}(\mu)=+1$). Then the wavefunctions are\foot{We
reversed the $\pm$ label of $\Psi_{P\pm}$ and changed the normalization of the
wavefunctions relative to \DouglasUP\ to have the coefficient of $e^{iP\phi}$
in the wavefunctions normalized to one.}
 \eqn\witso{\eqalign{
 &\langle \phi_1 \pm |\phi_2\pm \rangle =\delta(\phi_1-\phi_2) \cr
 &\langle \phi_1 0 |\phi_2 0 \rangle =\delta(\phi_1-\phi_2) \cr
  &\langle P_1 \pm |P_2\pm \rangle =2\pi \delta(P_1-P_2) \cr
 &\langle P_1 0 |P_2 0 \rangle =2\pi \delta(P_1-P_2) \cr
 &\Psi_{P\pm}(\phi)=\langle \phi \pm |P\pm \rangle =  { 2\over
 \Gamma \left(-{iP\over b}+ {1\over 2}\right)}\left({|\mu|\over4}
 \right)^{-{iP\over b}} \sqrt z \left(K_{{iP\over b} -{1\over
 2}}(z) \pm K_{{iP\over b} + {1\over 2}}(z)  \right)\cr
 &\qquad =  e^{iP \phi}\left(1+\CO(z)\right)\pm
 {\Gamma\left({iP\over b} + {1\over 2}\right) \over \Gamma
 \left(-{iP\over b} + {1\over 2}\right)} \left({|\mu|\over 4}
 \right)^{-{2iP\over b}} e^{-iP\phi}\left(1+\CO(z) \right) \cr
 &\Psi_{P0}= \langle \phi 0 |P0 \rangle={2\over
 \Gamma\left(-{iP\over b} \right) } \left({|\mu|\over 4}
 \right)^{-{iP\over b}} K_{iP\over b}(z) \cr
 &\qquad = e^{iP\phi}\left(1+\CO(z)\right) - {\Gamma(1+
 {iP\over b} ) \over \Gamma\left(1-{iP\over b} \right) }
 \left({|\mu|\over 4} \right)^{-{2iP\over b}}
 e^{-iP\phi}\left(1+\CO(z)\right)\ .
 }}
The wavefunctions satisfy
 \eqn\suweq{\eqalign{
 &(z \partial_z \pm z) \Psi_{P\pm}(z) = {i P\over b}\Psi_{P\mp}(z)\cr
 &\left( - (z \partial_z)^2 + z^2 - {P^2\over b^2}
 \right)\Psi_{P0}(z)=0\ .
 }}

When $\mu$ changes sign $z$ which is defined in terms of $|\mu|$ does not
change.  Therefore the equations \suweq\ are unchanged. However, since these
equations are derived from the action of the supercharges, the term linear in
$z$ must change sign. This means that the wavefunctions $\Psi_{P\pm}(z)$ have
the same functional form when $\mu$ changes sign but they occur for the states
with opposite fermion number; i.e.
 \eqn\prow{\langle \phi  \pm |P\pm\rangle = \Psi_{P,\pm \zeta}(z)\ .
 }

In order to study D-branes we need eigenstates of $\tilde \phi$. In the R
sector they are $|\tilde \phi \pm\rangle$ and in the NS sector $|\tilde \phi
0\rangle$.  After integrating out the various fermions as in \DouglasUP, we
have
 \eqn\suphiti{\eqalign{
  &\langle \phi 0| \tilde \phi 0\rangle =
 e^{- z \cosh (b\tilde \phi)} \cr
 &\langle \phi \pm | \tilde \phi \pm \rangle = \sqrt z
 e^{- z \cosh(b \tilde \phi)} \left( e^  {b\tilde \phi \over 2}
 \pm \zeta e^ {- {b\tilde \phi \over 2}}\right) \cr
 &\langle \tilde \phi 0|P0\rangle = {2 \over b}\left({|\mu|\over
 4}\right)^{-{iP\over b}}\Gamma\left({iP\over b}
 \right) \cos(P\tilde \phi) \cr
 &\langle \tilde \phi \pm |P\pm \rangle = {2 \over b}\left({|\mu|
 \over 4}\right)^{-{iP\over b}}\Gamma\left({iP\over b} + {1\over
 2} \right) \left(e^{-iP\tilde \phi} \pm \zeta e^{iP\tilde \phi}
 \right)\ .
 }}
{}From these we derive the decompositions
 \eqn\decoms{\eqalign{
 &|\tilde \phi 0\rangle = {1\over b\pi } \int _0^\infty dP\ \cos
 (P\tilde \phi) \Gamma\left(-{i P\over b}\right)\left({|\mu| \over
 4}\right)^{{i P\over b}} |P 0 \rangle \cr
 &|\tilde \phi \pm \rangle = {1\over b\pi } \int _0^\infty
 dP\  \left(e^{iP\tilde \phi} \pm \zeta e^{-iP\tilde \phi} \right)
 \Gamma\left(-{i P\over b} +{1\over 2} \right) \left({|\mu| \over
 4} \right)^{{iP\over b}} |P \pm \rangle\ .
 }}
These formulas for the boundary states agree with the semiclassical limit of
\superFZZTg.

\appendix{B}{Tests with the one-matrix model}

In this appendix, we will study in detail the $(2,2m-1)$ bosonic minimal
models, which can be described very simply in terms of the $m$th critical point
of a one-matrix model. This will serve as a useful check of the general
analysis of the bosonic $(p,q)$ models in sections 3 and 4 and the discussion
of the matrix model in section 9. In the one matrix model, there is only one
resolvent, so we will focus on the FZZT brane and not its dual. The resolvent
is given by \onematbos\ with $q=2m-1$:
\eqn\onematres{
    y_m=(\sqrt{\mu})^{m-1/2}{T_{m}(x)+T_{m-1}(x)\over
    \sqrt{1+x}}\ ,
}
where here we have restored the overall power of $\mu$ in front of $y$ for the
purposes of the present discussion. Our goal will be to explicitly confirm
\onematres\ using the one-matrix model. The method will be as follows. Turning
on $\mu$ in the minimal string theory corresponds to turning on a specific set
of scaling perturbations in the matrix model. Using the techniques of \MooreIR,
we can perform the change of basis from $\mu$ to scaling perturbations of the
matrix model, and then explicitly compute the resolvent at non-zero $\mu$,
confirming \onematres.

Exactly at the $m$th critical point of the one-matrix model, the resolvent
takes the form $y=(\sqrt{\mu})^{m-1/2} x^{m-1/2}(x-b)^{1/2}$. A general
perturbation around the $m$th critical point can be written as
\eqn\ymperturb{
y=\sum_{j=1}^{m}t_j y_j\ ,
}
where the $t_j$ are the couplings to the $j$th matrix model scaling operator,
and the $y_j$ are given by (see e.g. section 2.2 of \DiFrancescoNW):
\eqn\ymscj{
    y_j(x)={b^{1/2}(\sqrt{\mu})^{j-1/2}\over B(j,{1\over2})}\int_x^a
    ds\,(s-x)^{-1/2}s^{j-1}\ .
}
Here we are using the standard ``Gelfand-Dikii" normalization of the couplings
$t_j$, so that the string equation takes the form
\eqn\stringeqn{
\sum_j t_j R_j[u]=0\ .
}
Now, in order to compare with the continuum description in the ``conformal
background," i.e.\ with a fixed cosmological constant $\mu$, we must express
the $t_j$ in terms of $\mu$. Explicit formulas for this change of basis were
derived in \MooreIR:
\eqn\tjtauj{\eqalign{
    t_{m-2p} &= {c_{m-2p}\over a_{m-2p}}\mu^p\cr
    c_{m-2p}&={(-1)^{m+1}\pi\over \sqrt{8}}{2^{m-2p}\over
    (m-2p)!\,p\,!\,\Gamma(p-m+{3\over2})}\cr
    a_{m-2p}&={(-1)^m\over 2^{2(m-2p)}} \left(\matrix{2m-4p-1\cr m-2p}\right)\ .\cr
}}
Here $a_k$ is the leading coefficient of the Gelfand-Dikii polynomial $R_k[u]$,
i.e.\ $R_k[u]=a_k u^k+\dots$. We must include this factor when using the
results of \MooreIR, because there a slightly different normalization of the
couplings $t_j$ was used, in which the (genus zero) string equation has the
form $\sum_i t_j u^j=0$. After some gamma function manipulations, we obtain the
following simplified form of $t_{m-2p}$:
\eqn\tjsimp{
    t_{m-2p}={(-1)^{m+1}\sqrt{2\pi}\over 4(m-p)^2-1}{2^{m-2p+1}\over
    B(p+1,-{1\over2}-m+p)}\mu^p\ .
}
Substituting our formula for $t_j$ in \ymperturb, we find the form of the
resolvent in the conformal background:
\eqn\ymconfbg{\eqalign{
    y(x)&=(-1)^{m+1}b^{1/2}(\sqrt{\mu})^{m-1/2}\sqrt{2\pi}\int_x^a
    ds\,(s-x)^{-1/2}\times \cr
    &\quad \sum_{p=0}^{\left[{m-1\over2}\right]}{1\over 4(m-p)^2-1}{2^{m-2p+1}\over
    B(m-2p,{1\over2})B(p+1,p-m-{1\over2})}s^{m-2p-1}\ .\cr
}}
Using the identity
\eqn\betaident{
{(-1)^{p+1}2^{2m-2p}\over B(m-2p,{1\over2})B(p+1,p-m-{1\over2})}=
\left(4(m-p)^2-1\right)\left(\matrix{2m-2p-2\cr m-1}\right)\left(\matrix{m-1\cr
p}\right)
}
in \ymconfbg, we obtain
\eqn\ymconfbgsimp{\eqalign{
y(x)&=(-1)^{m}b^{1/2}(\sqrt{\mu})^{m-1/2}\sqrt{2\pi}\int_x^a
    ds\,(s-x)^{-1/2}\times \cr
    &\quad {1\over 2^{m-1}}\sum_{p=0}^{\left[{m-1\over2}\right]}(-1)^{p}
    \left(\matrix{2m-2p-2\cr m-1}\right)\left(\matrix{m-1\cr p}\right)s^{m-2p-1}\ .\cr
}}
We recognize the sum in \ymconfbgsimp\ as an explicit form of the Legendre
polynomial $P_{m-1}(x)$. Furthermore, the integral can be evaluated using
standard tables assuming $a=-1$. Thus we arrive at the final form of the matrix
model curve:
\eqn\ymconfbgfin{\eqalign{
y(x)&=b^{1/2}(-1)^{m+1}\sqrt{2\pi}(\sqrt{\mu})^{m-1/2}\int_{-1}^x
    ds\,(s-x)^{-1/2}P_{m-1}(s)\cr
    &={(-1)^m\sqrt{8\pi}\over 2m-1}i b^{1/2}(\sqrt{\mu})^{m-1/2}
    \left({T_{m}\left({x}\right)+
    T_{m-1}\left({x}\right)\over\sqrt{1+{x}}}\right)\ ,
}}
which is indeed in exact agreement with the continuum prediction \onematres, up
to an irrelevant overall normalization that can be absorbed into the definition
of $b$.

\listrefs
\end